\documentclass[smallextended]{svjour3}       
\smartqed  
\usepackage{graphicx}
\journalname{}
\usepackage{listings, color}
\usepackage{wrapfig}
\usepackage{hyperref}
\usepackage{xspace}

\definecolor{backcolour}{rgb}{0.95,0.95,0.92}
\lstdefinestyle{smallStyle}{
  numbers=left,
  backgroundcolor=\color{backcolour},
  breaklines=true,
  basicstyle=\scriptsize,
  breakatwhitespace=true,
  xleftmargin=3mm,
  numbersep=4pt
}

\AtBeginDocument{%
	\mathchardef\mathcomma\mathcode`\,
	\mathcode`\,="8000
}
{\catcode`,=\active
	\gdef,{\mathcomma\discretionary{}{}{}}
}
\mathchardef\breakingcomma\mathcode`\,
{\catcode`,=\active
	\gdef,{\breakingcomma\discretionary{}{}{}}
}

\newcommand\revision[1]{#1\xspace}

\graphicspath{ {./figures/} }

\begin{document}

\title{Multi-Dimensional Event Data in Graph Databases
}
\subtitle{Representing, Querying, and Process Mining}


\author{Stefan Esser \and
Dirk Fahland}


\institute{Stefan Esser \at
              INFORM GmbH, Aachen, Germany \\
              \email{stefan.esser@inform-software.de}           
           \and
           Dirk Fahland\at
              Eindhoven University of Technology, Eindhoven, the Netherlands \\
              \email{d.fahland@tue.nl}
}

\date{Received: date / Accepted: date}

\maketitle

\begin{abstract}
Process event data is usually stored either in a sequential process event log or in a relational database. While the sequential, single-dimensional nature of event logs aids querying for (sub)sequences of events based on \emph{temporal relations} such as ``directly/eventually-follows'', it does not support querying \emph{multi-dimensional} event data of multiple related entities. Relational databases allow storing multi-dimensional event data but existing query languages do not support querying for sequences or paths of events in terms of temporal relations. In this paper, we propose a general data model for multi-dimensional event data based on \emph{labeled property graphs} that allows storing structural and temporal relations in a single, integrated graph-based data structure in a systematic way. We provide semantics for all concepts of our data model, and generic queries for modeling event data over multiple entities that \emph{interact synchronously and asynchronously}. The queries allow for efficiently converting large real-life event data sets into our data model and we provide 5 converted data sets for further research. We show that typical and advanced queries for retrieving and aggregating such multi-dimensional event data  can be formulated and executed efficiently in the existing query language Cypher, giving rise to several new research questions. Specifically aggregation queries on our data model enable process mining over multiple inter-related entities using off-the-shelf technology.

\keywords{process mining \and event log \and multi-dimensional processes \and querying \and labeled property graphs \and graph databases}
\end{abstract}

\section{Introduction}\label{sec:introduction}


Retrieving and aggregating subsets of event data of a particular characteristic is a recurring activity in process analysis and process mining~\cite{van2016process}. Each \emph{event} is thereby defined by an \emph{event classifier} such as the \emph{activity} or state that was recorded, a \emph{case identifier} referring to the object or entity where the activity was carried out, and a \emph{timestamp} or \emph{ordering} attribute defining the order of events.

If all events use the same single case identifier attribute, then the event data has a {\emph{single behavioral dimension}}. It can be stored in an \emph{event log} as one \emph{sequence of events} per case according to the data model of the XES-Standard~\cite{ieee_xes_standard}. Such sequences can be easily queried for \emph{behavioral properties} such as \emph{event (sub-)sequences} or \emph{temporal relations} such as ``directly/eventually-follows'' in combination with other data attributes~\cite{DBLP:journals/eswa/BottrighiCLMT16,DBLP:conf/icdt/DeutchM09,Liu:2009:SEC:1944968.1944974,DBLP:conf/otm/RaimCMMM14,DBLP:conf/edoc/SongWWWTK11,DBLP:journals/information/TangMS18}. \emph{Aggregating} directly/eventually-follows relations between events 
is fundamental for discovering process models from event logs~\cite{van2016process,DBLP:journals/tkde/AugustoCDRMMMS19,DBLP:journals/is/WeerdtBVB12}.

Most processes in practice however involve multiple inter-related entities. In such data, each event is directly or indirectly linked to multiple different case identifiers and is part of multiple behaviors or dynamics~\cite{DBLP:conf/bpm/JansS17}. The data has \emph{multiple behavioral dimensions}. Sequential event logs cannot represent these multiple behavioral dimensions correctly~\cite{DBLP:journals/tsc/LuNWF15}. Relational databases (RDBs) can store \emph{1:n and n:m relations between events and case identifiers} and among different case identifiers. {However, tables in an RDB cannot represent sequences of events of arbitrary length explicitly. The explicit behavioral information is lost and hence cannot be queried in a natural way~\cite{DBLP:conf/bpm/MurillasRA16,DBLP:journals/sosym/MurillasRA19}}.

\paragraph{State of the art.}
Although sequential event logs and RDBs enjoy standardized data models built on a few basic concepts, they cannot easily be combined for correct information representation \emph{and} intuitive querying.
Extracting event data from RDBs into sequential event logs requires large non-intuitive queries~\cite{DBLP:conf/bpm/JansS17,DBLP:journals/sosym/MurillasRA19} and introduces false information~\cite{DBLP:journals/tsc/LuNWF15}. Direct behavioral queries on RDBs are limited to a single behavioral dimension~\cite{DBLP:journals/dpd/DijkmanGSDGH20,DBLP:conf/caise/SchonigRCJM16}. Extensions that integrate relational data into the event log or relate case identifiers of multiple sequential event logs to each other ~\cite{DBLP:conf/sefm/Aalst19,DBLP:conf/caise/LiMCA18,DBLP:journals/tsc/LuNWF15,DBLP:journals/ijcis/PopovaFD15,DBLP:journals/sosym/MurillasRA19} cannot be queried across multiple behavioral dimensions due to the strict sequential nature of event logs.

\emph{Graph-based data models} for event data~\cite{DBLP:journals/spe/BeheshtiBM18,DBLP:conf/simpda/BertiA19,esser2019storing,DBLP:journals/tsc/WernerG15} have been proposed as alternative to overcome these limitations. Graphs can describe relations between various entities \emph{and} sequential information as paths in one data structure. While all existing graph-based data models share similarities, no work systematically identified a \emph{minimal} set of core concepts and the necessary semantics needed to model, query, and aggregate event data over multiple behavioral dimensions; see Sect.~\ref{sec:background} for details.

\paragraph{Research problem.} In this paper, we approach the problem of identifying a generally applicable model of event data in a multi-dimensional setting. The specific problem is to \emph{identify a minimal set of core concepts for a data model of multi-dimensional process event data with clearly defined semantics} to fully (1) model, (2) query, and (3) aggregate all kinds of real-life process event data suitable for process analysis. {That data model and queries have to support known requirements for analyzing multi-dimensional event data~\cite{Gonzalez_2019_pm_db_phd,DBLP:journals/tsc/LuNWF15,DBLP:conf/bpm/MurillasRA16} and be realizable in existing off-the-shelf \emph{graph database systems}~\cite{robinson2013graph}.}

\paragraph{Method.} First, we determined the process event data concepts and requirements any data model had to support based on literature (see Sect.~\ref{sec:background:event_log_concepts} and Sect.~\ref{sec:background:multi-dim-req}). We then analyzed existing data models regarding these requirements (see Sect.~\ref{sec:background:multi-dim-literature}). All recent works that succeed in modeling (some) aspects of multi-dimensional event data employ graph properties. We therefore took the most complete proposal~\cite{esser2019storing} based on \emph{labeled property graphs} (LPGs, see Sect.~\ref{sec:background:lpg}) as a starting point.

{To ensure our model would support real-life process event data, we identified from a collection of public real-life event logs\footnote{\url{https://data.4tu.nl/repository/collection:event\_logs\_real}} 5 data-sets with unique multi-dimensional characteristics that can serve as benchmark: multiple entities interact via shared common entity (BPIC14~\cite{BPIC2014}); multiple entities interact asynchronously, based on click-stream data (BPIC16~\cite{BPIC2016}), based on a case management system (BPIC17~\cite{BPIC2017}\footnote{BPIC18~\cite{BPIC2018} originates from a similar kind of system and has the same characteristics as BPIC17 and hence was excluded}), based on ERP system data (BPIC19~\cite{BPIC2019}); multiple event logs of the same processes executed in different organizations, BPIC15~\cite{BPIC2015}. A data model has to allow modeling, querying, and aggregating at least these 5 data sets in their multi-dimensional nature.}

We then iteratively developed a data model and queries that could support all 5 benchmark data sets on the existing graph database system Neo4J (\url{neo4j.com}); Neo4j was chosen for LPG storage and querying due to off-the-shelf availability and suitable performance. {We used an \emph{Extract-Load-Transform}~\cite{ELT_10.14778/1687553.1687576,MARINORTEGA2014667} approach:}
\begin{enumerate}{
    \item We \emph{extracted} the event data into a standard event table where each record describes one event and its properties, including references to all entities involved. All event data can be extracted into this format; see Sect.~\ref{sec:background:multi-dim-literature}.
    \item We \emph{loaded} all events into the GDB as an LPG of unrelated raw event nodes.
    \item We identified the domain concepts stored in each data set, specifically entities and relations between entities and events stored in the event attributes.
    \item We iteratively identified semantic node types and relationship types for LPGs to abstract the domain concepts identified in step 3.  Our semantic concepts for nodes and relations thereby had to serve as adequate abstractions so that all event data could be queried and analyzed through these semantic abstractions only.
    \item We then developed queries to \emph{transform} the raw events of step 1 into a graph that uses the identified LPG concepts of step 4.}
    \item In case a suitable solution (node types, relationship types, queries) was found for one data set, we applied it on all other data sets. If the solution could not be applied on one data set, we identified the cause and generalized the concepts and queries and repeated steps 3-5 for all data sets.
\end{enumerate}
We conducted over 100 iterations of the above process over all 5 data sets until reaching a fixed point.

\paragraph{Contribution and Results.} We contribute a generally applicable, minimal, integrated data model for {\emph{event data in multiple behavioral dimensions} in labeled property graphs. We identified
\begin{itemize}
    \item[1.] \emph{4 semantic node types} for multi-dimensional event data in LPGs: (i) events, (ii) entities, (iii) logs, and (iv) event classes;
    \item[2.] \emph{3 semantic structural relations} for relating each event (i) to one or more entities, (ii)  to exactly one log, and (iii) to one or more event classes;
    \item[3.] and \emph{2 semantic behavioral relations} describing (i) directly-follows between two events (along a chosen entity), and its congruent (ii) directly-follows relation between event classes (summarizing event-level directly follows on class level)
\end{itemize}}
See Sect.~\ref{sec:represent} for the concepts and Sect.~\ref{sec:semantics} for their semantic definition in terms of LPGs. Our model allows querying and aggregating the modeled multi-dimensional event data and thereby subsumes and exceeds several prior works. {
\begin{itemize}
    \item[4.] We identified queries to extract entities and relations between entities from raw events, and to reify relations into composite entities,
    \item[5.] to correlate events to entities,
    \item[6.] to derive directly-follows relations for all events of an entity or relation.
\end{itemize}}
All the available multi-dimensional event data could be transformed into our model using a standard set of queries; see Sect.~\ref{sec:storing}. The resulting graphs are available for further research~\cite{BPIC2014_graph,BPIC2015_graph,BPIC2016_graph,BPIC2017_graph,BPIC2019_graph}.
\begin{itemize}
    \item[7.] {We show that existing query languages for labeled property graphs using our event data model satisfy all requirements for querying event data over multiple behavioral dimensions known from literature~\cite{Gonzalez_2019_pm_db_phd,DBLP:conf/bpm/MurillasRA16}.}
\end{itemize}
We evaluated the query results to be correct against a manually constructed ground truth. {Query execution times are on par or faster compared to commercially available event data pre-processing techniques on single-dimensional event data. Query execution times perform hand-written algorithms on multi-dimensional event data. We evaluated the expressive power of the query language used against existing process query languages~\cite{DBLP:journals/is/PolyvyanyyPH20,DBLP:journals/corr/abs-1909-09543} and identify potential for further research; see Sect.~\ref{sec:querying}.} We identified{
\begin{itemize}
    \item[8.] queries to aggregate event-level directly-follows relations into a directly-follows graph per entity, including filtering, thereby enabling basic process discovery;
    \item[9.] queries to derive and aggregate directly-follows relations between related entities into a directly-follows graph describing entity interaction (through a reified entity).
    \item[10.] The directly-follows graphs of different entities and relations share nodes. We thereby obtain a method for discovering process models which describe the behavior of a network of related entities and their interactions as a single \emph{multi-viewpoint model}~\cite{DBLP:conf/simpda/BertiA19}.
\end{itemize}}
In Sect.~\ref{sec:mining} we demonstrate discovery of artifact-centric models of multiple entities with asynchronous and synchronous interactions~\cite{DBLP:journals/tsc/LuNWF15}, also called multi-viewpoint models.

All queries realizing the transformations into our data model, querying the data, and discovering process models through graph databases are available at GitHub\footnote{\url{https://github.com/multi-dimensional-process-mining/graphdb-eventlogs}} and at~\cite{graphdataset}. We discuss limitations and alleys for future work in Sect.~\ref{sec:conclusion}.

\section{Background}\label{sec:background}

Information Systems (IS) create and update information records in structured transactions or \emph{activities} throughout process executions. Each update can be recorded as an \emph{event}. Event attributes describe the activity carried out and the \emph{timestamp} (or \emph{ordering} of updates). {Furthermore, each event is linked to one or more \emph{entities} on which the updates occurred via unique \emph{entity identifier} attributes, for example a credit application order and related credit offers. The entities themselves are related to each other via structural 1:1, 1:n, and n:m relations}~\cite{van2016process,DBLP:journals/tsc/LuNWF15}.

We first recall the foundational concepts of single-dimensional process event logs in Sect.~\ref{sec:background:event_log_concepts}. After discussing challenges and requirements for analyzing multi-dimensional event data in Sect.~\ref{sec:background:multi-dim-req}, we discuss the state of the art on modeling, querying, and analyzing multi-dimensional event data in Sect.~\ref{sec:background:multi-dim-literature}, before we recall the data model of labeled property graphs and the query language Cypher in Sect.~\ref{sec:background:lpg}.

\subsection{Modeling Single-Dimensional Event Logs}
\label{ssec:processEventData}
\label{sec:background:event_log_concepts}

A \emph{process event log} is a collection of recorded events $E$ structured into a specific \emph{view} on an information system from the of perspective handling of \emph{one} specific entity, e.g., handling a credit application. Table~\ref{tab:2_BPIC17ExampleCaseTable} shows a simplified event log taken from the BPIC17~\cite{BPIC2017} data set describing the handling of a credit application (identified by attribute \emph{Appl}.).

\begin{table}[]
\revision{
    \begin{center}
    \begin{tabular}{r|l|l|l|l|l|l|l}
    \hline
     & \textbf{Appl.} & \textbf{Activity}  & \textbf{Timestamp}& \textbf{Res.} & \textbf{Amount} & \textbf{oID} & \textbf{Origin} \\\hline
    $e_1$ & 1             & {\bf C}reate {\bf A}ppl.       & 29.08.19 10:30    & 10            & 1000            &                &  A      \\
    $e_2$ & 1             & {\bf A}ppl. {\bf R}eady        & 29.08.19 10:35    & 10            & 1000            &                &   A    \\
    $e_3$ & 1             & {\bf H}andle {\bf L}eads       & 29.08.19 10:40    & 42            & 1000            &                &   W     \\
    $e_4$ & 1             & {\bf C}reate {\bf O}ffer       & 29.08.19 13:14    & 11            & 1000            & 1              & O           \\
    $e_5$ & 1             & {\bf S}end {\bf O}ffer         & 29.08.19 13:35    & 11            & 1000            & 1              & O          \\
    $e_6$ & 1             & {\bf C}reate {\bf O}ffer       & 30.08.19 08:49    & 12            & 1000            & 2              & O           \\
    $e_7$ & 1             & Offer {\bf Ca}nceled     & 30.08.19 09:05    & 10            & 1000            & 1              & O           \\
    $e_8$ & 1             & {\bf S}end {\bf O}ffer         & 30.08.19 09:20    & 12            & 1000            & 2              & O           \\
    $e_9$ & 1             & {\bf C}ontact {\bf C}ust.        & 30.08.19 10:15    & 44            & 1000            &                & W       \\
    $e_{10}$ & 1             & Offer {\bf Ret}urned     & 30.08.19 15:20    & 16            & 1000            & 2              & O
    \end{tabular}

    \end{center}
    \caption{Simplified BPIC'17 Example}
    \label{tab:2_BPIC17ExampleCaseTable}
}
\end{table}

The following process-specific concepts are part of process event logs~\cite{van2016process}.{
\begin{itemize}
    \item[E1] \textbf{Event, activity, and timestamp.} Each event $e \in E$ in an event log records an \emph{atomic} observation of an activity name $e.\mathit{activity}$ that has been observed at a particular point in time $e.time$ (or an \emph{ordering attribute that orders the events in the log}), e.g., \emph{Activity} and \emph{Timestamp} in Tab.~\ref{tab:2_BPIC17ExampleCaseTable}.
    \item[E2] \textbf{Entity identifier, correlation, case.} Each event records at least one identifier $e.\mathit{entityid}$ of some entity, e.g. a specific credit application (\emph{Appl.}) or credit offer (\emph{oID}) in Tab.~\ref{tab:2_BPIC17ExampleCaseTable}. The set of events \emph{correlated} to the same entity identified by $id$ is $\{e_1,\ldots,e_n\} = \{ e \in E \mid e.\mathit{entityid} = id \}$. All events correlated to the same entity belong to the same \emph{case} (or execution) of the process.
    \item[E3] \textbf{Life-cycle information.} An optional attribute $e.\mathit{lifecycle}$ records states of long-running behavior, e.g. an activity \emph{started} or \emph{completed}.
    \item[E4] \textbf{Resource.} An optional attribute $e.\mathit{resource}$ records whether an actor or resource was involved in the event, e.g., \emph{Res.} in Tab.~\ref{tab:2_BPIC17ExampleCaseTable}.
    \item[E5] \textbf{Trace.} The sequence $\langle e_1,\ldots,e_n\rangle$ of all events correlated to an entity ordered by time (or the ordering attribute) is called a \emph{trace} (of this entity), e.g., $\langle e_1,\ldots,e_{10} \rangle$ is  the trace of ``Application 1'' in Tab.~\ref{tab:2_BPIC17ExampleCaseTable}.
    \item[E6] \textbf{Event class.} Besides the activity also other attributes or combinations of attributes can be designated as ``representative'' for an event. The notion of \emph{event class} generalizes the notion of the activity attribute (E1); the analyst can designated one (or more) event attributes as event class $e.\mathit{class}$ depending on the analysis, e.g., we can set $e.\mathit{class} \mathrel{:=} e.\mathit{Res}$ in Tab.~\ref{tab:2_BPIC17ExampleCaseTable} as event class when analyzing handover-of-work behavior between users.
\end{itemize}}
The IEEE XES-Standard~\cite{ieee_xes_standard} materializes these concepts in a tree-structure that specifically pre-determines a unique case identifier and available event classes.

{The behavior recorded in a sequential event log can be analyzed by querying and aggregating the \emph{directly-follows} relation over events $E$~\cite{van2016process}.
\begin{itemize}
    \item[E7] \textbf{Directly-follows (events).} Event $e_b$ \emph{directly follows} $e_a$, $e_a \to e_b$ iff there is a trace $\langle \ldots,e_a,e_b,\ldots \rangle$, e.g., $e_7$ directly follows $e_6$ in Tab.~\ref{tab:2_BPIC17ExampleCaseTable}.
    \item[E8] \textbf{Directly-follows (event classes).} We can aggregate the directly-follows relation over event classes: event class $b$ directly follows event class $a$, $a \to b$ iff there is a trace $\langle \ldots,e_a,e_b,\ldots \rangle$ with $e_a.\mathit{class} = a$ and $e_b.\mathit{class} = b$. For example, \emph{Offer Canceled} directly follows \emph{Create Offer} for $e.\mathit{class} = e.\mathit{Activity}$ and $10$ directly follows $12$ for $e.\mathit{class} = e.\mathit{Res}$.
\end{itemize}}
Nearly all process discovery algorithms create for each event class one activity node~\cite{van2016process}. Then dependencies between {activity nodes in the model} can be derived from the directly-follows relation between activities in the event log~\cite{DBLP:journals/tkde/AugustoCDRMMMS19,DBLP:journals/is/WeerdtBVB12}. 

\subsection{Requirements for Analyzing Multi-Dimensional Event Data}
\label{sec:background:multi-dim-req}

An Information System usually hosts multiple uniquely identifiable entities~\cite{DBLP:journals/tsc/LuNWF15}, e.g., credit applications and offers. For example, the BPIC'17 data (shown in Tab.~\ref{tab:2_BPIC17ExampleCaseTable}) identifies four entities: credit applications (identified by \emph{Appl.}, events with $\mathit{Origin}=A$), credit offers (identified by \emph{oID} with $\mathit{Origin}=O$) with a 1:n relation to Applications, the Workflow (identified by \emph{Appl.} with $\mathit{Origin}=W$); and the actors working on the case (identified by \emph{Res.}) with an n:m relation to Application, Workflow, and Offers.

{Relational databases allow recording events, their correlations to entities, and the 1:n and n:m relations between entities as partly illustrated in Fig.~\ref{fig:event_data_models}(3) and are therefore the default storage for event data in Information Systems. Analysis of the behavior however requires extraction of the data into a sequential format~\cite{DBLP:conf/bpm/JansS17}.}
Extracting a single-dimensional event log (Fig.~\ref{fig:event_data_models}(1)) correlates all events under a single entity (case identifier), e.g. the Application or the Offer document, and \emph{flattens} the data accordingly~\cite{DBLP:conf/bpm/JansS17} leading to false behavioral information called \emph{convergence} and \emph{divergence}~\cite{DBLP:conf/sefm/Aalst19,DBLP:journals/tsc/LuNWF15}.

\textbf{Convergence.} Flattening the data in Fig.~\ref{fig:event_data_models}(3) under \emph{Application} de-normalizes the 1:n relation to \emph{Offer} and results in the event log of Tab.~\ref{tab:2_BPIC17ExampleCaseTable} and Fig.~\ref{fig:event_data_models}(1):  {We observe \emph{Cancelled} ($e_7$) directly followed by \emph{Send Offer} ($e_8$), suggesting that an offer was sent after it has been cancelled. However, this sequence of activities never happened for any entity in the data: $e_7$ refers to $\mathit{oID} = 1$ whereas $e_8$ refers to $\mathit{oID} = 2$. The entire BPIC17 event log contains 15\% such false directly-follows pairs of events while missing over 50\% of the actual directly-follows pairs. Some events are only involved in false directly-follows pairs; see Appendix~\ref{app:df_bpic17}. The false directly-follows pairs cannot be removed from sequential event logs~\cite{DBLP:conf/sefm/Aalst19} and can amount to over 50\% of all pairs rendering the analysis useless~\cite{DBLP:journals/tsc/LuNWF15}}.

\textbf{Divergence.} Flattening the data in Fig.~\ref{fig:event_data_models}(3) under \emph{Offer} (\emph{oID}) via the n:1 relation replicates events on the 1-side for each entity on the n-side, see Fig.~\ref{fig:event_data_models}(1). The two traces for $o1$ and $o2$ contain the events $e_1,e_2,e_3,e_9$ which are only indirectly correlated to the offer (via their parent application). The resulting event log contains multiple copies of the same event and more directly-follows pairs than exist in reality~\cite{DBLP:conf/sefm/Aalst19}.

To avoid both phenomena, we formulate requirement \textbf{(R0)}: Any event data model for event data over multiple related entities has to describe directly-follows relations between pairs of events $(e_1,e_2)$ only along the entities to which $e_1$ and $e_2$ are both correlated.



A recent literature survey of 95 studies~\cite{Gonzalez_2019_pm_db_phd,DBLP:conf/bpm/MurillasRA16} established further requirements for \emph{modeling and querying} event data. {From this survey, we identified the following requirements that specifically address \emph{querying for structure and behavior in multi-dimensional event data} ~\cite[pp.133]{Gonzalez_2019_pm_db_phd}:} \textbf{(R1)} to query and analyze events (E1 of Sect.~\ref{sec:background:event_log_concepts}), and  \textbf{(R2)} to consider relations between multiple data entities (as in RDBs). The technique shall support \textbf{(R3)} storing and querying business process-oriented concepts (E3-E4) and \textbf{(R4)} capture information about how events are correlated to different entities (E2) to avoid convergence and divergence (generalize E5-E8 so that R0 is satisfied).

According to~\cite{Gonzalez_2019_pm_db_phd,DBLP:conf/bpm/MurillasRA16}, queries should \textbf{(R5)} be expressed as graphs to specify the behavior of interest in a natural way, \textbf{(R6)} allow to query paths (or sequences) of events (connected by some relation), \textbf{(R7)} allow to select individual cases based on partial patterns, \textbf{(R8)} allow to query temporal properties (such as directly/eventually-follows), \textbf{(R9)} correlate events related to the same entity, \textbf{(R10)} allow querying aspects related to several entities or processes at the same time on the same data set, and \textbf{(R11)} allow to query multiple event logs and combine results.

Prior work on analyzing multi-dimensional event data~\cite{DBLP:journals/tsc/LuNWF15,DBLP:conf/apn/Fahland19} identified four major aggregation operations for discovering so-called \emph{artifact-centric process models}. The technique has to support \textbf{(R12)} aggregating events {into \emph{user-defined activities} viz.\ event classes (see E6 in Sect.~\ref{sec:background:event_log_concepts}) based on event properties (e.g., to distinguish different types of ``update'' activities based on the nature of the update). Moreover, it has to support \textbf{(R13)} materializing a relation between two events as a new derived entity, as this allows to model, query and aggregate \emph{interactions} between different entities. Further, it has to allow} \textbf{(R14)} aggregating directly-follows relations from the event level to the \emph{event class} level \emph{per entity type}, and \textbf{(R15)} relating or synchronizing aggregated behavior of different entity types.

Altogether, \emph{a user shall be able to query and aggregate for individual events (and their properties), for different entities/case notions, for behavioral and structural relations, and for patterns of multiple events (within and across entities)}.

\subsection{Related work}
\label{sec:background:multi-dim-literature}

We review 5 existing types of data models for event data against the requirements of Sect.~\ref{sec:background:multi-dim-req} showing that \emph{no existing data model or query language on sequential event logs or RDBs satisfies R0-R15.} {Moreover, we discuss related work on behavioral  process query languages.}

\begin{figure}[t]
    \centering
    \includegraphics[width=\linewidth]{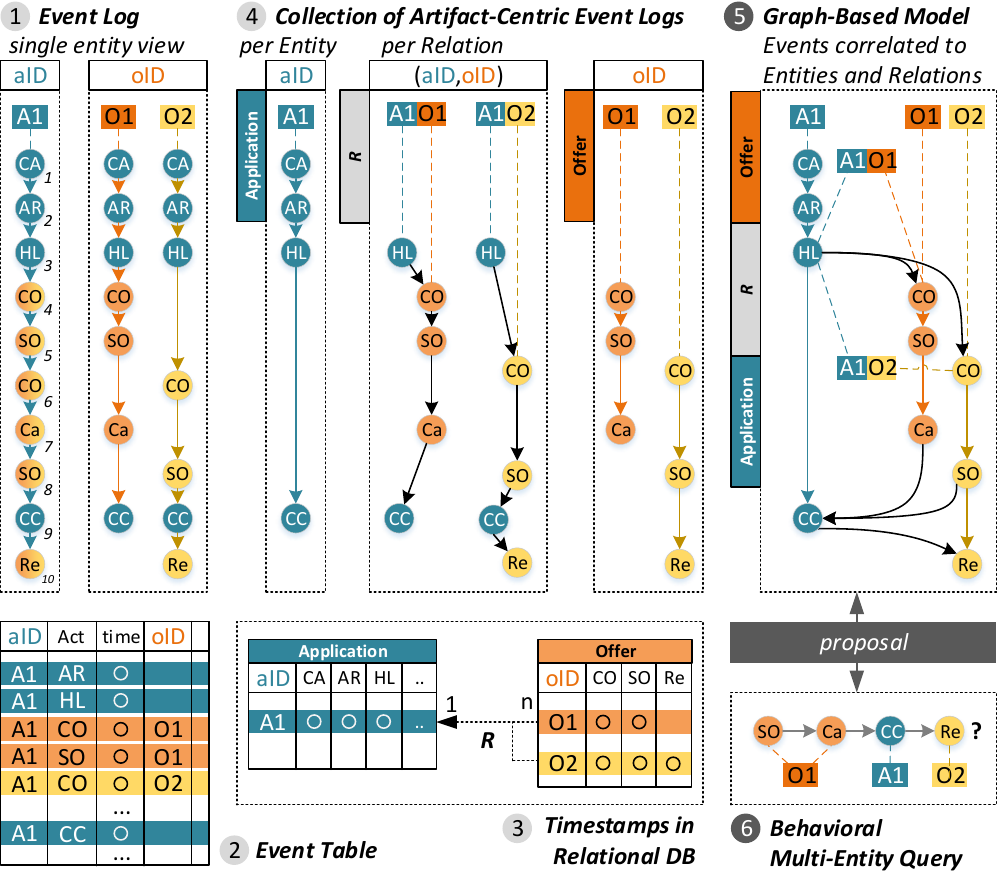}
    \caption{Illustration of Event Data Models on the example events of Tab.~\ref{tab:2_BPIC17ExampleCaseTable}}
    \label{fig:event_data_models}
\end{figure}

\paragraph{\#1. Single event log for a single, selected entity.} Event logs as described in Sect.~\ref{sec:background:event_log_concepts} and illustrated in Fig.~\ref{fig:event_data_models}(1) cannot correctly model or aggregate behavior over events related to multiple entities due to convergence and divergence as discussed in Sect.~\ref{sec:background:multi-dim-req}{, thus violating R0}. Sequential event logs can be stored and queried using files~\cite{ieee_xes_standard} or through RDBs~\cite{DBLP:conf/simpda/SyamsiyahDA16b}. {XES event logs fall into this type of data models~\cite{ieee_xes_standard}.}

Of the 95 works surveyed~\cite[pp.133]{Gonzalez_2019_pm_db_phd}, several approaches exist to retrieve cases from event logs for temporal properties~\cite{Liu:2009:SEC:1944968.1944974,DBLP:conf/edoc/SongWWWTK11}, for most frequent behavior~\cite{DBLP:conf/icdt/DeutchM09}, for sequences of activities~\cite{DBLP:journals/eswa/BottrighiCLMT16} or algebraic expressions of sequence, choice, and parallelism over activities~\cite{DBLP:journals/information/TangMS18}, or to check whether a temporal-logic property holds~\cite{DBLP:conf/otm/RaimCMMM14}. Several techniques support graph-based queries~\cite{DBLP:conf/sc/Cuevas-VicenttinDWSL12,DBLP:conf/icde/HuangBDMY15,Liu:2009:SEC:1944968.1944974}.
These techniques satisfy R7 and R8. However, they only support a single fixed case notion and thus fail R2, R10, R11.

\paragraph{\#2. Event table with multiple entity identifiers.} {The model
is a variant of Tab.~\ref{tab:2_BPIC17ExampleCaseTable} and Fig.~\ref{fig:event_data_models}(2) and defines a single table.} Each record is an event with \emph{multi-valued} entity identifier attributes. The model was first introduced by Popova et al.~\cite{DBLP:journals/ijcis/PopovaFD15} and later formalized by Aalst et al. as \emph{object-centric log}~\cite{DBLP:conf/sefm/Aalst19}. \emph{Redo} event logs~\cite{DBLP:conf/bpm/MurillasHR17} include database operations.
\emph{XOC} logs~\cite{DBLP:conf/caise/LiMCA18} even include database snapshots.
The BPAF~\cite{bpaf_standard} format is a precursor that allows querying event data of different processes~\cite{DBLP:conf/ic3k/BaqueroM12a} but not based on properties of specific events (violates R7). All these models only describe correlation (and data operations) of an event to multiple entities, but leave sequential ordering per entity implicit, see Fig.~\ref{fig:event_data_models}(2), {which violates R0} and prevents R6 and R8. They are usually transformed to other formats for analysis~\cite{DBLP:conf/simpda/BertiA19,DBLP:journals/ijcis/PopovaFD15}.

\paragraph{\#3. Event data in a relational database.} Events are stored as time-stamped attributes and can be related to various entities through primary and foreign keys as shown in Fig.~\ref{fig:event_data_models}(3). Dijkman et al.~\cite{DBLP:journals/dpd/DijkmanGSDGH20} show a native, efficient relational algebra operator to query directly-following events. Also pre-defined behavioral patterns can be queried efficiently~\cite{DBLP:conf/caise/SchonigRCJM16}. However, these operators are fixed to one entity identifier and querying paths requires unbounded joins (violates R4,R6,R10).

\paragraph{\#4. Multiple logs, one per entity and per relation.} Convergence and divergence can be avoided by extracting one log per entity providing multiple views~\cite{DBLP:conf/sefm/Aalst19,DBLP:journals/tsc/LuNWF15,DBLP:journals/sosym/MurillasRA19,DBLP:journals/ijcis/PopovaFD15}, thus satisfying R0. {For example, the Fig.~\ref{fig:event_data_models}(4) shows a log for the Application and a log for both Offers}. Sets of logs with disjoint sets of events can be extracted automatically by partitioning the relational schema~\cite{DBLP:journals/tsc/LuNWF15,DBLP:journals/ijcis/PopovaFD15}.

Interactions between entities can be modeled by extracting a sequential log per relation: per record in the relation, order all events correlated to the entities in the relation by time, the obtained trace describes the entities' interaction~\cite{DBLP:journals/tsc/LuNWF15}. {For example in Fig.~\ref{fig:event_data_models}(4), the log for the relation $R$ contains traces for the interaction $(A1,O1)$ between the Application and Offer 1 and for the interaction $(A1,O2)$ between the Application and Offer 2}. Extraction of an interaction log corresponds to reifying the relation into an entity which overlaps and synchronizes with other entities~\cite{DBLP:conf/apn/Fahland19}. The approach of~\cite{DBLP:journals/sosym/MurillasRA19} provides a meta-model to extract event logs of different perspectives from user-defined, composite entities~\cite{DBLP:journals/sosym/MurillasRA19} that also may overlap. However, the separation into multiple event logs violates R9 and R10, and if logs do not overlap also R15.

DAPOQ~\cite[Ch.7]{Gonzalez_2019_pm_db_phd} generalizes various prior query languages to query and extract events in the context of their relational data model for behavior properties, but does not support retrieving individual cases (R7) or specifying behavioral and structural patterns (R8).

\paragraph{\#5. Graph-based, events as nodes related to multiple entities.}
{One of the first graph-based data models for event data is built on RDF~\cite{DBLP:journals/spe/BeheshtiBM18,DBLP:journals/dpd/BeheshtiBM16,DBLP:conf/bpm/BeheshtiBNS11}. The data model expresses events (as nodes) and relations between events and entities, and between different entities (as edges). This graph can be queried using SPARQL, satisfying R1-R5 and R9,R10. A scalable service-oriented architecture allows to retrieve event logs through a regular-expression describing the behavior of interest wrt. various entities~\cite{DBLP:journals/spe/BeheshtiBM18} which satisfies R7 and R8. OLAP operations allow exploring multiple data dimensions~\cite{DBLP:journals/dpd/BeheshtiBM16}. The technique however always materializes the result as paths wrt.\ a single entity identifier (violates R0). Our proposed data model bears many similarities to this data model. The main difference is that we explicitly store events and directly-follows pairs  between events wrt.\ the entities they are correlated to. We thereby obtain an explicit partial-order event data model that treats events and entities as first-class citizens as shown in Fig.~\ref{fig:event_data_models}(5). This allow us to model and query multiple \emph{behavioral dimensions} at once, for example a path \textit{Send Offer} $\to$ \textit{Offer Cancelled} $\to$ \textit{Contact Customer} $\to$ \emph{Offer Returned} over 3 different entities as shown in Fig.~\ref{fig:event_data_models}(6). In contrast, the classical notion of multi-dimensionality in event data refers to attributes of entities and events~\cite{DBLP:journals/dpd/BeheshtiBM16}. Finally, we use labeled property graphs (LPGs) instead of RDF, which allows to reduce the graph size as LPGs allow storing key/value pairs within a node or relationship.}

Werner et al.~\cite{DBLP:journals/tsc/WernerG15} modeled behavior over two entity types in financial auditing as a \emph{graph over events} describing ``directly-follows'' per entity or relation (satisfying R0-R4), but their model was not generalized or used for querying (violates R5-R11). Our graph-based model~\cite{esser2019storing} shown in Fig.~\ref{fig:event_data_models}(c) generalizes the model of Werner et al.~\cite{DBLP:journals/tsc/WernerG15} to standard process concepts (Sect.~\ref{sec:background:event_log_concepts}) using labeled property graphs and Cypher~\cite{francis2018cypher}.

Berti et al.~\cite{DBLP:conf/simpda/BertiA19} convert object-centric logs (format \#2) into two separate graphs: one describes correlation of events to entities, and one describes the directly-follows relation between any two events per entity similar to~\cite{esser2019storing,DBLP:journals/tsc/WernerG15}.
Assuming that all relations between entities have been reified into entities (see \#4), they aggregate the directly-follows relations per entity and satisfy R14,R15. However, in most event data in practice~\cite{BPIC2017,DBLP:journals/tsc/LuNWF15}, such as Fig.~\ref{fig:event_data_models}, relations are not reified yet, limiting the applicability of~\cite{DBLP:conf/simpda/BertiA19} as it does not support R13. Further, the model does not support querying the event data prior to aggregation (violates R6, R7, R8, R10) because correlation and directly-follows relations are stored in separate graphs.

{In a prior exploratory case study~\cite{esser2019storing}, we presented  an integrated data model for BPIC17~\cite{BPIC2017} based on \emph{labeled property graphs}. We used edges to correlate events to entities and to model ``directly-following'' events per entity as shown in Fig.~\ref{fig:event_data_models}(5). We used \emph{graph query languages} of existing \emph{graph database systems}~\cite{robinson2013graph} to answer behavioral multi-entity queries (Fig.~\ref{fig:event_data_models}(6)) and for aggregating directly-follows relations for social network analysis, satisfying R0-R11 and R13. However, our data-model did not provide a generic data model for various real-life data sets and did not support R11, R12, and R14.}

In this paper, we generalize our previously explored integrated, graph-based model~\cite{esser2019storing} shown in Fig.~\ref{fig:event_data_models}(5) to be applicable to all kinds of real-life data sets while satisfying R0-R11 and additionally R12-R15, thereby subsuming prior works.

\paragraph{Behavioral process query languages.} {The problem of querying behavior has also been studied for the use case of retrieving a process model $M$ for a model collection $\mathcal{M}$. The query language PQL allows specifying behavior in the form of example patterns~\cite{DBLP:journals/is/PolyvyanyyPH20} or behavioral relations~\cite{DBLP:journals/corr/abs-1909-09543}. The query returns all models $M$ whose behavior intersects with the specified behavior. In contrast, querying event logs means to retrieve \emph{all} sub-graphs  over events, i.e., all concrete process executions, that match the specified behavior. Moreover, in our use case the behavior is explicitly stored whereas in process model querying the behavior first has to be (partially) computed from the process model syntax. Although both use cases are significantly different, the behavioral query constructs developed in PQL are of general nature and could be applied to this setting. However, PQL is currently not designed to be evaluated against an event data model, whereas Cypher and Neo4J are available off-the-shelf techniques. We discuss how our query language compares to PQL conceptually in Sect.~\ref{sec:querying}.}

\subsection{Labeled property graphs and querying}
\label{ssec:LPGandGraphDB}
\label{sec:background:lpg}

We recall basic concepts and notation for labeled property graphs and for the query language Cypher, which we use in subsequent sections.

A \emph{Labeled Property Graph} (LPG) is a data structure used in graph databases (GDBs)~\cite{robinson2013graph}. {Let $K$ be a set of \emph{keys}, $V$ be a set of \emph{values} and $\mathit{Label}$ be a set of labels, $\mathit{Label} \cap V = \emptyset$.} An LPG $G = (N,R,\mathit{label},\mathit{prop})$  consists of \emph{nodes} $N$ (vertices) and \emph{relationships} $R$ (edges) where each relationship $r\in R$ defines a directed edge $\overrightarrow{r} = (n_1,n_2) \in N \times N$ between two nodes. The labeling function $\mathit{label} : N \cup R \to 2^\mathit{Label}$ assigns to each node and each relationship a non-empty set of labels designating their type{, e.g., $\mathit{label}(n) = \{\mathit{Person}, \mathit{Student}\}$}. Function $\mathit{prop} : (N \cup R) \times K \to V$ assigns each node or relationship an arbitrary number of key-value pairs, called \emph{properties}. For the value $\mathit{prop}(x,k) = v$ of a property key $k$ of a node or relationship $x \in N \cup R$, we also write $x.k = v$, and $x.k = \perp$ if $k$ is undefined for $x$.

\begin{wrapfigure}{r}{.55\linewidth}
    \includegraphics[scale=.6]{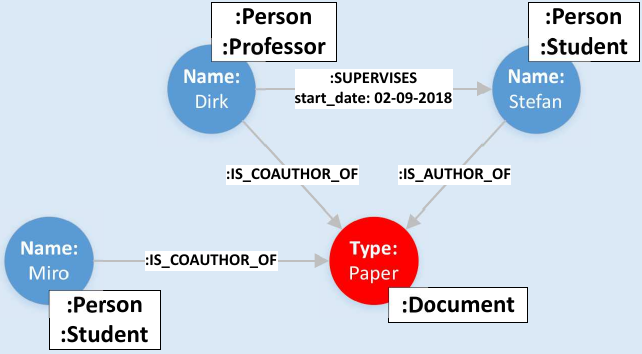}
    \caption{Labeled Property Graph}
    \label{fig:Example}
\end{wrapfigure}
The example LPG in Fig.~\ref{fig:Example} models the relationships between a professor and two students. The example contains nodes with the \emph{labels} \textit{:Person}, \textit{:Professor}, \textit{:Student} and \textit{:Document}. The document you are currently reading is authored by Stefan, a student supervised by Dirk who co-authors this document and say Miro is another student contributing to this paper. The ``Name'' of each person is a property of the \textit{:Person} nodes; ``Type'' is a property of \textit{:Document} nodes. The described relationships between the nodes can also hold properties like the starting date of a supervision. 
Neo4j supports multiple labels for nodes while relationships have exactly one label.

\emph{Cypher} is a language for querying LPGs~\cite{francis2018cypher} and supported by Neo4j. Cypher queries use pattern matching to select sub-graphs of interest. The Cypher pattern $(\mathit{n : \mathit{Label}}\ \{\mathit{Prop}: \mathit{Value}\})$ matches any node labeled
$\mathit{\mathord{:}Label}$ that has property $\mathit{Prop}$ set to $\mathit{Value}$. Pattern $(n)-[r\mathord{:}\mathit{Label}]\mathord{-}\mathord{>}(m)$ matches any relationship labeled $\mathit{\mathord{:}Label}$ from a node $n$ to a node $m$. Any combination of nodes and relations $(n,r,m)$ that match the pattern are included in the result set; if any variable $n,r,m$ is already bound, then only combinations including the bound nodes/relationships will be returned. We explain the Cypher query concepts used in this paper by a single (albeit inefficient) example query. For the graph in Fig.~\ref{fig:Example}, we query for the longest path between ``Dirk'' and a student, other than ``Stefan'', who also works on a document that ``Dirk'' co-authors.
\begin{lstlisting}[style=smallStyle]
MATCH path = (s:Student)-[*]-(p:Professor {Name: "Dirk"})
WHERE NOT s.Name = "Stefan"
WITH s AS student, p AS professor, path AS paths
MATCH (d:Document)<-[:IS_COAUTHOR_OF]-(professor:Professor)
WHERE (student:Student)--(d)
RETURN student, d, paths, length(paths) AS pLength
ORDER BY pLength DESC
LIMIT 1
\end{lstlisting}

The $\mathit{MATCH}$ clause retrieves pairs of nodes $s$ and $p$ and a $path$ between $s$ and $p$ that match the pattern in line 1: a Student $s$ related to Professor $p$ by a path  $-[*]-$ of arbitrary relationships, direction, and length. The $\mathit{WHERE}$ clause in line 2 restricts the pattern such that the student's name cannot be ``Stefan''. By defining the professors' name property to be ``Dirk'' in line 1 we also restrict the pattern. $\mathit{WITH}$ in line 3 formats and renames the result set (e.g., $s$ renamed to $\mathit{student}$) that is passed to the next query from line 4 on: variable $\mathit{student}$ in lines 4-6 may only take values retrieved for variable $s$ in lines 1-2, e.g., ``Miro'' but not ``Stefan.'' Line 4 matches the documents Dirk coauthors and line 5 restricts the results to documents that have a direct relationship to a student.

The $\mathit{RETURN}$ statement in line 6 formats the result set of lines 4-5 as output. For the example graph, the $student$ is ``Miro'' and the document $d$ is the ``paper''; variable $\mathit{paths}$ contains the 2 possible paths between Miro and Dirk. One walks over Stefan and one does not. ``Length()'' is a function of Cypher that returns the hops needed to walk a path. Lines 7-8 sort the results by path lengths (\emph{ORDER\ BY} clause) in descending order ($\mathit{DESC}$) and return only the first path of this ordered list (\emph{LIMIT} 1).

Instead of $\mathit{RETURN}$, a Cypher query may also end with a statement $\mathit{CREATE} (student) \mathord{<}\mathord{-}[r\mathord{:}\mathit{FOUND}]\mathord{-}(d)$ to add a new relationship of label \emph{:Found} from node $d$ to node $student$; statement $\mathit{MERGE}$ only creates the specified node/relationship if it doesn't exist yet; see~\cite{Esser2019cs_tue} for more details.

\section{Representing Multi-Dimensional Event Data in Labeled Property Graphs}\label{sec:represent}

Labeled property graphs introduced in Sect.~\ref{ssec:LPGandGraphDB} allow versatile data modeling of various concepts and relations between concepts. In this section, we propose how to model the central concepts and relations of process event data of Section~\ref{sec:background:event_log_concepts} in labeled property graphs. Figure~\ref{fig:complete_schema} summarizes our proposal which we explain in detail below.
In Section~\ref{sec:semantics}, we \emph{constrain} the way how the concepts and relations may occur in a labeled property graph describing event data, thereby defining the \emph{semantics} for process concepts in terms of LPGs. In that section, we also discuss how to \emph{refine} the proposed node and relationship types to aid in the analysis.

\subsection{Modeling Events Related to Multiple Entities or Logs}

\begin{figure}[tb]
    \centering
    \includegraphics[width=\linewidth]{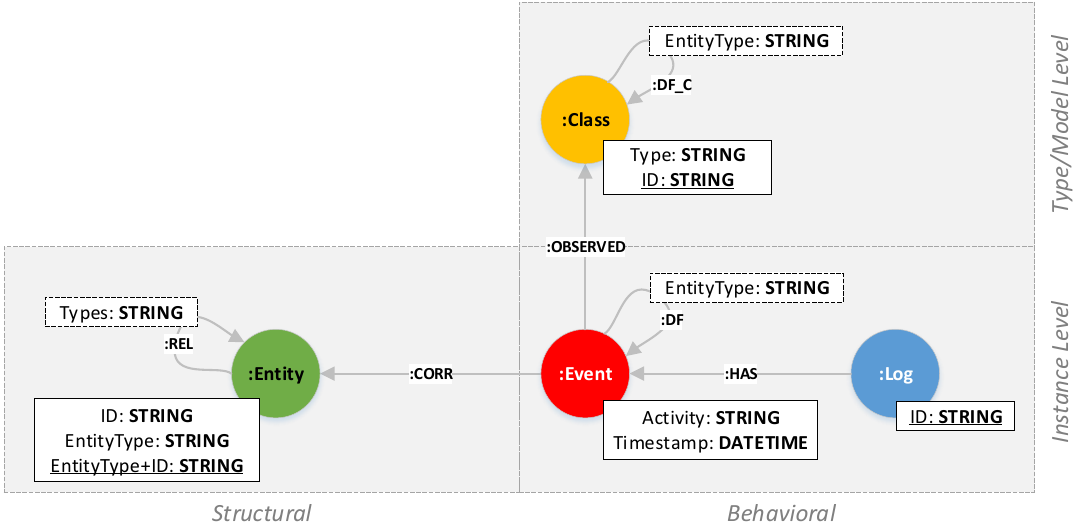}
    \caption{{Schema of node and relationship types for modeling multi-dimensional event data in labeled property graphs}}
    \label{fig:complete_schema}
    \label{fig:3event_node}
    \label{fig:3entityNodeType}
    \label{fig:3timestamp}
    \label{fig:3logNodeType}
    \label{fig:3attributeEvent}
\end{figure}

We introduce node and relation types for the concepts of event, entity, and event log; together they describe the \emph{instance-level} concepts of Fig.~\ref{fig:complete_schema}, i.e., concrete entities or recorded events.

\par
\begin{figure}
    \centering
    \includegraphics[width=\linewidth]{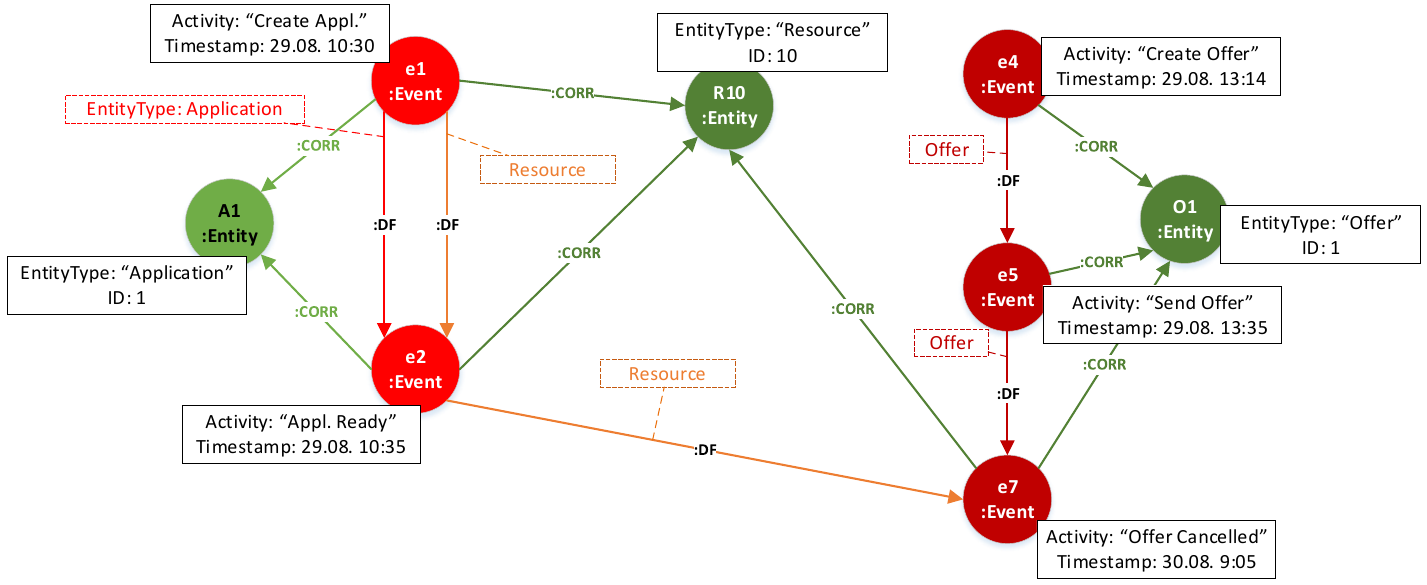}
    \caption{Graph describing the events of Application 1, Offer 1, and Resource 10 of Tab.~\ref{tab:2_BPIC17ExampleCaseTable}}
    \label{fig:example_entity_events_df}
\end{figure}

\vspace{.5em}\noindent\textbf{Event.} We represent the core element of each event log, the \emph{event}, as a node with label $\mathit{\mathord{:}Event}$ as shown in Figure~\ref{fig:3event_node}. Of the three mandatory event attributes (cf. Sect.~\ref{sec:background:event_log_concepts}), we only model \emph{activity} and \emph{timestamp} as properties to event nodes, having datatypes \emph{STRING} and \emph{DATETIME}, respectively. We describe correlation to multiple entity identifiers through the graph structure as we explain next. The graph in Fig.~\ref{fig:example_entity_events_df} models 5 events of Table~\ref{tab:2_BPIC17ExampleCaseTable}.

\par\vspace{.5em}\noindent\textbf{Entity.} Single-dimensional event logs fix a single entity identifier (called case identifier, cf. Sect.~\ref{sec:background:event_log_concepts}) to which each event is correlated. Our model abandons the notion of a case identifier in favor of the more general \emph{Entity} concept. We model each entity as a node with the label $\mathit{\mathord{:}Entity}$ as shown in Fig.~\ref{fig:3entityNodeType}. Its property \emph{EntityType} describes the type of the entity. Property \emph{ID} is the entity identifier. We require the combination of \emph{EntityType} and \emph{ID}, stored as property \emph{uID}, to be unique in the entire graph similar to a primary key value in a relational database (indicated by \emph{uID} being underlined in Fig.~\ref{fig:3entityNodeType}). We limit our model to plainly describing entity nodes and their relations and refer to existing work~\cite{robinson2013graph} for modeling their semantic nature further.

The graph in Fig.~\ref{fig:example_entity_events_df} models 3 entities of 3 different entity types. Each \emph{:Entity} node represents a concrete entity related to the process, such as \emph{Application} $1$  and \emph{Offer} $1$ of our running example of Tab.~\ref{tab:2_BPIC17ExampleCaseTable}.

\par\vspace{.5em}\noindent\textbf{Correlation.}
We model \emph{correlation} of an event to an entity by the dedicated relationship label \emph{:CORR}. Through \emph{:CORR}-relationships, we can correlate any event to any number of entities of different types, allowing for multi-dimensional correlation of events to entities as shown in Fig.~\ref{fig:example_entity_events_df}. For example, event $e1$ is correlated to \emph{Application 1} and \emph{Resource 10}, whereas event $e7$ is correlated to \emph{Offer 1} and \emph{Resource 10}.

\par\vspace{.5em}\noindent\textbf{Event Log.}
Some event data sets, such as BPIC'15\cite{BPIC2015}, consist of multiple event logs. To support multiple logs in one graph event log instance, we introduce a separate node type for logs with the label \emph{:Log} as shown in figure~\ref{fig:3logNodeType}. Similar to the entities, we model which event belongs to which log with the \emph{:HAS} relationship from \emph{:Log} to \emph{:Event} nodes.

\par\vspace{.5em}\noindent\textbf{Attributes.}
Figure~\ref{fig:3attributeEvent} shows all properties our graph data model expects to be present. Additionally, any event, entity or log node can carry any other property. 

{
\subsection{Modeling Behavior as Paths}
\label{ssec:DFgeneral}
\label{sec:semantics:df_refined}
}

\par\vspace{.5em}\noindent\textbf{Directly-Follows (Events).}
Events are ordered by time from the viewpoint of an entity they are correlated to (cf. Sect.~\ref{sec:background:event_log_concepts}). As our model allows events to be correlated to multiple entities, each event may have multiple ``next'' events, depending on the entity. We model that two \emph{:Event} nodes $x$ and $y$ are temporally ordered (from the perspective of an entity) by a \emph{:DF}-relationship from $x$ to $y$. Each \emph{:DF} relationship goes forward in time and holds the \emph{EntityType} for which this relationship holds as a property. Moreover, the events $x$ and $y$ both have to correlated to the same entity node via \emph{:CORR} relationships.

In Fig.~\ref{fig:example_entity_events_df}, events $e1$, $e2$, and $e9$ are connected by \emph{:DF} relationships from $e1$ to $e2$ and from $e2$ to $e9$ carrying the property \emph{EntityType} = \emph{Application}. Moreover, $e1$, $e2$, and $e9$ are all correlated to entity \emph{Application 1} through the \emph{:CORR} relationships. Thus, the \emph{:DF} relationships form a trace $\langle e1,e2,e9 \rangle$ from the perspective of \emph{Application 1}. In the same manner, $\langle e4,e5,e7 \rangle$ form a trace from the perspective of entity \emph{Offer 1}, and $\langle e1,e2,e7 \rangle$ is a trace from the perspective of \emph{Resource 10}. Note that this trace shares events $e_1$ and $e_2$ with the first trace and event $e_7$ with the second trace. Also note, that the graph contains a second \emph{:DF} relationship from $e1$ to $e2$ with property \emph{EntityType} = \emph{Resource}.

\par\vspace{.5em}\noindent\textbf{Design decisions for modeling directly-follows.}
Our data model requires that each \emph{:DF} relationship is specific to exactly one entity type. This constraint ensures that queries for behavioral paths can be easily written and efficiently evaluated.

Suppose we modeled only a single \emph{:DF} relationship between $e_1$ and $e_2$ that describes behavior for both \emph{Application 1} and \emph{Resource 10}. If we try to query a path from $e1$ to some other event along one specific entity $n$ (to which e1 is correlated), then we would have to test along each \emph{:DF} relationship that the events are correlated to the same entity $n$ as $e1$. This becomes infeasible when querying multi-entity paths (such as Fig.~\ref{fig:event_data_models}(6)). Also, the query evaluation becomes inefficient as all possible paths along all reachable \emph{:DF} relationships have to be explored.

The most fine-grained way of modeling would be to require a dedicated \emph{:DF} relationship per entity identifier (with a corresponding \emph{EntityID} property). However, this hinders querying for paths based on entity \emph{types}, e.g., all paths for \emph{Offer} entities.

By modeling one \emph{:DF} relationship per entity type, we enable explicitly querying for behavioral paths per entity types. We explicitly exploit this in query Q6 in Sect.~\ref{sec:querying}. As a consequence, our data model explicitly allows to model multiple behavioral paths through the event data, each path describes a trace from the perspective of one entity. Events can be part of multiple such traces if they are related to multiple entities.

A possible alternative to describing the entity type as a property of the \emph{:DF} relationship is to \emph{refine} the \emph{:DF} label into a set of labels, e.g., \emph{:DF\_Application}, \emph{:DF\_Offer}, \emph{:DF\_Resource}. While this choice simplifies retrieving paths, it requires more case distinction in aggregation queries.

\subsection{Modeling Relations and Interactions between Entities}
\label{sec:represent:reify}

\begin{figure}[t]
    \centering
    \includegraphics[width=\linewidth]{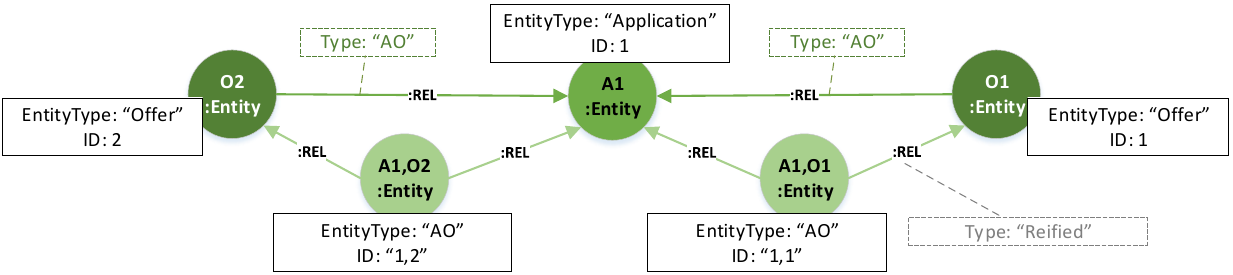}
    \caption{Graph describing relations between entities Application 1, Offer 1, and Offer 2, and the reification of these relations into composite entities.}
    \label{fig:example_entity_relation_reify}
\end{figure}

\textbf{Relations between entities.} Our data model includes a generic relationship type \emph{:REL} between entities to describe how entities relate to each other. The nature or name of the relation is stored in property \emph{Type}. For example, Fig.~\ref{fig:example_entity_relation_reify} describes that \emph{Offer 1} and \emph{Offer 2} are related to \emph{Application 1} through a relation of type \emph{AO}.

\par\vspace{.5em}\noindent\textbf{Reification of relations.} Any \emph{:DF} relationship between two events $e1$ and $e2$ requires an \emph{:Entity} to which both $e1$ and $e2$ are correlated via an \emph{:CORR} relationship. This ensures that any behavioral path is ``observed'' from the perspective of an entity. If we want to analyze the interaction between two related entities, we have to ``observe'' the behavior from the perspective of the relation between them. For example, to describe how \emph{Application 1} and \emph{Offer 1} interact, we have to observe the behavior from the perspective of the \emph{AO} relation between them~\cite{DBLP:journals/tsc/LuNWF15}. A plain graph-based data model does not allow us to correlate an \emph{:Event} node to the \emph{:REL} relationship between \emph{Application 1} and \emph{Offer 1}. A standard technique in data modeling is to reify the relationship in this case, that is, to introduce a \emph{derived :Entity node} which represents the relationship. We can then correlate events to this \emph{:Entity} node to model interactions between entities~\cite{DBLP:conf/apn/Fahland19}.

Figure~\ref{fig:example_entity_relation_reify} illustrates the reification of the \emph{:REL} relationships between \emph{Offer 1}, \emph{Offer 2}, and \emph{Application 1} in derived entities \emph{A1,O1} and \emph{A1,O2} of type \emph{AO} (the same as the original \emph{:REL} relationship). The derived entities are related to their original entities through \emph{:REL} relationships of type \emph{Reified}. Note that also n-ary relationships can be reified in this way.

\begin{figure}[t]
    \centering
    \includegraphics[width=\linewidth]{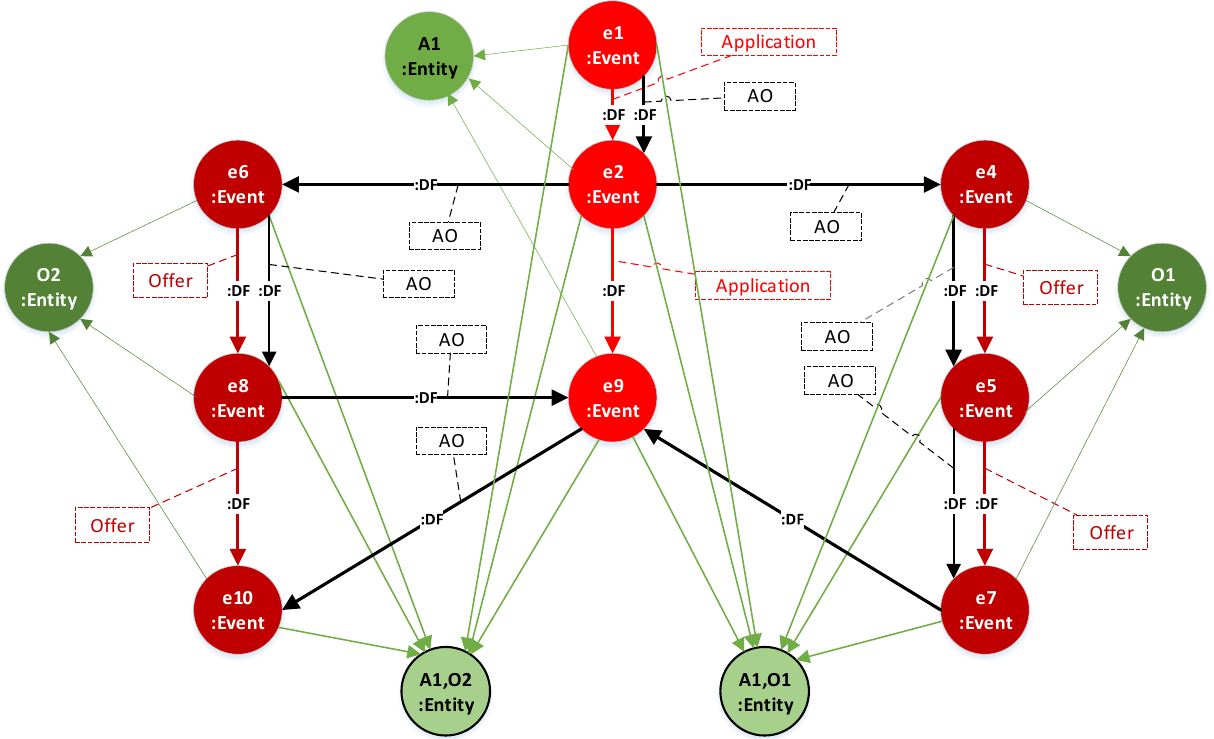}
    \caption{Graph with additional directly-follows relations along entities A1,O1 and A1,O2 describe interactions between Application 1, Offer 1, and Offer 2 of Tab.~\ref{tab:2_BPIC17ExampleCaseTable}.}
    \label{fig:example_entity_relation_reify_df}
\end{figure}

\par\vspace{.5em}\noindent\textbf{Correlating events to reified relations.} We can describe the interaction between two related entities $E1$ and $E2$ by correlating the events of $E1$ and of $E2$ to the derived entity. For example, in Fig.~\ref{fig:example_entity_relation_reify_df}, events $e1,e2,e9$ are correlated to \emph{A1} and $e6,e8,e10$ are correlated to \emph{O2}. Together, all these events are correlated to the derived entity \emph{A1,O2}, resulting in the \emph{:DF}-path $\langle e1,e2,e6,e8,e9,e10 \rangle$ for entity type \emph{AO}. The path $\langle e1,e2,e4,e5,e7,e9 \rangle$ describes the interaction between \emph{A1} and \emph{O1}. Note that Fig.~\ref{fig:example_entity_relation_reify_df} corresponds to Fig.~\ref{fig:event_data_models}(5).

In this way, all directly-follows relations of the events of Tab.~\ref{tab:2_BPIC17ExampleCaseTable} can be modeled correctly and in a single data structure, making this the only data model to satisfy R0 compared to the other data models discussed in Fig.~\ref{fig:event_data_models}.

\subsection{Modeling Aggregations of Events and Behavior}\label{sec:represent:class}

\par\vspace{.5em}\noindent\textbf{Event classes.} While we assume each event to have the mandatory attribute \emph{Activity}, events can be classified in other ways as well (cf. E6 in Sect.~\ref{sec:background:event_log_concepts}). In our model of Fig.~\ref{fig:complete_schema}, each event class is described by a node \emph{:Class} defined by a unique \emph{ID} and a \emph{Type} property that is the same for all event classes defined in the same way, e.g., based on the \emph{Activity} attribute only or a combination of attributes. Each event can be associated to zero or more event classes by relationships with label \emph{:OBSERVED} from an \emph{:Event} node to a \emph{:Class} node.

\begin{figure}[t]
    \centering
    \includegraphics[width=0.8\linewidth]{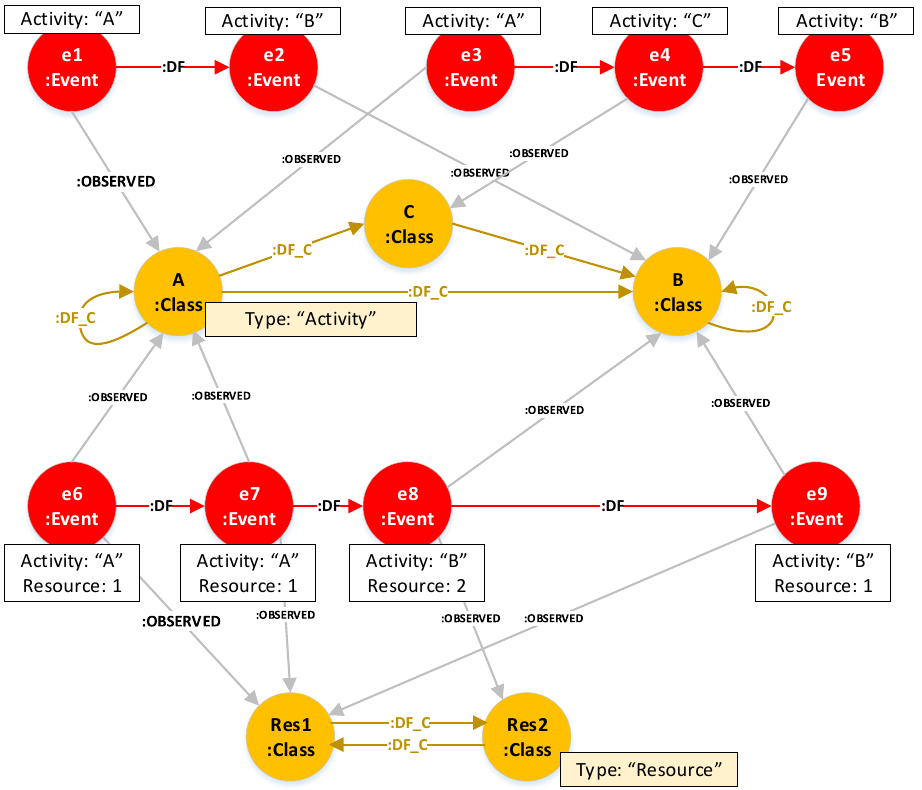}
    \caption{Graph with multiple event classes and aggregated directly-follows relationships}
    \label{fig:example_class_dfc}
\end{figure}

{
Figure~\ref{fig:example_class_dfc} shows an example. Two event classes of type ``Resource'' are defined by the Resource entities occurring in the data, e.g., $Res1$ and $Res2$; events $e6$, $e7$, and $e8$ observed class $Res1$. Further, three event classes of type ``Activity'' are defined; events $e6$ and $e7$ also observed class $A$ of this type.}

\par\vspace{.5em}\noindent\textbf{Directly-follows on event classes.} Event data analysis aggregates directly-follows relations between events to directly-follows relations between event classes (cf. Sect.~\ref{sec:background:event_log_concepts}). Our model of Fig.~\ref{fig:complete_schema} provides relationship label \emph{:DF\_C} between \emph{:Class} nodes. As for \emph{:DF}, each \emph{:DF\_C} relationship lists in attribute \emph{EntityType} for which entity type the aggregated directly-follows relationship holds. This allows to describe aggregated behavior per entity type. The graph in Fig.~\ref{fig:example_class_dfc} shows the aggregation of \emph{:DF} to \emph{:DF\_C}, assuming a single entity type. The two sub-graphs induced by \emph{:DF\_C} describe a hand-over of work social network (bottom)~\cite{DBLP:journals/cscw/AalstRS05} and a classical directly-follows graph~\cite{DBLP:journals/sosym/AalstRVDKG10} (top) of the event data in the same model.

\section{Semantics of Entities and Events in Labeled Property Graphs}\label{sec:semantics}

\begin{figure}[t]
    \centering
    \includegraphics[scale=0.6]{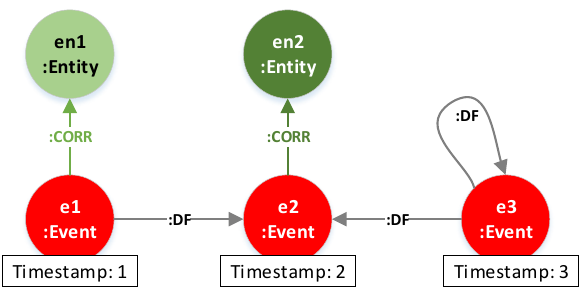}
    \caption{Incorrect semantic pattern of :CORR and :DF relationships}
    \label{fig:wrongGraph}
\end{figure}

The node and relationship types introduced in Sect.~\ref{sec:represent} allow us to model multi-dimensional event data in LPGs. However, LPGs allow for unrestricted use of any node and relationship types which would allow for creating LPGs that do not capture the semantics of event data.
For example, the LPG in Figure~\ref{fig:wrongGraph} only uses the node and relationship types of Sect.~\ref{sec:represent} but the graph violates the semantics the types shall encode:
\emph{:DF} does not order events $e2$ and $e3$ according to their timestamp, events $e1$ and $e2$ are ordered by \emph{:DF} but belong to different entities, and event $e3$ even directly-follows itself.

In this section, we formulate constraints on how the nodes and edges over the types of Sect.~\ref{sec:represent} may be related, thereby giving them semantics. In the following, we formulate such constraints for any labeled property graph $G = (N,R,\mathit{label},\mathit{prop})$ (see Sect.~\ref{sec:background:lpg}).

\subsection{Strictly Typed Nodes}
We formalize the semantics of the node labels $\mathit{Entity}$, $\mathit{Event}$, $\mathit{Log}$, $\mathit{Class}$ and of the relationship labels $\mathit{REL}$ (entity to entity), $\mathit{CORR}$ (event to entity), $\mathit{HAS}$ (event to log), $\mathit{DF}$ (directly-follows on events), $\mathit{OBSERVED}$ (event to event class), and $\mathit{DF\_C}$ (directly-follows on event classes).

Each node/relationship may have one of these types (i.e., no node or relationship may have two different semantic roles). Formally, $\mathit{Label}_N = \{ \mathit{Entity}, \mathit{Event}, \mathit{Log}, \mathit{Class}\}$,  $\mathit{Label}_R = \{ \mathit{REL}, \mathit{CORR}, \mathit{HAS}, \mathit{DF}, \mathit{OBSERVED}, \mathit{DF\_C} \}$ and
for each $n \in N$, $|\mathit{label}(n) \cap \mathit{Label}_N| \leq 1$ and $|\mathit{label}(n) \cap \mathit{Label}_R| = 0$ and for each $r \in R$, $|\mathit{label}(r) \cap \mathit{Label}_N| = 0$ and $|\mathit{label}(r) \cap \mathit{Label}_R| \leq 1$.

As each node has a unique label in $\mathit{Label}_N$ or $\mathit{Label}_R$, we write $n.\mathit{label}$ and $r.\mathit{label}$ for their labels in the following.
Further, we write $L$ for the set $\{ n \in N \mid n.\mathit{label} = L \}$ of nodes and of relations $\{ r \in R \mid r.\mathit{label} = L \}$ carrying label $L$, respectively. For example, $n \in \textit{Entity}$ and $df \in \textit{DF}$.

{
\subsection{Semantics of Entity-Entity Relations}\label{sec:semantics:rel}
The Entity-Entity relationship \emph{REL} is a placeholder for all kinds of structural relationships between entities which may carry various semantics. For the scope of this paper, we only require that any \emph{REL} relationship is between two entities, i.e., for any $rel \in \mathit{REL}, \overrightarrow{rel} = (n_1,n_2)$ holds $n_1,n_2 \in \textit{Entity}$. }

\subsection{Semantics of Event-Entity Relations}\label{sec:semantics:e_en}
\begin{figure}\centering
\includegraphics[scale=0.6]{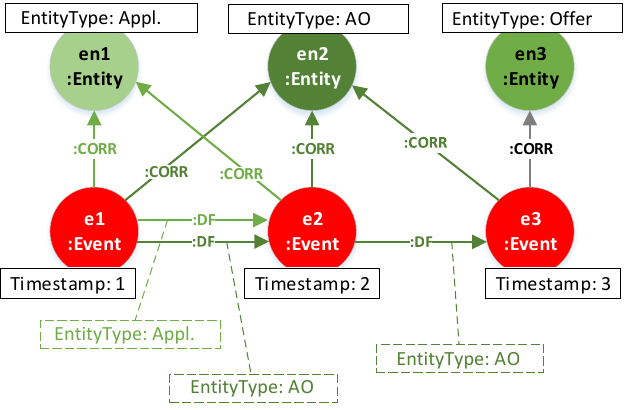}
\caption{Correct semantic pattern of :CORR and :DF relationships}
\label{fig:pattern:df}\label{fig:pattern:e_en}
\end{figure}

The Event-Entity relationship $CORR$ correlates an event to its process entities. While each event $e$ can be related to multiple different entities, for example an Application and a Resource, there must not be two $CORR$ relationships from one event to the same entity. Furthermore, each event is correlated to some entity, and vice versa, as shown in Fig.~\ref{fig:pattern:e_en}.

Formally, the following properties have to hold:
\begin{enumerate}
    \item Between any pair of $e \in \textit{Event}$ and $n \in \textit{Entity}$ is at most one relation $r \in \textit{CORR}, \overrightarrow{r} = (e,n)$. As a shorthand, we write $\textit{CORR} \subseteq \textit{Event} \times \textit{Entity}$, and $(e,n) \in \textit{CORR}$.
    \item Each event $e \in \textit{Event}$ is correlated to at least one entity: there exists $(e,n) \in \textit{CORR}$
    \item Each entity $n \in \textit{Entity}$ is correlated to at least one event: there exists $(e,n) \in\textit{CORR}$
\end{enumerate}

\subsection{Semantics of Log-Event Relations}\label{sec:semantics:l_e}

\begin{figure}\centering
\includegraphics[scale=0.58]{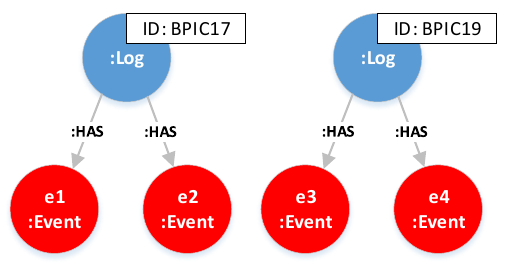}
\caption{Correct semantic pattern of :HAS relationships}\label{fig:pattern:l_e}
\end{figure}

Log-Event relationship \emph{HAS} explicitly encodes which event belongs to which log. Every event must be in exactly one log, and each log must have at least one event, as shown in Fig.~\ref{fig:pattern:l_e}.
Formally, the following properties have to hold:
\begin{enumerate}
    \item $\textit{HAS} \subseteq \textit{Log} \times \textit{Event}$
    \item Each event $e \in \textit{Event}$ is in exactly one event log: there exists exactly one $r \in \textit{HAS}, \overrightarrow{r} = (l,e)$.
    \item Each event log $l \in \textit{Log}$ has at least one event: there exists at least one $r \in \textit{HAS}, \overrightarrow{r} = (l,e)$ .
\end{enumerate}

{
\subsection{Semantics of Directly-Follows Relation}
\label{sec:semantics:df}
}

We model temporal relations as paths of \emph{:DF} relationships over \emph{:Event} nodes. Each  \emph{:DF} relationship must go forward in time from the point of view of an \emph{:Entity} node correlated to \emph{both} events involved as shown in Fig.~\ref{fig:pattern:df}. Overall, all \emph{:DF} relationships induce a \emph{partial order}.
Formally, the following properties have to hold:
\begin{enumerate}
    \item For any $df \in \mathit{DF}, \overrightarrow{df} = (e_1,e_2)$ holds $e_1,e_2 \in \textit{Event}$; note that there can be multiple $df$ relations between the same two events.
    \item For every $df \in \textit{DF}, \overrightarrow{df}=(e_1,e_2)$, exists a log $l \in \textit{Log}$ with $(e_1,l)$ and $(e_2,l) \in \textit{HAS}$.
    \item For every $df \in \textit{DF}, \overrightarrow{df}=(e_1,e_2)$, $e_1.\textit{Timestamp} \leq e_2.\textit{Timestamp}$ holds, i.e., events are ordered by time.
    \item For every $df \in \textit{DF}, \overrightarrow{df}=(e_1,e_2)$, exists an entity $n \in \textit{Entity}$ with $(e_1,n),(e_2,n) \in \textit{CORR}$ such that $n.EntityType = df.EntityType$, and there exists no event $e_x$ correlated to $n$, $(e_x,n) \in \textit{CORR}$, such that $e_1.\textit{Timestamp} < e_x.\textit{Timestamp} < e_2.\textit{Timestamp}$ holds, i.e., $e_2$ directly-follows $e_1$ from the perspective of entity $n$.
    \item For all events $e_1$, there is no $df \in \textit{DF}$ with $\overrightarrow{df} = (e_1,e_1)$, i.e., \emph{:DF} is irreflexive.
    \item For all events $e_0,e_n \in \textit{Event}$ exists no cycle $df_0,\ldots,df_n \in \textit{DF}$ with $\overrightarrow{df_i} = (e_{i},e_{i+1}), i=1,\ldots,n$, $\overrightarrow{df_n} = (e_{n},e_{0})$, i.e., \emph{:DF} is acyclic and hence the transitive closure of \emph{:DF} is a partial order.
\end{enumerate}

\subsection{Semantics of Event-Class relation}\label{sec:semantics:e_c}

Similar to \emph{:CORR}, relationship \emph{:OBSERVED} relates each event to at least one class of the same type, and vice versa. Formally, the following properties have to hold:
\begin{enumerate}
    \item $\textit{OBSERVED} \subseteq \textit{Event} \times \textit{Class}$
    \item Each event $e \in \textit{Event}$ has at least one event class: there exists $(e,c) \in \textit{OBSERVED}$, and there are no two classes $(e,c_1),(e,c_2) \in \textit{OBSERVED}$ with $c_1\neq c_2$ and $c1.\mathit{Type} = c2.\mathit{Type}$.
\end{enumerate}

{
\subsection{Semantics of Class-level Directly-Follows Relation}\label{sec:semantics:df_c}

\begin{figure}[t]
    \centering
    \includegraphics[width=\linewidth]{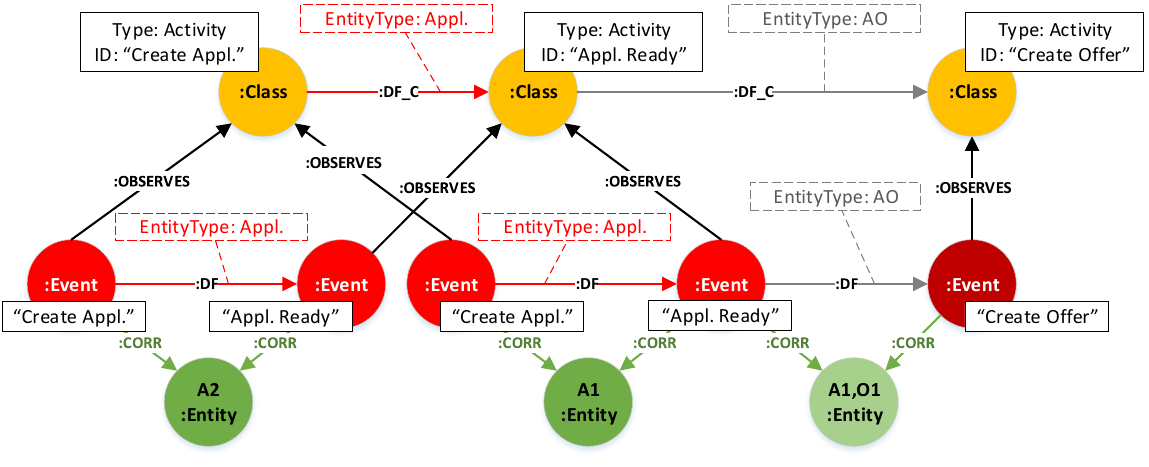}
    \caption{Correct Semantic Pattern of :OBSERVES and :DF\_C Relationship}
    \label{fig:pattern:dfc}
\end{figure}

The directly-follows relation between \emph{:Class} nodes aggregates existing directly-follows relations between \emph{:Events} nodes wrt.\ an entity type.
The class-level directly follows relationship \emph{:DF\_C} is only defined between \emph{:Class} nodes of the same type. A \emph{:DF\_C} relationship relation between two class nodes  $c_1$ and $c_2$ for an entity type $T$ may only aggregate \emph{:DF} relationships between events that observed $c_1$ and $c_2$ and are correlated to the same entity of type $T$, as shown in Fig.~\ref{fig:pattern:dfc}.} Formally, the following properties have to hold:
\begin{enumerate}
    \item $\textit{DF\_C} \subseteq \textit{Class} \times \textit{Class}$
    \item Any two related classes $(c_1,c_2) \in \textit{DF\_C}$ are of the same type $c_1.Type = c_2.Type$
    \item {For any two related classes $dfc \in \textit{DF\_C}, \overrightarrow{dfc} = (c_1,c_2)$ exist events $e_1,e_2$ of these classes $(e_1,c_1), (e_2,c_2) \in \textit{OBSERVED}$ ordered in the same way as $c_1$ and $c_2$ for the same entity type: there exists $df \in \textit{DF}, \overrightarrow{df} = (e1,e2)$ and $dfc.\textit{EntityType} = df.\textit{EntityType}$.}
\end{enumerate}

\section{Translating Event Logs to Labeled Property Graphs}\label{sec:storing}

We now present a semi-automatic procedure for translating event tables with multiple entity identifiers all stored {in a single event table} (cf. Sect.~\ref{sec:background:multi-dim-literature}) into the graph data structure introduced in Sect.~\ref{sec:represent} satisfying the semantic constraints of Sect.~\ref{sec:semantics}. This section covers creating the instance-level concepts \emph{:Log}, \emph{:Event}, \emph{:Entity}, \emph{:HAS}, \emph{:DF}, \emph{:CORR}, and \emph{:REL} of Sect.~\ref{sec:represent}, while the type-level concepts \emph{:Class}, \emph{:OBSERVES}, and \emph{DF\_C} are covered in Sect.~\ref{sec:mining}.

In a nutshell, our method has the following steps.
(\ref{sec:storing:source_format}) We assume the event data to be given in the form of an event table where each record describes one event.
(\ref{sec:storing:import_events}) We translate each record with all its attributes to an \emph{:Event} node in the LPG with corresponding properties, obtaining a graph of unrelated \emph{:Event} nodes.
(\ref{sec:storing:log_nodes}) We create \emph{:Log} nodes for each log in the source data set and relate them to the respective \emph{:Event} nodes.
(\ref{sec:storing:correlate}) We provide query templates to extract \emph{:Entity} nodes from \emph{:Event} properties (e.g. identifiers) and to correlate \emph{:Event} nodes to all their \emph{:Entity} nodes.
(\ref{sec:storing:df}) A generic query derives the entity-specific directly-follows \emph{:DF} relationships between events.
(\ref{sec:storing:relations}) We provide query templates to extract \emph{:REL} relationships between entities.
(\ref{sec:storing:reify}) Finally, we provide queries to reify relations between entities into derived entities; the queries of step (\ref{sec:storing:df}) then infer the \emph{:DF}-relationships for the derived entity allowing to study interactions between entities. We explain the queries on the running example of Tab.~\ref{tab:2_BPIC17ExampleCaseTable}.

We demonstrate the types of graphs obtained on the full BPIC17 dataset~\cite{BPIC2017} in Sect.~\ref{sec:storing:demo_bpic17} and report on a quantitative evaluation of all data sets in Sect.~\ref{sec:storing:evaluation}.

\subsection{Source Event Data Format}\label{sec:storing:source_format}

We expect the event data of the source log to be in event table format (see Sect.~\ref{sec:background:multi-dim-literature}) defining columns \emph{Activity} and \emph{Timestamp} and multiple columns \emph{Attribute1},\ldots,\emph{AttributeN} that also contain entity identifiers. In case the data comes from multiple logs, also a \emph{LogID} column is required.


\subsection{Import the Events}\label{sec:storing:import_events}

\begin{figure}
    \centering
    \includegraphics[width=\linewidth]{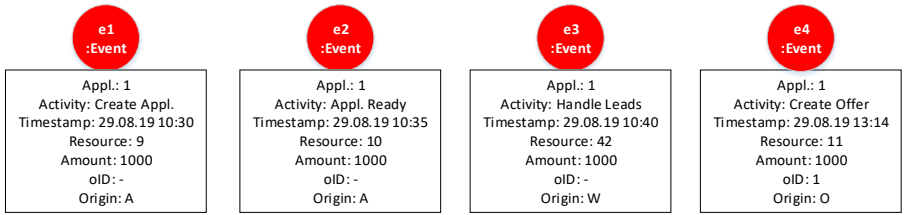}
    \caption{Graph after step 1: event nodes with properties}
    \label{fig:4_3_improted_events}
\end{figure}

From the source event table (Sect.~\ref{sec:storing:source_format}), we import each row $r$ as an :Event node $e$ and set for each attribute (column) $A$ of $r$ the property $e.A = r.A$. The following Cypher query implements this step for an event table given as a CSV file.
\begin{lstlisting}[escapechar=\#,style=smallStyle]
LOAD CSV WITH HEADERS FROM "file:///#\textit{filename}#.csv" as line
CREATE (e:Event {LogID: line.LogID, Activity: line.Activity, Timestamp: datetime(line.Timestamp), #\textit{\$Attribute1}#: line.#\textit{\$Attribute1}#, ..., #\textit{\$AttributeN}#: line.#\textit{\$AttributeN}# })
\end{lstlisting}
Importing the first four rows of Tab.~\ref{tab:2_BPIC17ExampleCaseTable} results in the graph shown in Fig.~\ref{fig:4_3_improted_events}. Importing event table data from other formats than a CSV file requires a corresponding query that applies line 2 of the query to any row in the event table.

\subsection{Create Logs}\label{sec:storing:log_nodes}

We assume each event $e$ carries attribute $e.\textit{LogID}$ describing in which log event $e$ was recorded (see Sect.~\ref{sec:storing:source_format}). For each unique value $e.\textit{LogID}$ we create exactly one \emph{:Log} node $\ell$ with $\ell.ID = e.\textit{LogID}$. The \emph{MERGE} command in the following Cypher query achieves this.
\begin{lstlisting}[style=smallStyle]
MATCH (e:Event)
MERGE (:Log {ID: e.LogID})
\end{lstlisting}
Then we create a \emph{:HAS} relationship from each event $e$ to the \emph{:Log} node $\ell$ with $\ell.ID = e.\textit{LogID}$; the following Cypher query implements this.
\begin{lstlisting}[style=smallStyle]
MATCH (e:Event)
MATCH (l:Log)
WHERE e.LogID = l.ID
CREATE (l)-[:HAS]->(e)
\end{lstlisting}
Applying these queries to the graph of Fig.~\ref{fig:4_3_improted_events} results in the structure shown at the top of Fig.~\ref{fig:4_4_logs} which conforms to the constraints of Section~\ref{sec:semantics:l_e}.


\subsection{Create Entities Nodes and Correlate Events (using Domain Knowledge)}\label{sec:storing:correlate}

We create entities and correlate events to entities in two steps. First we identify and create the set of all entities that occur in the data by creating \emph{:Entity} nodes. Then we correlate each event to all its entities by creating \emph{:CORR} relationships.

The general assumption in process event data is that if an event $e$ is correlated to an entity of a specific \emph{Type}, then $e$ has a specific entity identifier attribute $e.\textit{EntityID} = id$ where $id$ identifies the entity. Recognizing which attribute is an entity identifier and of which entity type requires \emph{domain knowledge}. Primary key identification techniques can be used to identify candidates entity identifiers~\cite{DBLP:journals/ijcis/PopovaFD15}. However, the same entity identifier attribute may refer to entities of different types~\cite{DBLP:journals/tsc/LuNWF15} so that the actual type can only be inferred from other event attributes. For instance in our running example, events can have two different entity identifiers: \emph{Appl} and \emph{oID} (see Tab.~\ref{tab:2_BPIC17ExampleCaseTable} or Fig.~\ref{fig:4_3_improted_events}).
The property \textit{Origin} distinguishes to which entity type \emph{Appl} or \emph{oID} refer, e.g., only an event $e$ with \textit{e.Origin = ``A''} is correlated to an \emph{Application} entity with id \textit{e.Appl}.

Assuming the user identified the \emph{Condition} when an event attribute \emph{EntityID} refers to an entity of type \emph{Type}, then each unique value for $e.\textit{EntityID} = id$ found in any event $e$ where $e.Condition$ holds indicates the existence of a distinct entity identified by $id$. We create for each such value $id$ a new \emph{:Entity} node $n$ with $n.id = id$ and $n.\textit{EntityType} = \textit{Type}$. The following Cypher query template with 3 parameters (\emph{Condition}, \emph{EntityID}, \emph{Type}) implements this.
\begin{lstlisting}[escapechar=\#,style=smallStyle]
MATCH (e:Event) WHERE #\textit{Condition}#
WITH e.#\textit{EntityID}# AS id, e.#\textit{Type}# as name
MERGE (en:Entity {ID:id, uID:(name+id), EntityType: name})
\end{lstlisting}
The query also sets property \emph{uID} as required in Sect.~\ref{sec:represent}.
Calling the template with  \textit{EntityProperty} $\equiv$ ``\textit{e.Origin = ``A''}, \textit{EntityID} $\equiv$ ``Appl'', \textit{Type} $\equiv$ ``\textit{Application}'' on the graph of Fig.~\ref{fig:4_3_improted_events} first matches event nodes $e1$ and $e2$ (only these have \textit{e.Origin = ``A''}) and then returns their value for \emph{e.Appl}, i.e., 1 and 1. Finally, one \emph{:Entity} node $n$ of type Application with $n.\textit{ID} = 1$ is created.

\begin{figure}[tb]\centering
    \centering
    \includegraphics[scale=0.5]{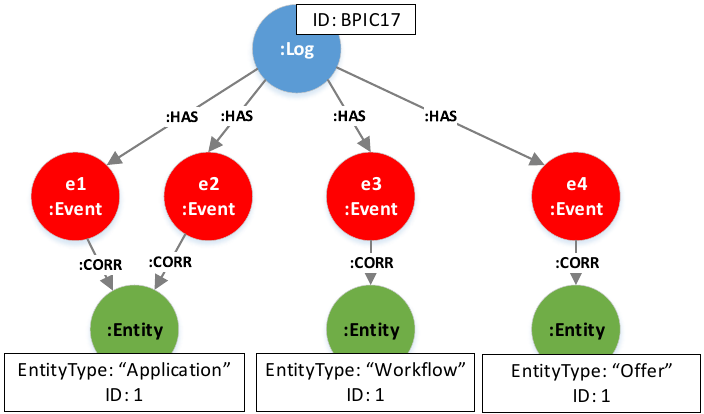}
    \caption{Graph after creating log and entity nodes}
    \label{fig:4_4_logs}\label{fig:4_5_entities}
\end{figure}

Now we can materialize that an event $e$ correlates to an entity $n$ of type $Type$ as a \emph{:CORR} relationship from $e$ to $n$ if $e$ satisfies the \emph{Condition} and refers to $n$ by $e.\textit{EntityID} = n.\textit{ID}$. The following Cypher query implements this:
\begin{lstlisting}[escapechar=\#,style=smallStyle]
MATCH (e:Event) WHERE #\textit{Condition}#
MATCH (n:Entity {EntityType: #\textit{Type}#}) WHERE e.#\textit{EntityID}# = n.ID
CREATE (e)-[:CORR]->(n)
\end{lstlisting}
The above query template has to be executed for each entity type in the data. Assuming each event contains correlation information for at least one entity, it conforms to the semantic requirements of Sect.~\ref{sec:semantics:e_en}. The event graph after applying both queries for Application, Workflow, and Offer entities in this way is shown in Fig.~\ref{fig:4_5_entities}.

\subsection{Create Entity-specific Directly-Follows Relation}\label{sec:storing:df}

We now can construct the behavioral \emph{:DF}-relationships from the perspective of each entity $n$ in the graph individually. We retrieve all events $e_1,\ldots,e_k$ correlated to $n$ (via \emph{:CORR}), order them by $e_i.\textit{timestamp}$ and create a \emph{:DF} relationship from event node $e_i$ to event node $e_{i+1}$. The following Cypher query implements this: Lines 1-4 collect all events correlated to entity node $n$ in an \emph{eventList} of length $k$ (ordered by their timestamp attribute). We then iterate over the 0-indexed $\textit{eventList} = \langle e_0,\ldots,e_{k-1} \rangle$ (lines 5-6) and create a \emph{:DF} relationship from $e_i$ to $e_{i+1}$ for each $i=0,\ldots,k-2$.
\begin{lstlisting}[style=smallStyle]
MATCH (n : Entity )
MATCH (n) <-[:CORR]- ( ev )
WITH n , ev as events ORDER BY ev.timestamp,ID(e)
WITH n , collect ( events ) as eventList
UNWIND range(0,size(eventList)-2) AS i
WITH n , eventList[i] as e1, eventList[i+1] as e2
MERGE ( e1 ) -[df:DF {EntityType:n.EntityType}]->( e2 )
\end{lstlisting}
The data may contain events with identical timestamps, typically due to coarse-grained or imprecise recording~\cite{DBLP:conf/bpm/LuFA14,DBLP:conf/bpm/PegoraroUA19}. To ensure that all directly-follows relations form a directed acyclic graph (see Sect.~\ref{sec:semantics:df}), we need to provide a globally consistent ordering for events with identical timestamps. We do so using the internal unique $\textit{ID}(e)$ of the \emph{:Event} nodes in line 3 to order events by $\textit{ID}(e)$ in case their timestamps are identical. As we import the events in the same order as in the source data $\textit{ID}(e)$ is consistent with the implicit ordering in the source data.

\begin{figure}[tb]\centering
    \includegraphics[scale=0.5]{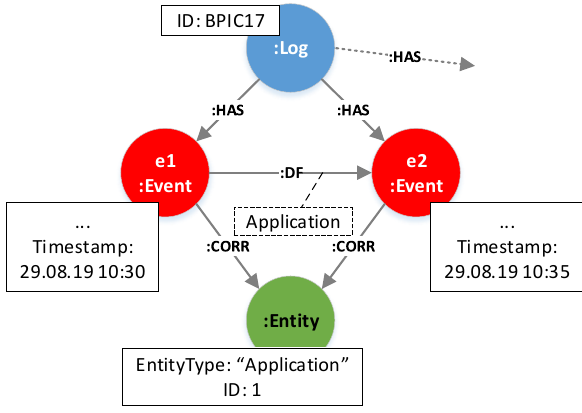}
    \caption{Graph after adding directly-follows relationships between Application events}
    \label{fig:4_5_df}
\end{figure}

The query creates \emph{:DF} relationships for events \emph{per entity} node in the graph; through \emph{MERGE} in line 7, we ensure that we only add relationships between different events per \emph{EntityType} as discussed in Sect.~\ref{sec:semantics:df_refined}. Applying the above query on the graph of Fig.~\ref{fig:4_5_entities} results in the \emph{:DF} relationship from $e1$ to $e2$ shown in Fig.~\ref{fig:4_5_df}. Creating \emph{:DF} relationships in this way conforms to the constraints of Sect.~\ref{sec:semantics:df}.

\subsection{Deriving relations between entities (using domain knowledge)}
\label{sec:storing:relations}

Entity creation and correlation may leave events of different entities unrelated if an event is not explicitly related to more than on entity. In our running example of BPIC17~\cite{BPIC2017}, events are correlated to either an Application, Offer, or Workflow entity as shown in Fig.~\ref{fig:4_5_entities}. Deriving directly-follows relations per entity as in Sect.~\ref{sec:storing:df} leaves their behaviors disconnected.

We cannot further correlate \emph{Offer} events to other existing entities such as \emph{Application}. If we would correlate $e3$ and $e4$ directly to \emph{Application 1} by entity identifier \emph{Appl}., we would ``pollute'' the directly-follows relation of \emph{Application 1} with events that are only remotely related to it, resulting in convergence errors (see Sect.~\ref{sec:background:multi-dim-req}).

To ensure requirement R0, we extend the data with structural relationships between entities. We can later reify these relationships into entities that describe interactions between other entities. Our data model starts from recorded events, thus we have to infer relations between entities from event attributes using domain knowledge. Alternatively, {foreign key identification techniques can be used to identify candidates relationships~\cite{DBLP:journals/ijcis/PopovaFD15}}.

\begin{figure}
    \centering
    \includegraphics[scale=0.5]{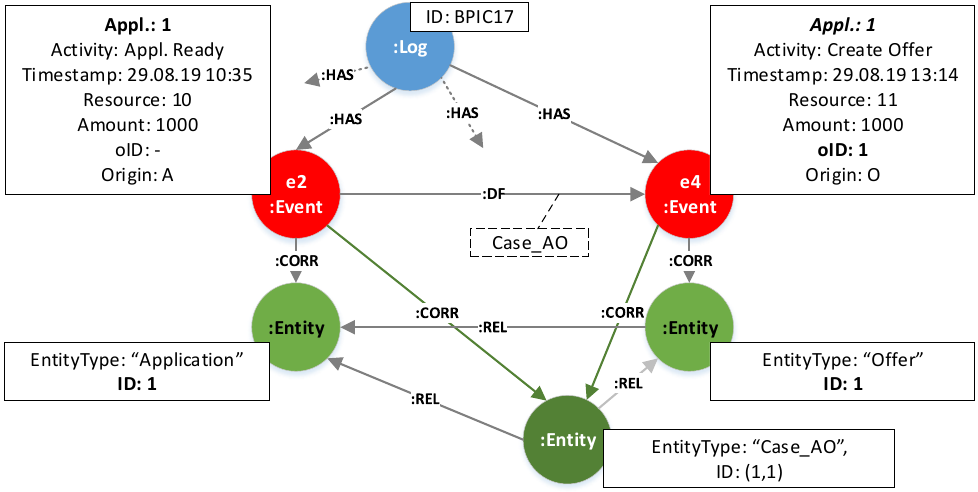}
    \caption{Graph after reifying relation between \emph{Application 1} and \emph{Offer 1} into composite entity \emph{Case\_AO (1,1)} }
    \label{fig:4_5_entities_composite}
\end{figure}

Assume two events \emph{(e1:Event) -[:CORR]-$\mathord{>}$ (n1:Entity)} and \emph{(e2:Event) -[:CORR]-$\mathord{>}$ (n2:Entity)} are correlated to different entities \emph{n1 $\mathord{<>}$ n2}. If \emph{e2} contains some property \emph{refto1} referencing the entity identifier \emph{ID} of \emph{n1}, i.e., a foreign key, we observe that \emph{n2} is related to \emph{n1} through event \emph{e2}. In our running example, we observe that \emph{Offer 1} is related to \emph{Application 1} through event \emph{e4} of \emph{Offer 1} via property \emph{Appl.}, see Fig.~\ref{fig:4_5_entities_composite}.

\begin{figure}[t]
    \centering
    \includegraphics[scale=0.5]{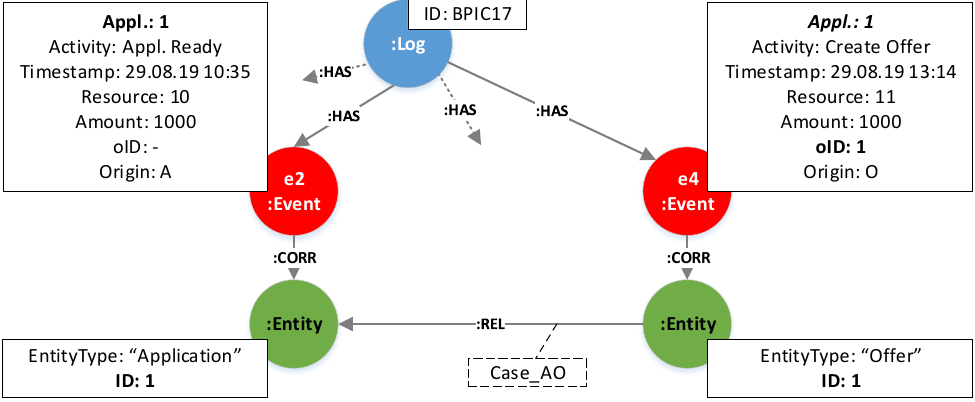}
    \caption{Graph after inferring the relation between \emph{Application 1} and \emph{Offer 1}}
    \label{fig:4_5_relation}
\end{figure}

We lift this observation to entity types. The relation $R$ from entity \emph{type1} to entity \emph{type2} is the set of all pairs $(n1.\mathit{ID},n2.\mathit{ID})$ where (1) there is an entity $n1$ of \emph{type1}, (2) some event \emph{(e2:Event) -[:CORR]-$\mathord{>}$ (n2:Entity)} is correlated to entity $n2$ of \emph{type2}, (3) with $e2.\textit{refto1} = n1.\textit{ID}$. Lines 1-4 of the query template below identify such a pair of entities $n1$ and $n2$  for chosen parameters \emph{type1}, \emph{type2}, and \emph{refto1} and line 5 construct this relation $R$ of \emph{typeR}.
\begin{lstlisting}[escapechar=\#,style=smallStyle]
MATCH (e1:Event) -[:CORR]-> (n1:Entity) WHERE n1.EntityType='#\emph{type1}#'
MATCH (e2:Event) -[:CORR]-> (n2:Entity) WHERE n2.EntityType='#\emph{type2}#'
  AND n1 <> n2 AND e2.#\emph{refto1}# = n1.ID
WITH DISTINCT n1.ID as n1_id, n2.ID as n2_id
WHERE n1_id <> 'Unknown' AND n2_id <> 'Unknown'
CREATE (n1) <-[:REL {Type:'#\emph{typeR}#'}]- (n2)
\end{lstlisting}
Applying the above query for \emph{type1} $\equiv$ \emph{Application}, \emph{type2} $\equiv$ \emph{Offer}, \emph{refto1} $\equiv$ \emph{Appl}, \emph{typeR} $\equiv$ \emph{Case\_AO}, on our running example results in the relationship of type \emph{Case\_AO} from \emph{Offer 1} to \emph{Application 1} shown in Fig.~\ref{fig:4_5_relation}.

{
\subsection{Reify relations between entities for describing interactions}\label{sec:storing:reify}
}

To model behavioral relations between two structurally related entities, we reify their relation into a new entity type. Each pair $(n1,n2)$ of entities in some relation $R$ gets materialized as a new entity $r$. By correlating the events of $n1$ and $n2$ to $r$ and ordering them over time, we can study the behavior or interaction between $n1$ and $n2$ along $r$.

The following query reifies each pair $(n1,n2)$ in a \emph{:REL}-relationship of ``\emph{typeR}'' into a new Entity $r$ and relates $r$ to $n1$ and to $n2$ as relationships of type ``Reified''.
\begin{lstlisting}[escapechar=\#,style=smallStyle]
MATCH (n1:Entity) -[rel:REL {Type:'#\emph{typeR}#'}]-> (n2:Entity)
CREATE (n1) <-[:REL {Type:'Reified'}]- ( r : Entity { ID:n1.id+'_'+n2.id, EntityType : '#\emph{typeR}#', uID:'#\emph{typeR}#_'+n1.id+'_'++n2.id) } ) -[:REL {Type: 'Reified'}]-> (n2)
\end{lstlisting}
Applying this query on the graph of Fig.~\ref{fig:4_5_relation} for \emph{typeR} $\equiv$ \emph{Case\_AO} yields the new entity $(1,1)$ of type \emph{Case\_AO} in Fig.~\ref{fig:4_5_entities_composite} which refers to both original entities \emph{Application 1} and \emph{Offer 1}.

Any event $e$ correlated to an entity $n$ to which the composite entity $r$ of \emph{typeR} refers by a ``Reified'' relation also correlates to $r$. The following query achieves this.
\begin{lstlisting}[escapechar=\#,style=smallStyle]
MATCH ( e:Event ) -[:CORR]-> (n:Entity) <-[:REL {Type:"Reified"}]- (r:Entity {EntityType: '#\emph{typeR}#'})
CREATE (e) -[:CORR]-> (r)
\end{lstlisting}
Applying this query on the graph obtained in the previous step for \emph{typeR} $\equiv$ \emph{Case\_AO} adds the \emph{CORR} relationship from $e2$ and $e4$ to entity $(1,1)$ of type \emph{Case\_AO}. Depending on the available domain knowledge, the correlation query can be made more specific by adding \emph{WHERE} clauses for only correlating events that satisfy specific properties.

We may now derive \emph{:DF} relationships for the composite entity \emph{typeR} using the queries of Sect.~\ref{sec:storing:df}, e.g., Fig.~\ref{fig:4_5_entities_composite} also shows relationship \emph{:DF} for \emph{Case\_AO}. By correlating events of related entities \emph{Application 1} and \emph{Order 1} to their own reified entity \emph{Case\_AO (1,1)}, we constructed a new \emph{:DF}-relationship for entity type \emph{Case\_AO} describing only the interaction between \emph{Application 1} and \emph{Offer 1}. The original directly-follows relations for Application and Offer remain as before and ``unpolluted''.

\subsection{Demonstration on BPIC17}\label{sec:storing:demo_bpic17}

We applied the above queries on the events of the full BPIC17 dataset~\cite{BPIC2017}. After importing all events\footnote{For the visualizations in this section, we filtered out events with life-cycle attribute \emph{suspend} or \emph{resume} for reducing the size of the figures.}, we derived entities for 3 types: \emph{Application}, \emph{Workflow}, \emph{Offer}. We reified the binary relations between the three entities into \emph{Case\_AO}, \emph{Case\_AW},  \emph{Case\_WO} and derived their \emph{:DF} relationships.

\begin{figure}
    \centering
    \includegraphics[width=\linewidth]{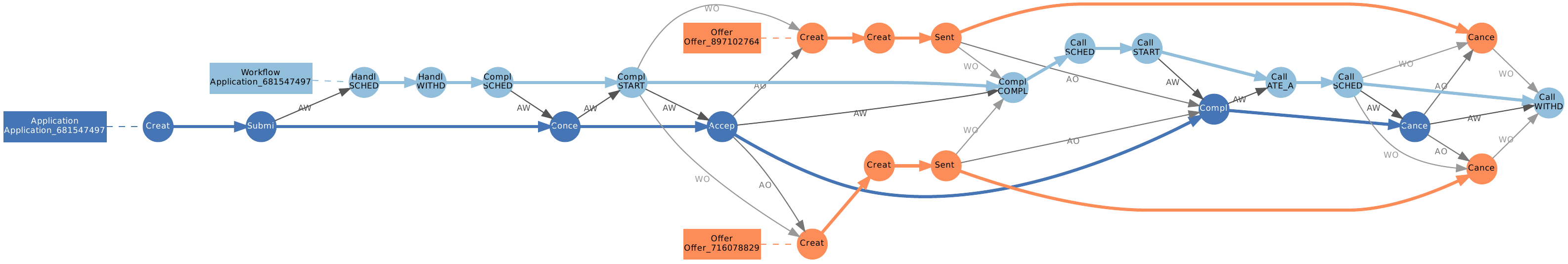}
    \includegraphics[width=\linewidth]{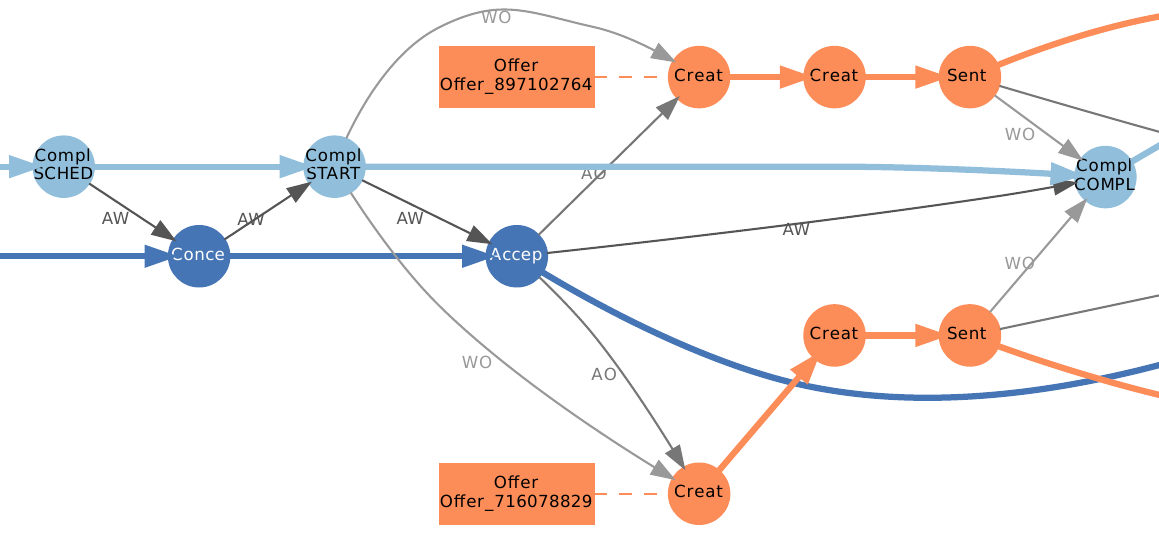}
    \caption{Graph of handling loan application 681547497 in BPIC17~\cite{BPIC2017} (top) with detail of two parallel offers (bottom)}
    \label{fig:bpic17_singlecase}
\end{figure}

Figure~\ref{fig:bpic17_singlecase} shows the graph of handling loan application 681547497 involving one Application entity (dark blue), one Workflow entity (light blue), and two Offer entities (orange). Interactions are shown through the grey \emph{:DF}-relationships of \emph{Case\_AO}, \emph{Case\_AW}, and \emph{Case\_WO}.\footnote{To simplify the visualization, the graph does not contain \emph{:DF\_Case\_AO}, \emph{:DF\_Case\_AW}, \emph{:DF\_Case\_WO} relationships which are in parallel to a \emph{DF\_Application}, \emph{DF\_Workflow}, \emph{DF\_Offer} relationship.} The graph shows how both Offers are created and handled concurrently to the application entity.

\begin{figure}
    \centering
    \includegraphics[width=\linewidth]{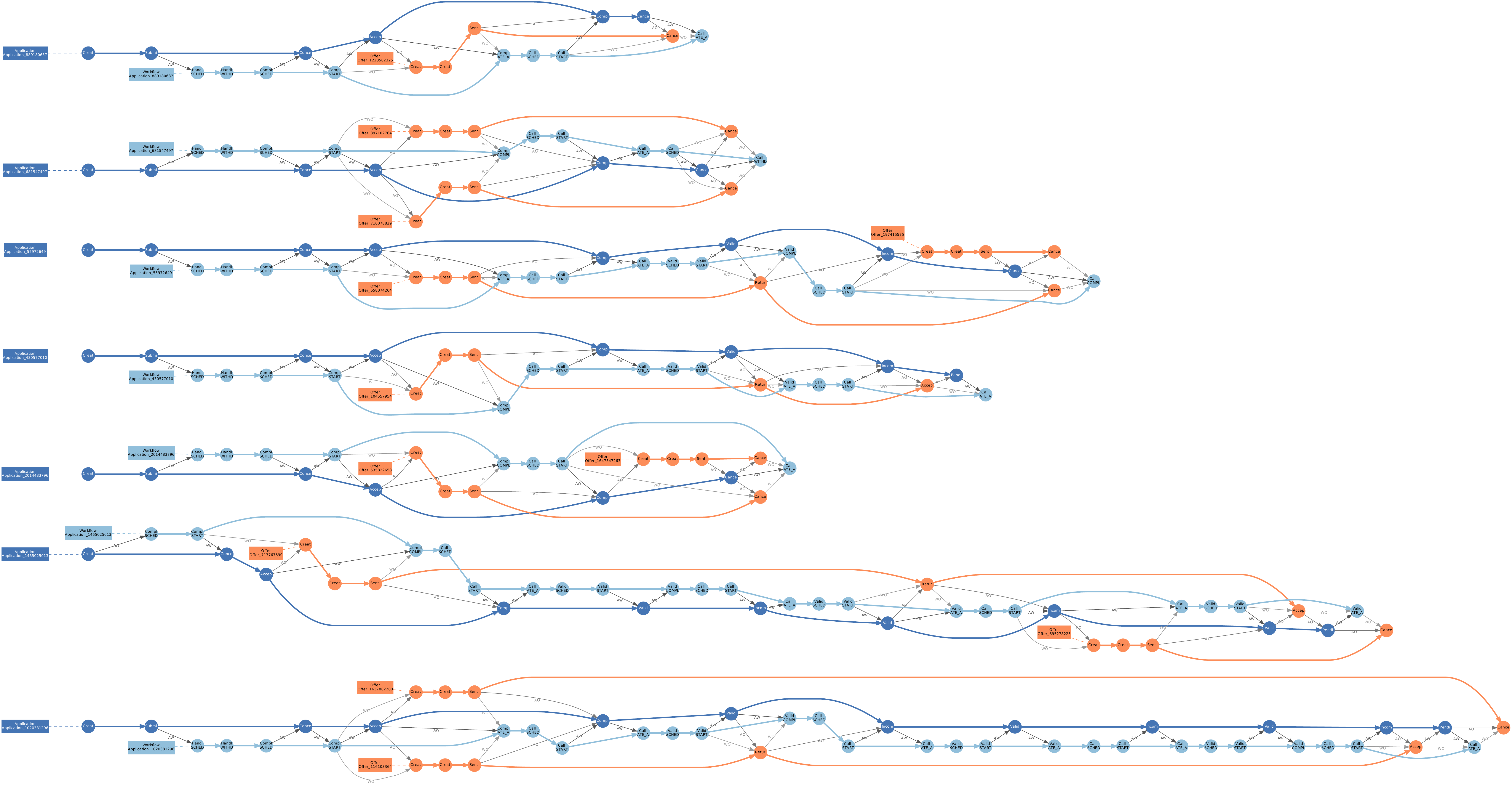}
    \caption{Graph of 7 randomly selected process executions of BPIC17~\cite{BPIC2017}}
    \label{fig:bpic17_multiple_cases}
\end{figure}

Figure~\ref{fig:bpic17_multiple_cases} visualizes 7 randomly selected process executions: the 1st and 4th involve only one Offer whereas all others involve two Offer entities; some executions in BPIC17 involve 5 or more Offer entities. Offers may be created in parallel (2nd, 7th) or with Application events in between (3rd, 5th, 6th). Offers may conclude in parallel (2nd, 3rd, 4th) or with Application events in between (6th, 7th).

\begin{figure}
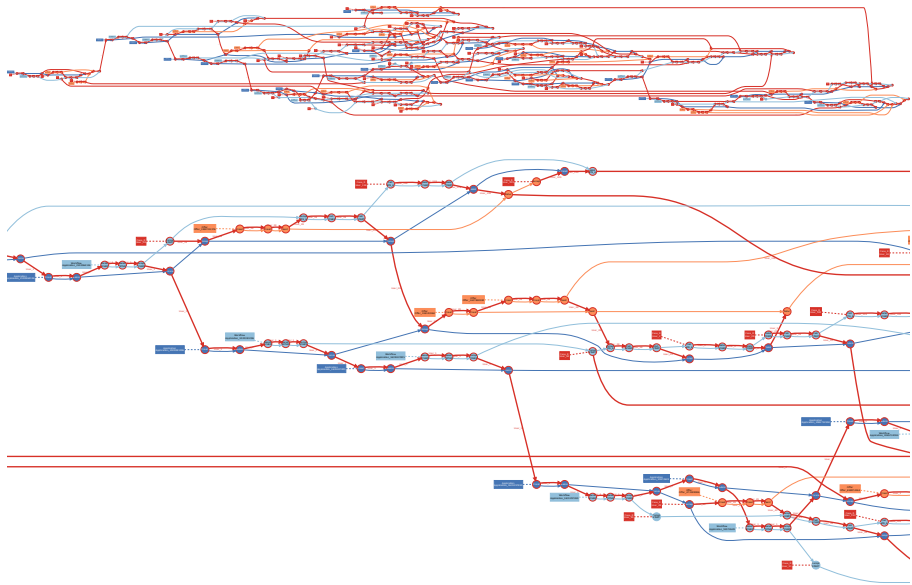

    \centering
    \includegraphics[width=\linewidth]{figures/bpic17_graph_resource_connects_all.pdf}
    \includegraphics[width=\linewidth]{figures/bpic17_graph_resource_connects_all_zoom_in.pdf}
    \caption{Behavior of resources overlaid on the 7 process executions of Fig.~\ref{fig:bpic17_multiple_cases}(top) and some details (bottom).}
    \label{fig:bpic17_resources_across_cases}
\end{figure}

We then derived \emph{Resource} as additional entity from the \emph{e.resource} property of events. While Application, Workflow, and Offer are local to a process execution, the \emph{Resource} entities describe works who persist in the system and work on many entities. We derived the \emph{:DF} relationships for \emph{Resource} entities. Querying the data for the events of the 7 process executions of Fig.~\ref{fig:bpic17_multiple_cases} and the \emph{:DF} relationships of all entities results in the graph in Fig.~\ref{fig:bpic17_resources_across_cases}, where \emph{Resource}-\emph{:DF} relationships are shown in red.

We can clearly see that all process executions and entities are tightly connected through the resources. Each resource is always involved for a sequence of several events of the same or related entities, and then moves to another entity in another process execution while handing the previous entity over to another resource.

\begin{figure}
    \centering
    \includegraphics[width=\linewidth]{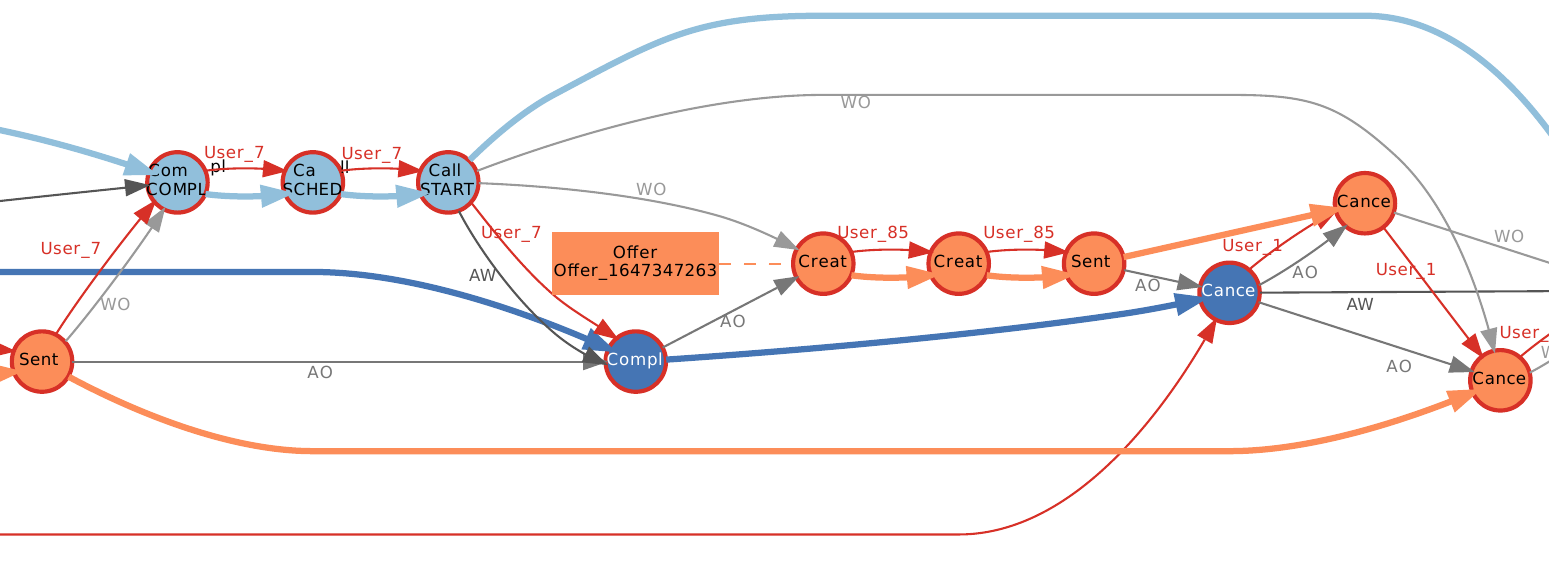}
    \caption{Resource behavior does not explain entity interactions}
    \label{fig:bpic17_resource_vs_derived}
\end{figure}

Overlaying the \emph{Resource}-\emph{:DF} relationships on the graphs also allows us to see that interactions between related Application, Workflow, and Offer entities of the same process execution are not explained by Resource entities. In the graph in Fig.~\ref{fig:bpic17_resource_vs_derived}, the first event of \emph{Offer\_1647347263} correlated to \emph{User\_85} follows after an Application event (via \emph{Case\_AO}) correlated to \emph{User\_7}, i.e., there is no resource explaining the ordering of Application and Offer. This confirms the importance of reifying relations between entities into composite entities\,---\,otherwise the graph would describe that \emph{Offer\_1647347263} would start concurrently to all preceding events.

\subsection{Evaluation}\label{sec:storing:evaluation}

We applied the above steps for importing and transforming the event data into our proposed graph-based data model on 5 real-life datasets~\cite{BPIC2014,BPIC2015,BPIC2016,BPIC2017,BPIC2019}\footnote{For BPIC2016, we omitted all click events without a session identifier as these could not be correlated.} using a Neo4j instance with 20GB of main memory allocated. The Cypher queries are available at \url{https://github.com/multi-dimensional-process-mining/graphdb-eventlogs} and at~\cite{graphdataset}. The resulting graphs are available at~\cite{BPIC2014_graph,BPIC2015_graph,BPIC2016_graph,BPIC2017_graph,BPIC2019_graph}.

Table~\ref{tab:logStatsAll:types} summarizes the domain knowledge we applied for each data set during the conversion in terms of the number of distinct event logs, entity types, relationship types between entities, and how many relationship types were materialized as derived entities. Although the source data for BPIC'14~\cite{BPIC2014}, and BPIC'16~\cite{BPIC2016} contains several CSV files, the event records in these CSV files refer to shared entities. We therefore did not import the events from these files under separate \emph{:Log} nodes but \emph{integrated} them under a single \emph{:Log} node. Only BPIC'15~\cite{BPIC2015} contains 5 distinct event logs which were also imported as separate \emph{:Log} nodes. Each data set contained multiple entity types. All data sets contain a resource entity type. BPIC'14~\cite{BPIC2014_graph} has no classical case identifier at all but describes a complex interplay of hard- and software \emph{Components} and their \emph{Configuration Items}, for which \emph{Incidents} are reported which are handled in \emph{Interactions} leading to \emph{Changes} on the \emph{Components} supported by \emph{Knowledge Base Entries}. BPIC'15~\cite{BPIC2015_graph} has a classical case notion and \emph{3 different notions of Resources} involved in an event. BPIC'16~\cite{BPIC2016} has a \emph{Customer}, \emph{Complaints} which are collected in \emph{Dossiers}, users seeking information on a website in \emph{Browsing Sessions} from an \emph{IP Address}. BPIC'17~\cite{BPIC2017_graph} has \emph{Loan Applications}, \emph{Offers}, and a \emph{Workflow} instance handling these documents, and entities derived from the relations between these. BPIC'19~\cite{BPIC2019_graph} has \emph{Purchase Orders} containing \emph{Items} supplied by \emph{Vendors}. Only BPIC'17 required materializing relationships to represent the data correctly.

\begin{table}[]\small\centering
\begin{tabular}{lrrrrr}
\hline
Data Set & :Log  & :Entity & :REL & :REL types\\
         & nodes & types   & types & materialized \\ \hline
BPIC'14~\cite{BPIC2014_graph}  & 1    & 7 & 2 & 0 \\
BPIC'15~\cite{BPIC2015_graph}  & 5    & 4 & 2 & 0 \\
BPIC'16~\cite{BPIC2016_graph}  & 1    & 7 & 0 & 0 \\
BPIC'17~\cite{BPIC2017_graph}  & 1    & 7 & 3 & 3 \\
BPIC'19~\cite{BPIC2019_graph}  & 1    & 4 & 1 & 0 \\ \hline
\end{tabular}
\caption{Type information of labeled property graphs of 5 real-life event data sets}
\label{tab:logStatsAll:types}
\end{table}

\begin{figure}
    \centering
    \includegraphics[width=\linewidth]{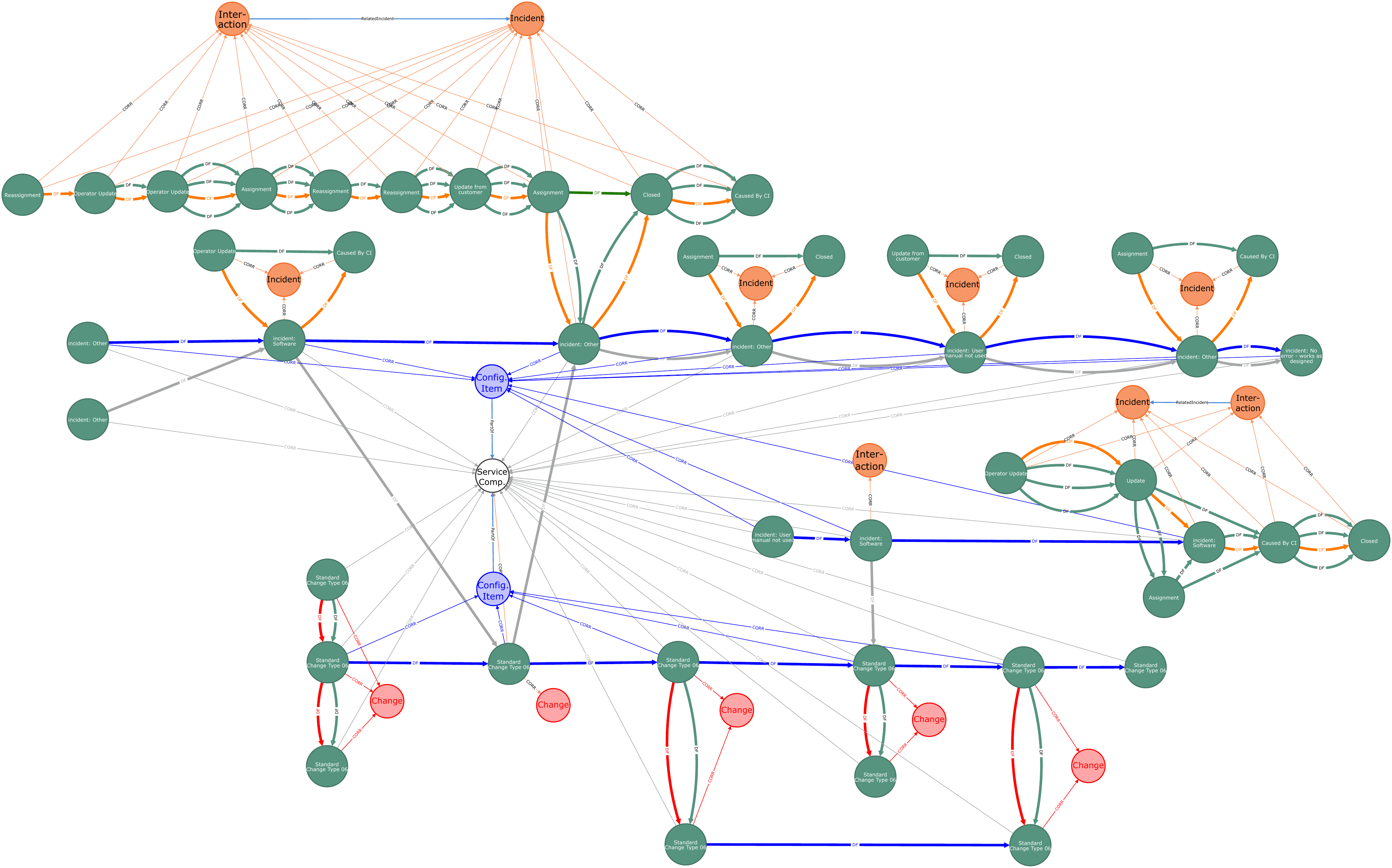}
    \caption{Part of BPIC'14 data showing a service component (white) with two related configuration items (blue), and related incidents and interactions (orange) and changes (red) and their events (green).}
    \label{fig:bpic14_sample_structure}
\end{figure}

All data sets could be converted using our method. We measured the size of the resulting graphs, the memory requirements for storing the data in Neo4j and the time required for the conversion. Table~\ref{tab:logStatsAll:graph_size} shows the number of nodes and relationships for the labels \emph{:Event}, \emph{:Entity}, \emph{:Class}, \emph{:DF}, \emph{:CORR}, and \emph{:REL}.
Where a classical event log of $n$ events and $m$ cases only contains $n-m$ \emph{:DF}-relationships, all event graphs have between $1.9\cdot n$ (BPIC'17) and $4.8 \cdot n$ (BPIC'16) \emph{:DF}-relationships due to the additional entity types (c.f.~Tab~\ref{tab:logStatsAll:types}). Consequently, almost all events have more than one incoming or outgoing \emph{:DF}-relationship. On average, in each graph, each event is related to four other entities (c.f., \emph{:CORR} vs \emph{:Entity}). By comparison, structural relations are much sparser in the graphs. Further, in each graph, each event has one \emph{:HAS} and one \emph{:OBSERVES} relationship. Overall the graphs have between 7.2 (BPIC'14, BPIC'17) and 10.7 (BPIC'16) edges per node. Figure~\ref{fig:bpic14_sample_structure} illustrates the typical structures found in these graphs by a (small) part of the BPIC'14 graph~\cite{BPIC2014_graph}.
The graph shows 17 entities of 5 different types that are all directly or indirectly connected.

\begin{table}[]\small
\begin{tabular}{lrrrrrr}
\hline
Data Set                       & \multicolumn{3}{c}{Nodes} & \multicolumn{3}{c}{Relationships} \\
                               & :Event    & :Entity & :Class & :DF       & :CORR      & :REL    \\ \hline
BPIC'14~\cite{BPIC2014_graph}  & 690,622   & 228,885 & 330    & 2,503,328  & 2,732,213  & 65,601  \\
BPIC'15~\cite{BPIC2015_graph}  & 262,628   & 5,649   & 356    & 1,044,650  & 1,050,512  & 107     \\
BPIC'16~\cite{BPIC2016_graph}  & 7,360,146 & 26,647  & 620    & 35,681,967 & 36,430,880 & 0       \\
BPIC'17~\cite{BPIC2017_graph}  & 1,202,267 & 255,170 & 66     & 2,298,372  & 5,244,794  & 352,497 \\
BPIC'19~\cite{BPIC2019_graph}  & 1,595,923 & 328,083 & 52     & 5,653,917  & 5,984,602  & 251,734 \\ \hline
\end{tabular}
\caption{Logical size of labeled property graphs of 5 real-life event datasets}
\label{tab:logStatsAll:graph_size}
\end{table}

Table~\ref{tab:logStatsAll:size_time} lists the resulting storage sizes of the Neo4J instances and conversion times.
Explicitly encoding the structural and behavioral information in the graph requires a factor 3 (BPIC'15) to 13.3 (BPIC'17) more space than the source data in CSV format. Relationships account for 35\%-50\% of the space while properties account for 40\%-61\% of the Neo4j storage.

\begin{table}[]\small
\begin{tabular}{lrrrrrrrr}
\hline
Data Set & source size &  \multicolumn{5}{c}{DB size (GB)}        & \multicolumn{2}{c}{time (mins)} \\
         &    (GB)  & nodes & prop. & strings & rel's & total  &  all  & no :REL\\ \hline
BPIC'14~\cite{BPIC2014_graph}  & 0.08     & 0.01  & 0.34  & $<$0.01 & 0.31  & 0.67   & 11.2 & 3.1  \\
BPIC'15~\cite{BPIC2015_graph}  & 0.11     & $<$0.01 & 0.20  & 0.01  & 0.12  & 0.33   & 11.7 & 1.3  \\
BPIC'16~\cite{BPIC2016_graph}  & 1.06     & 0.12  & 3.42  & 0.76    & 4.27  & 8.60   & 58.6 & 58.6 \\
BPIC'17~\cite{BPIC2017_graph}  & 0.29     & 0.80  & 2.37  & 0.05    & 1.36  & 3.87   & 8.2  & 8.0  \\
BPIC'19~\cite{BPIC2019_graph}  & 0.52     & 0.02  & 0.96  & 0.24    & 0.90  & 2.13   & 9.7  & 8.7  \\ \hline
\end{tabular}
\caption{Resulting database size and conversion time for converting 5 real-life event datasets into labeled property graphs}
\label{tab:logStatsAll:size_time}
\end{table}

All translations, except for BPIC'15~\cite{BPIC2015} succeeded within several minutes. We observed execution time to depend on two particulars. (1) On the product $n \cdot k$ of the number $n$ of events and $k$ of entity types: our conversion approach iterates over all $n$ events at most $2\cdot k$ times (to create an entity from an event attribute and to derive the \emph{:DF} relationship for this entity). The large number of events and entity types thus causes the higher running time for BPIC'15. (2) On the number $m$ of entity nodes when deriving structural relationships; in the worst case (almost) all $m^2$ pairs of entities nodes have to be checked for the presence of a relation, which was the case in BPIC'15 and for BPIC'14, where execution time without constructing \emph{:REL} relationships is significantly lower.

In general, the size of the source log is not a reliable indicator for the size of a graph event log. For example, the BPIC'14 source data~\cite{BPIC2014} is small in size but defines several related entities resulting in a large graph. More importantly, any graph can be adapted to the needs of a particular research question, e.g., by deriving only a limited number of (composite) entities.


\section{Querying Multi-Dimensional Event Data}\label{sec:querying}

In the following we present 6 classes of analysis questions that we formulated to evaluate requirements R5-R11 of  Sect.~\ref{sec:background:multi-dim-req} for querying multi-dimensional event data on the LPGs of Sect.~\ref{sec:storing}. In Sect.~\ref{sec:mining}, we evaluate the aggregation requirements R12-R15.

We conducted the querying experiment on the BPIC'17 dataset for which we additionally derived the \emph{Case\_AWO} entity type which corresponds to the original case notion, i.e., all events sharing the same \emph{e.case} attribute. We did this in order to be able to verify the correctness of our results against classical process mining software which works with the original case notion only. For each analysis question we provide a Cypher query and report results and the query processing times (measured on a 6 Core Intel i7-9850H @ 2.60 GHz windows machine with 32 GB RAM @ 2667 MHz with Neo4j Browser).

\subsection{Graph Queries}

\par\vspace{.5em}\noindent\textbf{Q1. Query Attributes of Events/Cases}.
We want to query for the first-class concepts of event logs: a case and an event based on event/case attributes by using a partial patterns to satisfy R7. The following query returns the event attribute ``timestamp'' and the case attribute ``LoanGoal'' of Case ``Application\_681547497''\footnote{The specific cases were randomly selected from all available cases in the data}. Note that all (event and entity) attributes are encoded as properties of event nodes.
\begin{lstlisting}[style=smallStyle]
MATCH (c:Entity {EntityType: 'Case_AWO'}) <-[:CORR]- (e:Event)
WHERE c.ID = "Application_681547497" AND e.Activity = "A_Submitted"
RETURN e.timestamp, e.LoanGoal
\end{lstlisting}
The query has been processed in 0.079 seconds. After modifying the query to consider all cases, i.e. remove the condition for a specific case in line 2, the query completed in 0.117 seconds.

\par\vspace{.5em}\noindent\textbf{Q2. Query Directly-Follows Relations}.
Q2 is focused on temporal aspects. Here we show a query that satisfies R8 by considering 2 consecutive events.
Directly-follows relations of events in a case are an important characteristic of event logs as they represent the case internal temporal order of events and many of today's process mining techniques rely on these relations. The query below returns the event directly following the node with the activity property ``O\_Created'' of a given offer entity by matching the \textit{:DF\_Offer} relationship.


\begin{lstlisting}[style=smallStyle]
MATCH (o:Entity {EntityType: 'Offer'}) <-[:CORR]- (e1:Event) <-[:DF {EntityType: 'Offer'}]- (e2:Event)
WHERE o.ID = "Offer_716078829" AND e1.Activity = "O_Created"
RETURN e1,e2
\end{lstlisting}

The query execution time for one specific offer was 0.160 seconds whereas querying the \textit{:DF\_Offer} relations with destination node ``O\_Created'' for all 42,995 offers took 0.146 seconds. Directly-follows relations of other entities (Application and Workflow) or across entities (Case\_AWO) can be queried by adjusting the query in the \textit{MATCH} and \textit{WHERE} clauses accordingly.

\par\vspace{.5em}\noindent\textbf{Q3. Query Eventually-Follows Relations}.
We want a query that satisfies R8 by considering the temporal relationship of any 2 events of a case. Eventually-follows relations are also related to the case internal order of events. Event $y$ \emph{eventually follows} event $x$ if $y$ occurs after $x$ in the same case, that is, if $x$ and $y$ are connected through a path of directly-follows relations of arbitrary length. We query the offer specific eventually-follows relationship between ``O\_Created'' and ``O\_Cancelled'' for a given offer as follows:


\begin{lstlisting}[style=smallStyle]
MATCH (o:Entity {EntityType: 'Offer'}) <-[:CORR]- (e1:Event) -[:DF* {EntityType: 'Offer'}]-> (e2:Event)
WHERE o.ID = "Offer_716078829" AND e1.Activity = "O_Created" AND e2.Activity = "O_Cancelled"
RETURN e1,e2
\end{lstlisting}

Even though the \textit{MATCH} clause looks similar to the one of the directly-follows query, the *-Operator changes the pattern from a direct relationship to a path of arbitrary length. Since we want to find the eventually-follows relationship of two specific activities we also added condition $\mathit{e2.Activity} = \textit{``O\_Cancelled''}$ to the \textit{WHERE} clause to define the endpoint of the paths we want to match in the graph. For the given offer the query took 0.161 seconds. For all 20,898 offers where ``O\_Created'' is eventually followed by ``O\_Cancelled'' we removed the condition for ``Offer\_716078829'' from the query which then took 0.140 seconds.

\par\vspace{.5em}\noindent\textbf{Q4. Case Variants}.
We want a query to return a case variant as path in the graph to satisfy R6.
A \emph{case variant} is the sequence of activities of a case. Case variants are for example used to detect frequent behaviour of a process. We can query the graph to retain the path of events of a case by walking over all of its directly-follows relationships from the first to the last event. For a given case (Case\_AWO) this can be done as follows:


\begin{lstlisting}[style=smallStyle]
MATCH (c:Entity {EntityType: 'Case_AWO'}) <-[:CORR]- (e1:Event) -[:DF* {EntityType: 'Case_AWO'}]-> (e2:Event)
WHERE NOT ()-[:DF {EntityType: 'Case_AWO'}]->(e1) AND NOT (e2)-[:DF {EntityType: 'Case_AWO'}]->() AND c.ID = 'Application_681547497'
RETURN (e1:Event) -[:DF* {EntityType: 'Case_AWO'}]-> (e2:Event) AS paths
\end{lstlisting}

The pattern of the match clause follows the same logic as the eventually-follows match pattern. For variants we limit the output to the first and last event of a case, i.e. the events that have no incoming or no outgoing ``:DF'' relationship with ``Case\_AWO'' entity type. The query completed in 0.250 seconds. Similarly, we can query the graph for variants of another entity such as Offer. The paths of events returned by the above query can be turned in a list of activity sequences by Cypher's list operators: \textit{UNWIND} processes each path in the \textit{paths} variable iteratively, function \textit{nodes()} translates the path into a list of nodes, and list comprehension maps each event node to its activity property. The resulting list of activities can be compared for equality with other lists, etc.

\par\vspace{.5em}\noindent\textbf{Q5. Query Duration/Distance between two specific Activities}.
The information on how much time or how many activities were needed to get an Offer from ``O\_Created'' to ``O\_Accepted'' for example can be used to measure process performance. For Q5 we want to query temporal relations in the form of durations and path lengths to satisfy R8.
Say we are interested in the offer entity that took the longest time to get accepted. We can query the eventually-follows relation of two given activities and use their timestamps to calculate the elapsed time between them:

\begin{lstlisting}[style=smallStyle]
MATCH (e1:Event) -[:CORR]-> (o:Entity {EntityType: 'Offer'}) <-[:CORR]- (e2:Event)
WHERE e1.Activity = "O_Created" and e2.Activity = "O_Accepted"
WITH e1, e2, duration.indays(e1.timestamp, e2.timestamp) AS days, duration.between(e1.timestamp, e2.timestamp) AS time, o
ORDER BY time DESC
LIMIT 1
RETURN e1, e2, days.days as days, time.hours  AS hours, time.minutes % 60 as minutes , o
\end{lstlisting}

The query matches all \textit{:CORR} relationships, filters for the given activities and then uses Cypher's duration function to calculate the time spans. Only the result with the longest duration is returned. In case we want to retrieve the distance wrt. the number of activities, we can aggregate over the nodes along the path between the two events with eventually-follows relation and count the hops with the ``Length()'' function as shown in~\cite{Esser2019cs_tue}. The query for the elapsed time completed in 1.072 seconds. Querying for the longest path took 2.176 seconds.


\par\vspace{.5em}\noindent\textbf{Q6. Query for Behavior across Multi-Instance Relations}.
Event logs such as BPIC'17 can contain multiple case identifiers. A case identifier may be a single entity, e.g. Offer, or any combination of entities such as the Case notion of BPIC'17 combining Application, Workflow and Offer entities. Querying the behavior across different instances of these entities typically requires multiple steps with traditional event logs such as custom scripts to be able to select, project, aggregate and combine the results accordingly. With Q6 we want to satisfy R9 by querying for events correlated to the same entity, R10 by combining data from different entities in the same query, and to satisfy R11 by querying 2 (sub)processes in a single query.
We defined a query that returns (1) all paths from ``A\_Create Application'' to ``O\_Cancelled'' (2) for only those BPIC'17 Cases that have more than one Offer with ``O\_Created'' directly followed by ``O\_Cancelled'' (from the perspective of the Offer entity, not the global case entity).

\begin{lstlisting}[style=smallStyle]
MATCH (o:Entity {EntityType: "Offer"})<-[:CORR]-(e1:Event {Activity: "O_Created"}) -[df:DF {EntityType: "Offer"}]-> (e2:Event {Activity: "O_Cancelled"})-[:CORR]->(o)
MATCH (e2)-[:CORR]->(c:Entity {EntityType: "Case_AWO"})<-[:CORR]-(e1)-[:CORR]->(o)
WITH c, count(o) AS ct
WHERE ct > 1
MATCH (o:Entity {EntityType: "Offer"})<-[:CORR]-(e1:Event {Activity: "O_Created"}) -[df:DF {EntityType: "Offer"}]-> (e2:Event {Activity: "O_Cancelled"})-[:CORR]->(o)
MATCH (e2)-[:CORR]->(c)<-[:CORR]-(e1)-[:CORR]->(o)
WITH e2 AS O_Cancelled,c
MATCH (A_Created:Event {Activity: "A_Create Application"})-[:CORR]->(c)<-[:CORR]-(O_Cancelled:Event {Activity: "O_Cancelled"})
MATCH p = (A_Created) -[:DF* {EntityType: "Case_AWO"}]-> (O_Cancelled)
RETURN p
\end{lstlisting}

The query demonstrates several unique aspects of querying multi-dimensional event data in labeled property graphs.

The first \textit{MATCH} clauses (lines 1-4) return all entity nodes $c$ of the original case notion ``Case\_AWO'' which are related to more than one Offer $o$ (via \textit{:CORR}) for which ``O\_Created'' is directly followed by ``O\_Cancelled'' (via \textit{:DF} for entity type ``Offer''). The second pair of \textit{MATCH} clauses (lines 5-7) return all ``O\_Cancelled'' events that directly succeed ``O\_Created'' (via \textit{:DF} for ``Offer'') and are correlated to one of the cases $c$ with multiple offers (found in Lines 1-4). The returned ``O\_Cancelled'' events are used in the last pair of \textit{MATCH} clauses (lines 8-9) to return paths from some ``A\_Create Application'' event (also corrlated to $c$) to one of the ``O\_Cancelled'' events returned previously. This way we get a unique path for every Offer that meets the criteria.

\begin{figure}[t]
    \centering
    \includegraphics[width=\linewidth]{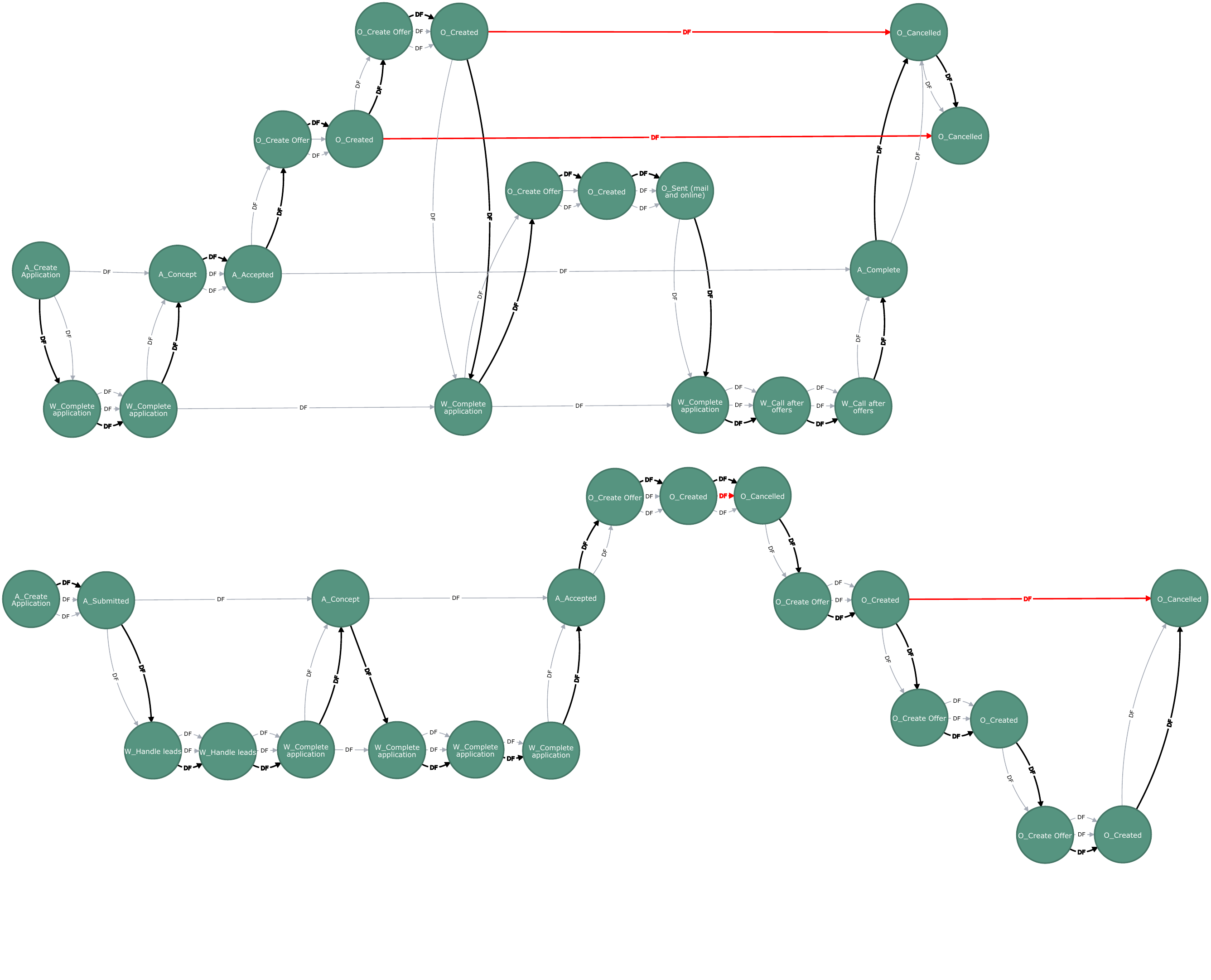}
    \caption{Q6 Output}
    \label{fig:Q6}
\end{figure}

The query's execution time was 1.246 seconds in Neo4j Browser. Figure~\ref{fig:Q6} shows 2 of the 218 paths of the query's output in Neo4j's graphical representation. The thick black edges show the returned path of events along the \emph{:DF}-relationships for \emph{Case\_AWO}. The grey edges show additional \emph{:DF}-relationships for \emph{Application}, \emph{Offer}, and \emph{Workflow} entities, revealing how the path walks across different entities. The red edges highlight where ``O\_Create Offer'' is directly followed by ``O\_Created'', which is the graph feature used by the query to identify these paths.

\subsection{Graph Query Validation and Performance}

We discuss the validity and performance of our Cypher queries against other implementations.

To validate the above queries again ground truth, we used process mining software such as ProM and Disco with standard operators as baseline implementation to filter the event data on BPIC'17 event log to generate the set of expected results for the graph queries. We used Disco's \emph{Attribute} filter for Q1, the \emph{Follower} filter for Q2 and Q3, the \emph{Attribute} filter on a case attribute for Q4, and the \emph{Follower} filter with constraints on the time between eventually following events for Q5. Q6 could not be queried with classical process mining tools but required a custom procedural algorithm using a single-pass search over the data.

Our Cypher queries obtained the same result as the baseline implementations. The graph analysis for Q1-Q6 required only Cypher queries with clauses and functions as described in~\cite{francis2018cypher} (except for the typecasts which are not part of Cypher but provided by Neo4j). A technical report provides full details~\cite{Esser2019cs_tue}.

We compare the performance of querying our event data model to Disco and to the specialized business intelligence software Qlik Sense, which provides a custom data engine specialized for efficient data analysis. In Neo4j we used the property graph database as described in Sect.~\ref{sec:storing} and took the reported query time from the database. The running time in Disco has been measured manually with a preloaded log and a pre-configured set of filtering operations As Qlik Sense has no native query language, we built a custom data model to transform and filter the data according to the query logic. Populating the data model is included in the measured runtime of the Qlik Sense queries which was measured using system time. We validated the correctness of the Qlik Sense queries against the ground truth developed in Disco.

\begin{table}[]\small
\revision{
\begin{tabular}{lrrrr}
\hline
Query           & Runtime Neo4j     & Runtime Qlik  & Runtime Disco      & Results Match     \\ \hline
Q1 single event & 0.079 s           & 0.571 s       & 1.10 s                  & yes          \\
Q1 full log     & 0.117 s           & 0.516 s       & 1.62 s                  & yes          \\
Q2 single event & 0.160 s           & 1.892 s       & 1.27 s                  & yes         \\
Q2 full log     & 0.146 s           & 5.866 s       & 3.28 s                  & yes       \\
Q3 single event & 0.161 s           & 3.203 s       & 1.40 s                  & yes         \\
Q3 full log     & 0.140 s           & 3.820 s       & 2.82 s                  & yes       \\
Q4              & 0.250 s           & 2.095 s       & 0.89 s                  & yes     \\
Q5 time         & 1.072 s           & 1.679 s       & 2.34 s                  & yes        \\
Q5 path         & 2.086 s           & 6.391 s       & 1.54 s                  & yes       \\
Q6              & 1.246 s           & 6.816 s       & 15 mins${}^*$              & yes       \\ \hline
                &                   &               & ${}^*$custom script     &        \\
\end{tabular}
\caption{Query performance and correctness overview}
\label{tab:performanceStatsAll}
}
\end{table}

Table~\ref{tab:performanceStatsAll} gives an overview of the corresponding results for BPIC'17. All approaches returned the same result. Neo4j was fastest in all queries except Q5 where Disco performed best. Our custom procedural algorithm to validate the correctness of Q6 required 15mins compared to less than 2 seconds for Neo4j.

{
\subsection{Expressive Power of Cypher for Behavioral Queries}
\label{sec:querying:expressive}
}

We finally discuss the expressive power of Cypher for behavioral queries. We specifically focus on Cypher's abilities to specify paths compared to theoretical graph query languages~\cite{DBLP:series/synthesis/2018Bonifati}] and to PQL~\cite{DBLP:journals/is/PolyvyanyyPH20,DBLP:journals/corr/abs-1909-09543}.

Regular path queries (RPQs)~\cite{DBLP:series/synthesis/2018Bonifati} are queries built from edges (as atoms), edge inversion, path concatenation (sequence), choice between two alternative paths, and the transitive closure. Conjunctive RPQs allow to query for paths which must overlap in named edges or nodes, e.g., the same start and end nodes. Cypher supports path conjunction but not the full extent of RPQs~[Ch.3]\cite{DBLP:series/synthesis/2018Bonifati} as not all forms of transitive closure are supported. This renders Cypher weaker than regular languages. However, ongoing standardization efforts subsume Cypher in the more expressive graph query language G-Core~\cite{DBLP:conf/sigmod/AnglesABBFGLPPS18}. Any developments in this field will be applicable to our data model on standard labeled property graphs.

PQL has two alternative approaches to behavioral querying. Scenario-based queries allow to specify a trace pattern with wildcards that has to be matched~\cite{DBLP:journals/is/PolyvyanyyPH20}; the SQL-like syntax of PQL also allows to specify attributes. This renders PQL comparable to behavioral querying with Cypher; though Cypher's conjunctive path queries are more expressive than sequences with wildcards. PQL also supports querying for behavioral relations~\cite{DBLP:journals/corr/abs-1909-09543} from the 4C spectrum~\cite{DBLP:conf/apn/PolyvyanyyWCRH14}. In terms of our data model, these behavioral relations make statements over \emph{Class} nodes, i.e., sets of events, for example ``all events of activity A precede all events of activity B''. Some behavioral relations of PQL can be expressed in Cypher. For instance, the following query returns all entities of type $X$ (process executions) where activity $A$ and $B$ are in conflict (never occur together).
\begin{lstlisting}[style=smallStyle]
MATCH (n:Entity {EntityType:"X"})
WHERE (n)<-[:CORR]-(:Event {Activity:"A"}) AND NOT (n)<-[:CORR]-(:Event {Activity:"B"})
OR NOT (n)<-[:CORR]-(:Event {Activity:"A"}) AND (n)<-[:CORR]-(:Event {Activity:"B"})
RETURN n
\end{lstlisting}
However, a more detailed comparison requires further research for two reasons: (1) PQL is aimed at retrieving models which specify a language while our Cypher-based query language is aimed at retrieving individual executions (words of a language). (2) Our current data model does not express concurrency within an entity, i.e., all events are ordered over time, so concurrency only emerges between different entities. PQL in turn is not designed to handle multiple entities in a process, hindering a comparison.

\section{Constructing Simple Models for Multiple Entities}\label{sec:mining}

We now show that our data model of Sect.~\ref{sec:represent} allows \emph{aggregating} the directly-follows relations between events to directly-follows relations between event classes\,---\,taking the notion of entities and entity types into account. We provide queries that satisfy R12-R15 of Sect.~\ref{sec:background:multi-dim-req}.

\subsection{Aggregating events into user-defined event classes}
\label{sec:mining:class}

An event \emph{:Class} is a node describing a \emph{set} of events with the same characteristics, e.g., having the same \emph{Activity} or other combination of data attributes. We can aggregate events into user-defined event classes using the same principles as deriving and correlating \emph{:Entity} nodes: we query for all distinct values of a particular (combination of) event attributes and create a new \emph{:Class} node per retrieved value(s). The following two queries illustrate the concept for two types of event classes: the combination of activity name and life-cycle attribute, and the resource attribute.
\begin{lstlisting}[style=smallStyle]
MATCH (e:Event) WITH distinct e.Activity AS actName,e.lifecycle AS lifecycle
MERGE ( c : Class { Name:actName, Lifecycle:lifecycle, Type:"Activity+Lifecycle", ID: actName+"+"+lifecycle})
\end{lstlisting}
\begin{lstlisting}[style=smallStyle]
MATCH ( e : Event ) WITH distinct e.resource AS name
MERGE ( c : Class { Type:"Resource", ID: name})
\end{lstlisting}
We then link each \emph{:Class} node to all events of this class when they match on the defining attributes, as for correlating events to entities. We show the query for \emph{Activity+Lifecycle}.
\begin{lstlisting}[style=smallStyle]
MATCH ( c : Class ) WHERE c.Type = "Activity+Lifecycle"
MATCH ( e : Event ) where e.Activity = c.Name AND e.lifecycle = c.Lifecycle
CREATE ( e ) -[:OBSERVED]-> ( c )
\end{lstlisting}
We may also derive event classes based on behavioral properties of events, e.g., based on an event \emph{(e2:Event) -[:DF]-$\mathord{>}$ (e)} preceding $e$. The above queries satisfy the semantic constraints for \emph{:OBSERVED} of Sect.~\ref{sec:semantics:e_c}.

\subsection{Aggregating directly-follows relations}
\label{sec:mining:df_c}

The \emph{(c1:Class) -[:DF\_C]-$\mathord{>}$ (c2:Class)} directly-follows relation between class $c1$ and class $c2$ aggregates all directly-follows relations \emph{(e1:Event) -[:DF]-$\mathord{>}$ (e2:Event)} between events $e1$ of $c1$ and $e2$ of $c2$, see Sect.~\ref{sec:semantics:df_c}.

We may only aggregate \emph{:DF} relationships between events correlated to the same entity $n$ (line 2) for which the \emph{:DF} relationship was also defined (line 3). Further classes $c1$ and $c2$ must be of the same \emph{Type} (line 3). This ensures that we satisfy R15. We aggregate by counting how many \emph{:DF} relationships (df) exist between $c1$ and $c2$ (line 4) and create a \emph{:DF\_C} relationship for this entity type between $c1$ and $c2$  and record the count as a property of the new \emph{:DF\_C} relationship.
\begin{lstlisting}[style=smallStyle]
MATCH (c1:Class) <-[:OBSERVED]- (e1:Event) -[df:DF]-> (e2:Event) -[:OBSERVED]-> (c2:Class)
MATCH (e1) -[:CORR] -> (n) <-[:CORR]- (e2)
WHERE n.EntityType = df.EntityType AND c1.Type = c2.Type
WITH n.EntityType as Type,c1,count(df) AS df_freq,c2
MERGE ( c1 ) -[rel2:DF_C {EntityType:Type}]-> ( c2 ) ON CREATE SET rel2.count=df_freq
\end{lstlisting}
Note that $c1$ and $c2$ can refer to the same node and thus self loops are also included in the graph.

Observe that the aggregation query builds only on concepts of our data model, \emph{Event}, \emph{Class}, and \emph{Entity} nodes and the relationships \emph{DF}, \emph{CORR}, \emph{OBSERVED}, and requires no further domain knowledge of the underlying event data. In other words, the aggregation query demonstrates that the data model provides the right abstraction concepts for event data over multiple entities.

Because of this abstraction, the above query can be applied on ``normal'' entities as well as composite entities of reified relations, satisfying R14.

\subsection{Demonstration on BPIC17}
\label{sec:mining:demonstration_bpic17}

We demonstrate the aggregation queries on the BPIC17 dataset for which have already derived entities and \emph{:DF} relationships for \emph{Application, Workflow, Offer, Case\_AO, Case\_WO, Case\_AW} (see Sect.~\ref{sec:storing:demo_bpic17}), Resource, and the original case notion \emph{Case\_AWO} (see Sect.~\ref{sec:querying}).

We consider two use cases: computing the handover-of-work social network between resources, and computing an ``artifact-centric'' directly-follows graph which distinguishes between different entity types.

\paragraph{Handover of Work}
We derived event \emph{:Class}es for \emph{Resource} as shown in Sect.~\ref{sec:mining:class}. Note that the \emph{:Class} nodes of type \emph{Resource} are semantically different from the \emph{:Entity} nodes of type \emph{Case\_R} created in Sect.~\ref{sec:storing:demo_bpic17}, although we have one node of each per \emph{e.resource} value.

We restricted the aggregation query of Sect.~\ref{sec:mining:df_c} to aggregate only the \emph{:DF} relationships of \emph{Case\_AWO} for \emph{:Class}es of type \emph{Resource}. We thereby obtain the Handover-of-Work social network; the \emph{count} property of the \emph{:DF} relationship describes how often a resource handed work (of a specific \emph{Case\_AWO}) to another resource. The aggregation query had an execution time of 8.954 seconds.

\begin{figure}
    \centering
    \includegraphics[width=.6\linewidth]{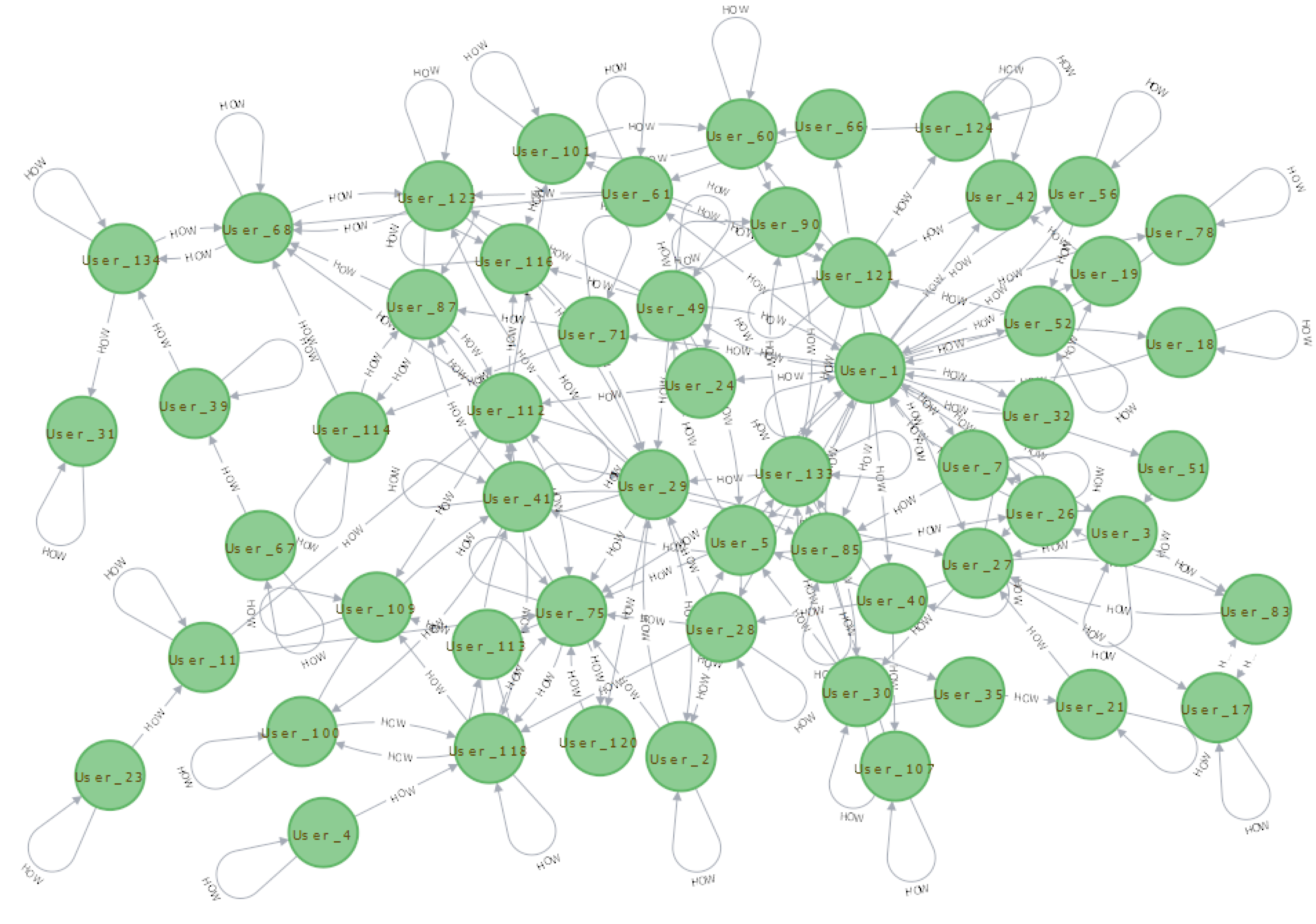}
    \caption{Handover of Work Network}
    \label{fig:HoW}
\end{figure}
We can retrieve this network with the query \emph{MATCH (c1) -[dfc:DF\_C]-$\mathord{>}$ (c2) WHERE c1.Type = ``Resource'' AND c2.Type = ``Resource'' AND dfc.EntityType = ``Case\_AWO''}. We verified the correctness of the query using the social network mining plugin of ProM (\url{www.promtools.org}), see~\cite{Esser2019cs_tue} for details. Figure~\ref{fig:HoW} shows the Neo4j graph output of the query above on a sample of 20 cases.

With traditional event logs, creating a handover of work network typically requires the use of a tool or programming language whereas Neo4j is capable of creating them by in-DB processing only.

\paragraph{Mining behavioral models over multiple entities}

In the same way an aggregated directly-follows graph can be obtained in-DB by aggregating \textit{:DF} relations. We aggregated the \emph{:DF} relationships of Application, Workflow, Offer, Case\_AO, Case\_WO, Case\_AW, Case\_AWO for event class \emph{Activity+Lifecycle}.

\begin{figure}
    \centering
    \includegraphics[height=8cm]{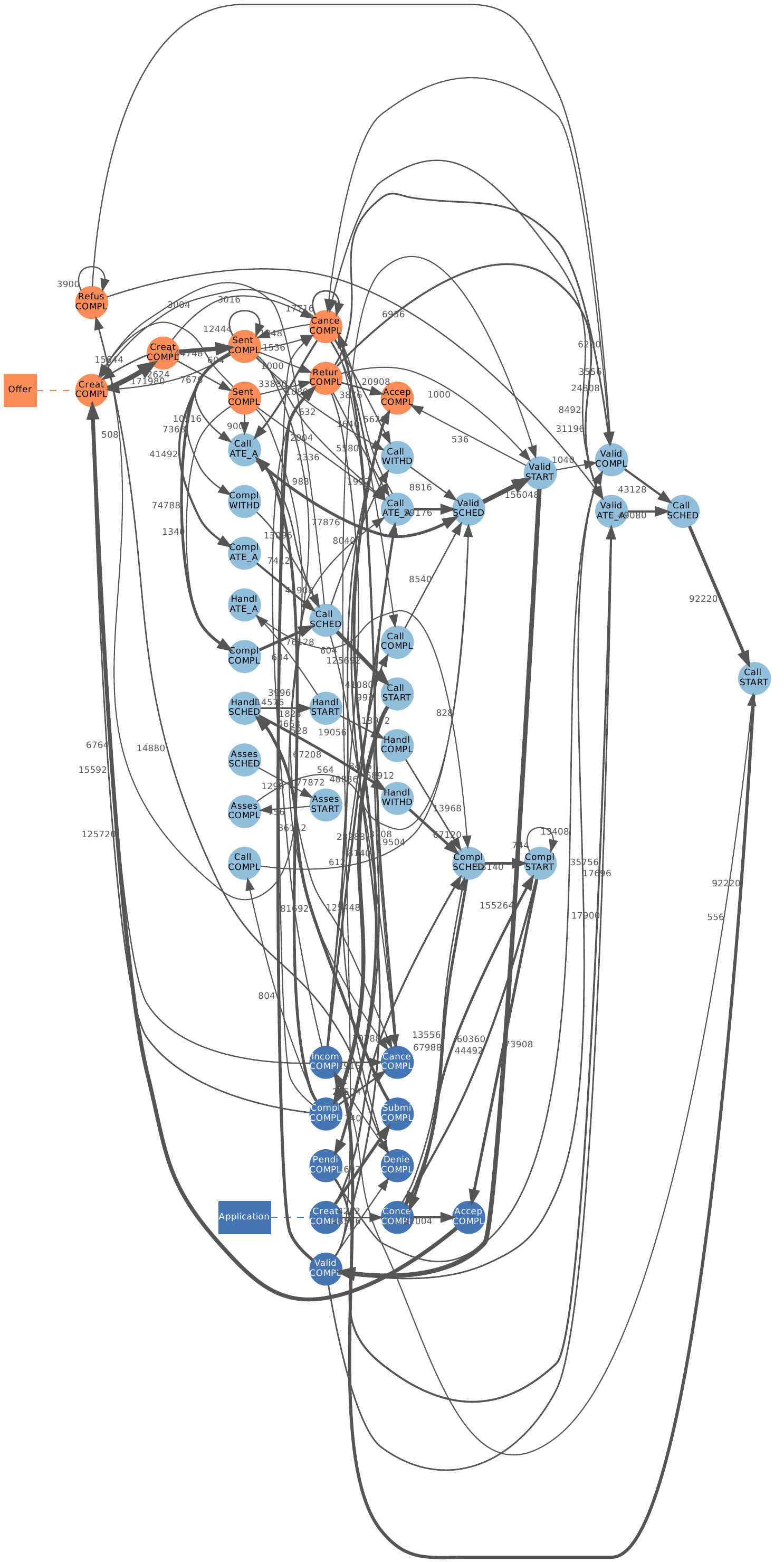}
    \includegraphics[height=8cm]{figures/bpic17_sm.pdf}
    \caption{Classical directly-follows graph (left) and process model discovered by Split Miner on classical event log of BPIC17~\cite{BPIC2017}.}
    \label{fig:mining:dfg}\label{fig:mining:split}
\end{figure}

Figure~\ref{fig:mining:dfg}(left) shows the classical directly-follows graph that can be obtained by aggregating all \emph{:DF} relationships for the global case notion of the original log (entity type \emph{Case\_AWO}). Each node is a \emph{:Class} node and each edge is a \emph{:DF\_C} relationship of type \emph{Case\_AWO}; we only queried relationships with $\mathit{count} \geq 500$. Figure~\ref{fig:mining:split}(right) shows a process model discovered by the Split Miner~\cite{DBLP:journals/kais/AugustoCDRP19} from the same event log; the model has a fitness of 95\%, i.e. cannot explain 5\% of the data.
However, both describe the behavior as a complex interleaving of steps of three different entities while the underlying log suffers from convergence and divergence, see Sect.~\ref{sec:background:multi-dim-req}.

\begin{figure}
    \centering
    \includegraphics[width=\linewidth]{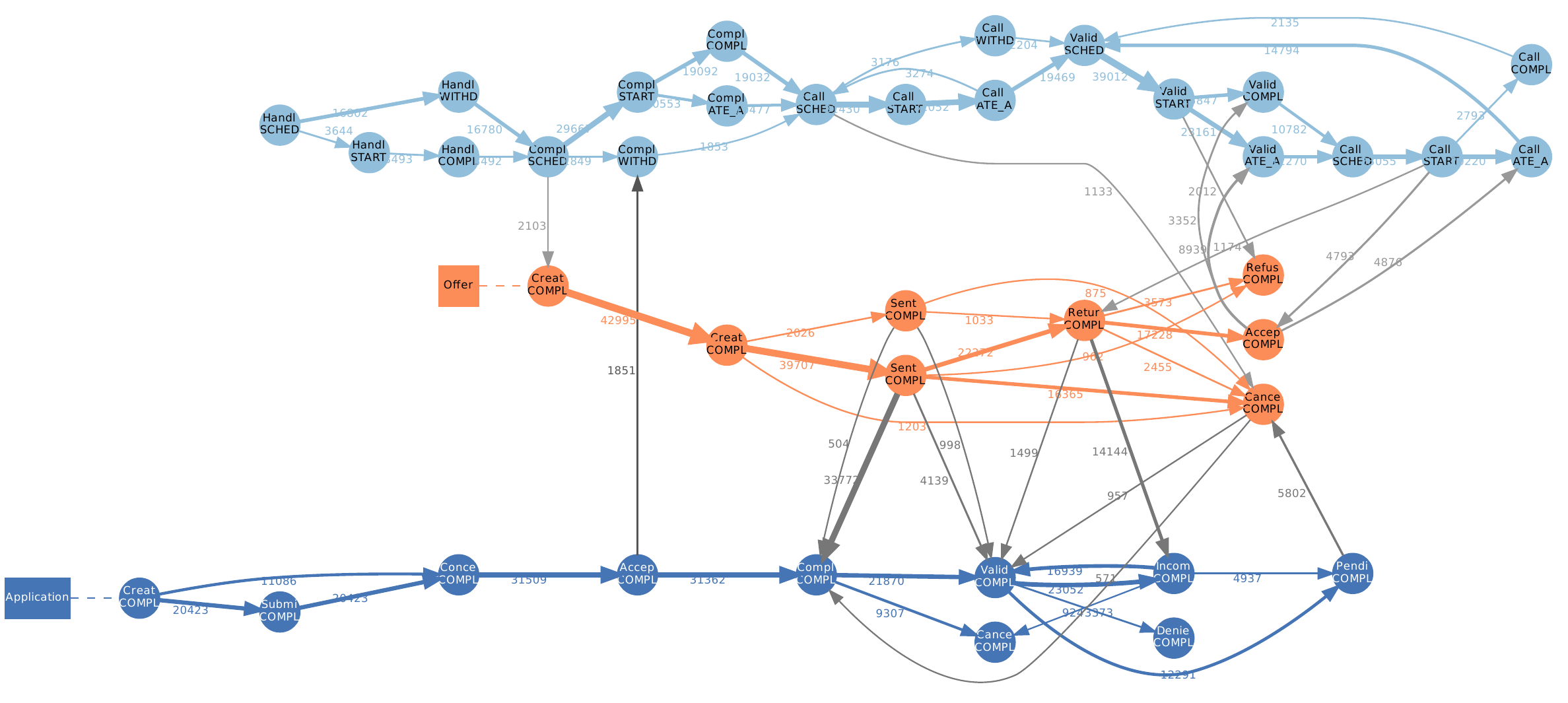}
    \caption{Entity-centric directly-follows graph for Application, Workflow, Offer, and their interactions.}
    \label{fig:mining:mvp}
\end{figure}

Figure~\ref{fig:mining:mvp} shows the graph we obtained by querying for the aggregated \emph{:DF\_C} relationships of types \emph{Application} (dark blue) \emph{Workflow} (light blue), \emph{Offer} (orange), and of the reified relations \emph{Case\_AO, Case\_WO, Case\_AW} (grey) with $\mathit{count} \geq 500$, i.e., $>98\%$ of all process executions. This graph describes directly-follows relations per entity and thus is similar to an artifact-centric model~\cite{DBLP:journals/tsc/LuNWF15} or a multiple viewpoint model~\cite{DBLP:conf/simpda/BertiA19}.
Compared to Fig.~\ref{fig:mining:dfg}, the graph of Fig.~\ref{fig:mining:mvp} explicitly describes the directly-follows behavior of each entity; the behavior of each entity is concurrent to the behavior of other entities up to the few explicit interactions shown by grey edges. In contrast, Fig.~\ref{fig:mining:dfg} shows few edges among the event classes of the same entity (Application, Workflow or Offer) and most edges in between event classes of different entities because the classical event log interleaved all events. The graph of Fig.~\ref{fig:mining:mvp} is significantly easier to understand and more precise as it was derived from data without convergence and divergence.

\section{Conclusion}\label{sec:conclusion}

We introduced a new data model for event data based on labeled property graphs. Our data model provides node types and relationship types (see Sect.~\ref{sec:represent}) with semantic constraints (see Sect.~\ref{sec:semantics}) for all first-class concepts of event logs: events, entities (generalizing the case notion), event classes (generalizing the activity and the resource attribute), and the directly-follows relation between events related to the same entity only, satisfying requirements R0, R1 and R3 of Sect.~\ref{sec:background:multi-dim-req}.
The semi-structured nature of graphs allowed us to represent multiple different, related entities (R2) and the relations between entities and events (R4) through dedicated correlation relationships. Thus, the data model can be seen as a \emph{multi-dimensional event log} or \emph{knowledge graph over events}, where events of each entity are ordered by ``their'' directly-follows relation leading to a \emph{partial order} of events. Our data model avoids all shortcomings of existing event data models including event tables, event logs, and relational databases, see Sect.~\ref{sec:background:multi-dim-literature}, while building on a standard data storage format. {It specifically generalizes over XES event logs~\cite{ieee_xes_standard} in supporting multiple perspectives on the event data from different case identifiers at once. The data model however has higher memory requirements and takes more time to load into a database than XES event log.}

We provided a succinct set of queries to efficiently convert data in event table format into our data model (see Sect.~\ref{sec:storing}. The queries are parameterized where user-provided domain-knowledge is required. We specifically provide queries to reify relations between entities into composite entities allowing to derive directly-follows relations describing interactions between entities (R13). The data model and queries allowed us to convert represent 5 different real-life dataset into our data model.

We demonstrated that the query language Cyper allows querying event data in our data model, see Sect.~\ref{sec:querying}. Queries and results are given as graphs, satisfying (R5). Queries Q4 and Q6 retrieve entire paths of events (R6) allowing to analyse the sequences. Q1-Q3 and Q6 select individual cases based on partial patterns (R7) allowing to ``query by example''. Q2,Q3,Q5 and Q6 query for temporal properties (R8) where Q5 specifically considers time; all queries correlate events related to a common entity (R9); Q7 queries aspects of multiple entities in the same query (R10) and allows to query behavior of multiple entities and combine results (R11). Altogether, we could demonstrate the queries over labeled property graphs satisfy R5-R11, which no existing query language on event data offers, see Sect.~\ref{sec:background:multi-dim-literature}.

Finally, we demonstrated that our data model and Cypher allow aggregating events to event classes (R12) and directly-follows relations to event classes per entity (R14,R15). The resulting graphs are simpler and describe the behavior more accurately than techniques using other data models, see Sect.~\ref{sec:mining}. \emph{The queries and data sets are publicly available for further research}.~\cite{graphdataset,BPIC2014_graph,BPIC2015_graph,BPIC2016_graph,BPIC2017_graph,BPIC2019_graph}.

The model has several limitations and requires further research. Our data model does not model properties of entities and semantics of relations between entities; practical applications require a more complete data model of this aspect as well. {Within the scope of this work, we only consider converting event tables to our data model whereas most event data is stored in relational databases; an automated techniques for conversion is desirable for practical adoption. Furthermore, we currently completely rely on domain knowledge when loading the data into the graph database system. It is an open question how to leverage available context information or properties of the data to (partly) automate this process. Service-oriented platforms for event data such as ProcessAtlas~\cite{DBLP:journals/spe/BeheshtiBM18} offer functionality that aid analysts in exploring the context information of the event data; a similar approach would be applicable for the proposed data model.}

When one event is correlated to multiple entities of the same type, then the current modeling of the directly-follows relation does not distinguish between different individual entities. As a result, queries become more complex and aggregations of directly-follows relations lose information about these multiplicities; further research is required to aggregate behavior that preserves multiplicities of entities in interactions.

Cypher is highly expressive but not specifically designed for querying event data\,---\,it takes expertise and patience to write the right queries; query patterns and best practices have to be established. While we demonstrated feasibility and obtained performance that allows for usage in practice, existing graph database systems are still significantly slower than relational databases or dedicated algorithms, specifically due to deficiencies in query optimization which may easily render queries practically infeasible. Further improvements on the performance of graph databases is required, possible specifically taking the partially-ordered nature of our data into account. {Moreover, existing behavioral query languages such as PQL~\cite{DBLP:journals/corr/abs-1909-09543} offer dedicated constructs for behavioral querying that are not available as primitives in Cypher. It is an open question whether PQL's behavioral constructs can be expressed in Cypher (see Sect.~\ref{sec:querying:expressive}) and which other higher-level primitives for querying multi-dimensional event data are desirable.}

Our model also enables new lines of research. Providing a more general standard event data model allows for development of new event data analysis and process mining techniques that explicitly consider the presence of multiple entities. The data format enables the adoption of knowledge graph and graph mining techniques for event data.

\begin{acknowledgements}
The results of this paper have been greatly influenced and shaped by discussions and inspirations over several years with Wil M.P. van der Aalst, Claudio di Ciccio, Marlon Dumas, Manuel Haug, Martin Klenk, Massimiliano de Leoni, Xixi Lu, Jan Mendling, Marco Montali, Alexander Rinke, and Stefan Sch\"{o}ning. The results would have been impossible without the careful preparation of the public BPI Challenge event data sets by Boudewijn van Dongen in preserving their multi-dimensional nature for this research, and without the regular availability of George Fletcher for introducing us to graph databases and providing insights into this field.
\end{acknowledgements}

\bibliographystyle{spmpsci}      
\bibliography{md-events-lpg}   

\begin{thebibliography}{10}
\providecommand{\url}[1]{{#1}}
\providecommand{\urlprefix}{URL }
\expandafter\ifx\csname urlstyle\endcsname\relax
  \providecommand{\doi}[1]{DOI~\discretionary{}{}{}#1}\else
  \providecommand{\doi}{DOI~\discretionary{}{}{}\begingroup
  \urlstyle{rm}\Url}\fi

\bibitem{ieee_xes_standard}
Ieee standard for extensible event stream (xes) for achieving interoperability
  in event logs and event streams.
\newblock IEEE Std 1849-2016 pp. 1--50 (2016)

\bibitem{van2016process}
Van~der Aalst, W.M.: Process mining: data science in action.
\newblock Springer (2016)

\bibitem{DBLP:conf/sefm/Aalst19}
van~der Aalst, W.M.P.: Object-centric process mining: Dealing with divergence
  and convergence in event data.
\newblock In: P.C. {\"{O}}lveczky, G.~Sala{\"{u}}n (eds.) Software Engineering
  and Formal Methods - 17th International Conference, {SEFM} 2019, Oslo,
  Norway, September 18-20, 2019, Proceedings, \emph{Lecture Notes in Computer
  Science}, vol. 11724, pp. 3--25. Springer (2019).
\newblock \doi{10.1007/978-3-030-30446-1\_1}.
\newblock \urlprefix\url{https://doi.org/10.1007/978-3-030-30446-1\_1}

\bibitem{DBLP:journals/cscw/AalstRS05}
van~der Aalst, W.M.P., Reijers, H.A., Song, M.: Discovering social networks
  from event logs.
\newblock Computer Supported Cooperative Work \textbf{14}(6), 549--593 (2005).
\newblock \doi{10.1007/s10606-005-9005-9}.
\newblock \urlprefix\url{https://doi.org/10.1007/s10606-005-9005-9}

\bibitem{DBLP:journals/sosym/AalstRVDKG10}
van~der Aalst, W.M.P., Rubin, V.A., Verbeek, H.M.W., van Dongen, B.F., Kindler,
  E., G{\"{u}}nther, C.W.: Process mining: a two-step approach to balance
  between underfitting and overfitting.
\newblock Software and Systems Modeling \textbf{9}(1), 87--111 (2010).
\newblock \doi{10.1007/s10270-008-0106-z}.
\newblock \urlprefix\url{https://doi.org/10.1007/s10270-008-0106-z}

\bibitem{DBLP:conf/sigmod/AnglesABBFGLPPS18}
Angles, R., Arenas, M., Barcel{\'{o}}, P., Boncz, P.A., Fletcher, G.H.L.,
  Gutierrez, C., Lindaaker, T., Paradies, M., Plantikow, S., Sequeda, J.F., van
  Rest, O., Voigt, H.: {G-CORE:} {A} core for future graph query languages.
\newblock In: G.~Das, C.M. Jermaine, P.A. Bernstein (eds.) Proceedings of the
  2018 International Conference on Management of Data, {SIGMOD} Conference
  2018, Houston, TX, USA, June 10-15, 2018, pp. 1421--1432. {ACM} (2018).
\newblock \doi{10.1145/3183713.3190654}.
\newblock \urlprefix\url{https://doi.org/10.1145/3183713.3190654}

\bibitem{DBLP:journals/tkde/AugustoCDRMMMS19}
Augusto, A., Conforti, R., Dumas, M., Rosa, M.L., Maggi, F.M., Marrella, A.,
  Mecella, M., Soo, A.: Automated discovery of process models from event logs:
  Review and benchmark.
\newblock {IEEE} Trans. Knowl. Data Eng. \textbf{31}(4), 686--705 (2019).
\newblock \doi{10.1109/TKDE.2018.2841877}.
\newblock \urlprefix\url{https://doi.org/10.1109/TKDE.2018.2841877}

\bibitem{DBLP:journals/kais/AugustoCDRP19}
Augusto, A., Conforti, R., Dumas, M., Rosa, M.L., Polyvyanyy, A.: Split miner:
  automated discovery of accurate and simple business process models from event
  logs.
\newblock Knowl. Inf. Syst. \textbf{59}(2), 251--284 (2019).
\newblock \doi{10.1007/s10115-018-1214-x}.
\newblock \urlprefix\url{https://doi.org/10.1007/s10115-018-1214-x}

\bibitem{DBLP:conf/ic3k/BaqueroM12a}
Baquero, A.V., Molloy, O.: Integration of event data from heterogeneous systems
  to support business process analysis.
\newblock In: {IC3K}, \emph{CCIS}, vol. 415, pp. 440--454. Springer (2012)

\bibitem{DBLP:journals/spe/BeheshtiBM18}
Beheshti, A., Benatallah, B., Motahari{-}Nezhad, H.R.: Processatlas: {A}
  scalable and extensible platform for business process analytics.
\newblock Softw. Pract. Exp. \textbf{48}(4), 842--866 (2018).
\newblock \doi{10.1002/spe.2558}.
\newblock \urlprefix\url{https://doi.org/10.1002/spe.2558}

\bibitem{DBLP:journals/dpd/BeheshtiBM16}
Beheshti, S., Benatallah, B., Motahari{-}Nezhad, H.R.: Scalable graph-based
  {OLAP} analytics over process execution data.
\newblock Distributed Parallel Databases \textbf{34}(3), 379--423 (2016).
\newblock \doi{10.1007/s10619-014-7171-9}.
\newblock \urlprefix\url{https://doi.org/10.1007/s10619-014-7171-9}

\bibitem{DBLP:conf/bpm/BeheshtiBNS11}
Beheshti, S., Benatallah, B., Nezhad, H.R.M., Sakr, S.: A query language for
  analyzing business processes execution.
\newblock In: {BPM} 2011, \emph{LNCS}, vol. 6896, pp. 281--297. Springer (2011)

\bibitem{DBLP:conf/simpda/BertiA19}
Berti, A., van~der Aalst, W.M.P.: Extracting multiple viewpoint models from
  relational databases.
\newblock In: P.~Ceravolo, M.~van Keulen, M.T.G. L{\'{o}}pez (eds.) Data-Driven
  Process Discovery and Analysis - 8th {IFIP} {WG} 2.6 International Symposium,
  {SIMPDA} 2018, Seville, Spain, December 13-14, 2018, and 9th International
  Symposium, {SIMPDA} 2019, Bled, Slovenia, September 8, 2019, Revised Selected
  Papers, \emph{Lecture Notes in Business Information Processing}, vol. 379,
  pp. 24--51. Springer (2020).
\newblock \doi{10.1007/978-3-030-46633-6\_2}.
\newblock \urlprefix\url{https://doi.org/10.1007/978-3-030-46633-6\_2}

\bibitem{DBLP:series/synthesis/2018Bonifati}
Bonifati, A., Fletcher, G.H.L., Voigt, H., Yakovets, N.: Querying Graphs.
\newblock Synthesis Lectures on Data Management. Morgan {\&} Claypool
  Publishers (2018).
\newblock \doi{10.2200/S00873ED1V01Y201808DTM051}.
\newblock \urlprefix\url{https://doi.org/10.2200/S00873ED1V01Y201808DTM051}

\bibitem{DBLP:journals/eswa/BottrighiCLMT16}
Bottrighi, A., Canensi, L., Leonardi, G., Montani, S., Terenziani, P.: Trace
  retrieval for business process operational support.
\newblock Expert Syst. Appl. \textbf{55}, 212--221 (2016)

\bibitem{ELT_10.14778/1687553.1687576}
Cohen, J., Dolan, B., Dunlap, M., Hellerstein, J.M., Welton, C.: Mad skills:
  New analysis practices for big data.
\newblock Proc. VLDB Endow. \textbf{2}(2), 1481–1492 (2009).
\newblock \doi{10.14778/1687553.1687576}.
\newblock \urlprefix\url{https://doi.org/10.14778/1687553.1687576}

\bibitem{DBLP:conf/sc/Cuevas-VicenttinDWSL12}
Cuevas{-}Vicentt{\'{\i}}n, V., Dey, S.C., Wang, M.L.Y., Song, T.,
  Lud{\"{a}}scher, B.: Modeling and querying scientific workflow provenance in
  the {D-OPM}.
\newblock In: 2012 {SC} Companion, pp. 119--128. {IEEE} Computer Society (2012)

\bibitem{DBLP:conf/icdt/DeutchM09}
Deutch, D., Milo, T.: {TOP-K} projection queries for probabilistic business
  processes.
\newblock In: {ICDT} 2009, \emph{{ACM} International Conference Proceeding
  Series}, vol. 361, pp. 239--251. {ACM} (2009)

\bibitem{DBLP:journals/dpd/DijkmanGSDGH20}
Dijkman, R.M., Gao, J., Syamsiyah, A., van Dongen, B.F., Grefen, P., ter
  Hofstede, A.H.M.: Enabling efficient process mining on large data sets:
  realizing an in-database process mining operator.
\newblock Distributed and Parallel Databases \textbf{38}(1), 227--253 (2020).
\newblock \doi{10.1007/s10619-019-07270-1}.
\newblock \urlprefix\url{https://doi.org/10.1007/s10619-019-07270-1}

\bibitem{BPIC2014}
van Dongen, B.: {BPI Challenge 2014. Dataset.}
\newblock
  \url{https://doi.org/10.4121/uuid:c3e5d162-0cfd-4bb0-bd82-af5268819c35}

\bibitem{BPIC2015}
van Dongen, B.: {BPI Challenge 2015. Dataset.}
\newblock
  \url{https://doi.org/10.4121/uuid:31a308ef-c844-48da-948c-305d167a0ec1}

\bibitem{BPIC2016}
van Dongen, B.: {BPI Challenge 2016. Dataset.}
\newblock
  \url{https://doi.org/10.4121/uuid:360795c8-1dd6-4a5b-a443-185001076eab}

\bibitem{BPIC2017}
van Dongen, B.: {BPI Challenge 2017. Dataset.}
\newblock
  \url{https://doi.org/10.4121/uuid:5f3067df-f10b-45da-b98b-86ae4c7a310b}

\bibitem{BPIC2018}
van Dongen, B.: {BPI Challenge 2018. Dataset.}
\newblock
  \url{https://doi.org/10.4121/uuid:3301445f-95e8-4ff0-98a4-901f1f204972}

\bibitem{BPIC2019}
van Dongen, B.: {BPI Challenge 2019. Dataset.}
\newblock
  \url{https://doi.org/10.4121/uuid:d06aff4b-79f0-45e6-8ec8-e19730c248f1}

\bibitem{Esser2019cs_tue}
Esser, S.: Using graph data structures for event logs.
\newblock Capita selecta research project., Eindhoven University of Technology
  (2019).
\newblock \url{https://doi.org/10.5281/zenodo.3333831}

\bibitem{BPIC2014_graph}
Esser, S., Fahland, D.: {Event Graph of BPI Challenge 2014. Dataset.}
\newblock \url{https://doi.org/10.4121/14169494}

\bibitem{BPIC2015_graph}
Esser, S., Fahland, D.: {Event Graph of BPI Challenge 2015. Dataset.}
\newblock \url{https://doi.org/10.4121/14169569}

\bibitem{BPIC2016_graph}
Esser, S., Fahland, D.: {Event Graph of BPI Challenge 2016. Dataset.}
\newblock \url{https://doi.org/10.4121/14164220}

\bibitem{BPIC2017_graph}
Esser, S., Fahland, D.: {Event Graph of BPI Challenge 2017. Dataset.}
\newblock \url{https://doi.org/10.4121/14169584}

\bibitem{BPIC2019_graph}
Esser, S., Fahland, D.: {Event Graph of BPI Challenge 2019. Dataset.}
\newblock \url{https://doi.org/10.4121/14169614}

\bibitem{esser2019storing}
Esser, S., Fahland, D.: Storing and querying multi-dimensional process event
  logs using graph databases.
\newblock In: C.D. Francescomarino, R.M. Dijkman, U.~Zdun (eds.) Business
  Process Management Workshops - {BPM} 2019 International Workshops, Vienna,
  Austria, September 1-6, 2019, Revised Selected Papers, \emph{Lecture Notes in
  Business Information Processing}, vol. 362, pp. 632--644. Springer (2019).
\newblock \doi{10.1007/978-3-030-37453-2\_51}.
\newblock \urlprefix\url{https://doi.org/10.1007/978-3-030-37453-2\_51}

\bibitem{graphdataset}
Esser, S., Fahland, D.: {Event Data and Queries for Multi-Dimensional Event
  Data in the Neo4j Graph Database (Version 1.0). Dataset.}
\newblock \url{https://doi.org/10.5281/zenodo.3865222} (2020)

\bibitem{DBLP:conf/apn/Fahland19}
Fahland, D.: Describing behavior of processes with many-to-many interactions.
\newblock In: S.~Donatelli, S.~Haar (eds.) Application and Theory of Petri Nets
  and Concurrency - 40th International Conference, {PETRI} {NETS} 2019, Aachen,
  Germany, June 23-28, 2019, Proceedings, \emph{Lecture Notes in Computer
  Science}, vol. 11522, pp. 3--24. Springer (2019).
\newblock \doi{10.1007/978-3-030-21571-2\_1}.
\newblock \urlprefix\url{https://doi.org/10.1007/978-3-030-21571-2\_1}

\bibitem{francis2018cypher}
Francis, N., Green, A., Guagliardo, P., Libkin, L., Lindaaker, T., Marsault,
  V., Plantikow, S., Rydberg, M., Selmer, P., Taylor, A.: Cypher: An evolving
  query language for property graphs.
\newblock In: Management of Data, pp. 1433--1445. ACM (2018)

\bibitem{Gonzalez_2019_pm_db_phd}
{Gonzalez Lopez de Murillas}, E.: Process mining on databases: extracting event
  data from real-life data sources.
\newblock Ph.D. thesis, Department of Mathematics and Computer Science (2019).
\newblock Proefschrift

\bibitem{DBLP:conf/icde/HuangBDMY15}
Huang, X., Bao, Z., Davidson, S.B., Milo, T., Yuan, X.: Answering regular path
  queries on workflow provenance.
\newblock In: {ICDE} 2015, pp. 375--386. {IEEE} Computer Society (2015)

\bibitem{DBLP:conf/bpm/JansS17}
Jans, M., Soffer, P.: From relational database to event log: Decisions with
  quality impact.
\newblock In: {BPM} 2017 Workshops, \emph{LNBIP}, vol. 308, pp. 588--599.
  Springer (2017)

\bibitem{DBLP:conf/caise/LiMCA18}
Li, G., de~Murillas, E.G.L., de~Carvalho, R.M., van~der Aalst, W.M.P.:
  Extracting object-centric event logs to support process mining on databases.
\newblock In: J.~Mendling, H.~Mouratidis (eds.) Information Systems in the Big
  Data Era - CAiSE Forum 2018, Tallinn, Estonia, June 11-15, 2018, Proceedings,
  \emph{Lecture Notes in Business Information Processing}, vol. 317, pp.
  182--199. Springer (2018).
\newblock \doi{10.1007/978-3-319-92901-9\_16}.
\newblock \urlprefix\url{https://doi.org/10.1007/978-3-319-92901-9\_16}

\bibitem{Liu:2009:SEC:1944968.1944974}
Liu, D., Pedrinaci, C., Domingue, J.: Semantic enabled complex event language
  for business process monitoring.
\newblock In: 4th Int. Workshop on Semantic Business Process Management, pp.
  31--34 (2009)

\bibitem{DBLP:conf/bpm/LuFA14}
Lu, X., Fahland, D., van~der Aalst, W.M.P.: Conformance checking based on
  partially ordered event data.
\newblock In: F.~Fournier, J.~Mendling (eds.) Business Process Management
  Workshops - {BPM} 2014 International Workshops, Eindhoven, The Netherlands,
  September 7-8, 2014, Revised Papers, \emph{Lecture Notes in Business
  Information Processing}, vol. 202, pp. 75--88. Springer (2014).
\newblock \doi{10.1007/978-3-319-15895-2\_7}.
\newblock \urlprefix\url{https://doi.org/10.1007/978-3-319-15895-2\_7}

\bibitem{DBLP:journals/tsc/LuNWF15}
Lu, X., Nagelkerke, M., van~de Wiel, D., Fahland, D.: Discovering interacting
  artifacts from {ERP} systems.
\newblock {IEEE} Trans. Services Computing \textbf{8}(6), 861--873 (2015)

\bibitem{MARINORTEGA2014667}
Marín-Ortega, P.M., Dmitriyev, V., Abilov, M., Gómez, J.M.: Elta: New
  approach in designing business intelligence solutions in era of big data.
\newblock Procedia Technology \textbf{16}, 667 -- 674 (2014).
\newblock \doi{https://doi.org/10.1016/j.protcy.2014.10.015}.
\newblock
  \urlprefix\url{http://www.sciencedirect.com/science/article/pii/S2212017314002424}

\bibitem{bpaf_standard}
zur Muehlen, M.: Workflow management coalition - business process analytics
  format specification.
\newblock Tech. rep., WfMC (2009)

\bibitem{DBLP:conf/bpm/MurillasHR17}
de~Murillas, E.G.L., Hoogendoorn, G.E., Reijers, H.A.: Redo log process mining
  in real life: Data challenges {\&} opportunities.
\newblock In: E.~Teniente, M.~Weidlich (eds.) Business Process Management
  Workshops - {BPM} 2017 International Workshops, Barcelona, Spain, September
  10-11, 2017, Revised Papers, \emph{Lecture Notes in Business Information
  Processing}, vol. 308, pp. 573--587. Springer (2017).
\newblock \doi{10.1007/978-3-319-74030-0\_45}.
\newblock \urlprefix\url{https://doi.org/10.1007/978-3-319-74030-0\_45}

\bibitem{DBLP:conf/bpm/MurillasRA16}
de~Murillas, E.G.L., Reijers, H.A., van~der Aalst, W.M.P.: Everything you
  always wanted to know about your process, but did not know how to ask.
\newblock In: BPM Workshops, \emph{LNBIP}, vol. 281, pp. 296--309 (2016)

\bibitem{DBLP:journals/sosym/MurillasRA19}
de~Murillas, E.G.L., Reijers, H.A., van~der Aalst, W.M.P.: Connecting databases
  with process mining: a meta model and toolset.
\newblock Software and System Modeling \textbf{18}(2), 1209--1247 (2019)

\bibitem{DBLP:conf/bpm/PegoraroUA19}
Pegoraro, M., Uysal, M.S., van~der Aalst, W.M.P.: Discovering process models
  from uncertain event data.
\newblock In: C.D. Francescomarino, R.M. Dijkman, U.~Zdun (eds.) Business
  Process Management Workshops - {BPM} 2019 International Workshops, Vienna,
  Austria, September 1-6, 2019, Revised Selected Papers, \emph{Lecture Notes in
  Business Information Processing}, vol. 362, pp. 238--249. Springer (2019).
\newblock \doi{10.1007/978-3-030-37453-2\_20}.
\newblock \urlprefix\url{https://doi.org/10.1007/978-3-030-37453-2\_20}

\bibitem{DBLP:journals/corr/abs-1909-09543}
Polyvyanyy, A., ter Hofstede, A.H.M., Rosa, M.L., Ouyang, C., Pika, A.: Process
  query language: Design, implementation, and evaluation.
\newblock CoRR \textbf{abs/1909.09543} (2019).
\newblock \urlprefix\url{http://arxiv.org/abs/1909.09543}

\bibitem{DBLP:journals/is/PolyvyanyyPH20}
Polyvyanyy, A., Pika, A., ter Hofstede, A.H.M.: Scenario-based process querying
  for compliance, reuse, and standardization.
\newblock Inf. Syst. \textbf{93}, 101563 (2020).
\newblock \doi{10.1016/j.is.2020.101563}.
\newblock \urlprefix\url{https://doi.org/10.1016/j.is.2020.101563}

\bibitem{DBLP:conf/apn/PolyvyanyyWCRH14}
Polyvyanyy, A., Weidlich, M., Conforti, R., Rosa, M.L., ter Hofstede, A.H.M.:
  The 4c spectrum of fundamental behavioral relations for concurrent systems.
\newblock In: G.~Ciardo, E.~Kindler (eds.) Application and Theory of Petri Nets
  and Concurrency - 35th International Conference, {PETRI} {NETS} 2014, Tunis,
  Tunisia, June 23-27, 2014. Proceedings, \emph{Lecture Notes in Computer
  Science}, vol. 8489, pp. 210--232. Springer (2014).
\newblock \doi{10.1007/978-3-319-07734-5\_12}.
\newblock \urlprefix\url{https://doi.org/10.1007/978-3-319-07734-5\_12}

\bibitem{DBLP:journals/ijcis/PopovaFD15}
Popova, V., Fahland, D., Dumas, M.: Artifact lifecycle discovery.
\newblock Int. J. Cooperative Inf. Syst. \textbf{24}(1), 1550001:1--1550001:44
  (2015).
\newblock \doi{10.1142/S021884301550001X}.
\newblock \urlprefix\url{https://doi.org/10.1142/S021884301550001X}

\bibitem{DBLP:conf/otm/RaimCMMM14}
R{\"{a}}im, M., Ciccio, C.D., Maggi, F.M., Mecella, M., Mendling, J.: Log-based
  understanding of business processes through temporal logic query checking.
\newblock In: OTM, \emph{LNCS}, vol. 8841, pp. 75--92. Springer (2014)

\bibitem{robinson2013graph}
Robinson, I., Webber, J., Eifrem, E.: Graph databases.
\newblock O'Reilly Media (2013)

\bibitem{DBLP:conf/caise/SchonigRCJM16}
Sch{\"{o}}nig, S., Rogge{-}Solti, A., Cabanillas, C., Jablonski, S., Mendling,
  J.: Efficient and customisable declarative process mining with {SQL}.
\newblock In: S.~Nurcan, P.~Soffer, M.~Bajec, J.~Eder (eds.) Advanced
  Information Systems Engineering - 28th International Conference, CAiSE 2016,
  Ljubljana, Slovenia, June 13-17, 2016. Proceedings, \emph{Lecture Notes in
  Computer Science}, vol. 9694, pp. 290--305. Springer (2016).
\newblock \doi{10.1007/978-3-319-39696-5\_18}.
\newblock \urlprefix\url{https://doi.org/10.1007/978-3-319-39696-5\_18}

\bibitem{DBLP:conf/edoc/SongWWWTK11}
Song, L., Wang, J., Wen, L., Wang, W., Tan, S., Kong, H.: Querying process
  models based on the temporal relations between tasks.
\newblock In: {EDOCW} 2011, pp. 213--222. {IEEE} Computer Society (2011)

\bibitem{DBLP:conf/simpda/SyamsiyahDA16b}
Syamsiyah, A., van Dongen, B.F., van~der Aalst, W.M.P.: {DB-XES:} enabling
  process discovery in the large.
\newblock In: P.~Ceravolo, C.~Guetl, S.~Rinderle{-}Ma (eds.) Data-Driven
  Process Discovery and Analysis - 6th {IFIP} {WG} 2.6 International Symposium,
  {SIMPDA} 2016, Graz, Austria, December 15-16, 2016, Revised Selected Papers,
  \emph{Lecture Notes in Business Information Processing}, vol. 307, pp.
  53--77. Springer (2016).
\newblock \doi{10.1007/978-3-319-74161-1\_4}.
\newblock \urlprefix\url{https://doi.org/10.1007/978-3-319-74161-1\_4}

\bibitem{DBLP:journals/information/TangMS18}
Tang, Y., Mackey, I., Su, J.: Querying workflow logs.
\newblock Information \textbf{9}(2), 25 (2018)

\bibitem{DBLP:journals/is/WeerdtBVB12}
Weerdt, J.D., Backer, M.D., Vanthienen, J., Baesens, B.: A multi-dimensional
  quality assessment of state-of-the-art process discovery algorithms using
  real-life event logs.
\newblock Inf. Syst. \textbf{37}(7), 654--676 (2012).
\newblock \doi{10.1016/j.is.2012.02.004}.
\newblock \urlprefix\url{https://doi.org/10.1016/j.is.2012.02.004}

\bibitem{DBLP:journals/tsc/WernerG15}
Werner, M., Gehrke, N.: Multilevel process mining for financial audits.
\newblock {IEEE} Trans. Serv. Comput. \textbf{8}(6), 820--832 (2015).
\newblock \doi{10.1109/TSC.2015.2457907}.
\newblock \urlprefix\url{https://doi.org/10.1109/TSC.2015.2457907}

\end{thebibliography}

\appendix

\section{{False positives and false negatives in the directly-follows relation in BPIC17}}
\label{app:df_bpic17}

We calculated the total number of directly-follows pairs $(e_A,e_B)$ in the BPIC17~\cite{BPIC2017} data set for all events related to an offer (activity name starts with ``O\_''), denoted as ``A df B for all events'' in Tab.~\ref{tab:df_false_edges_bpic17}.

We calculated the total number of directly-follows pairs $(e_A,e_B)$ in the BPIC17 data set for events related to an offer (activity name starts with ``O\_'') where $e_A$ and $e_B$ refer to \emph{different} offers; denoted as ``A df B for events of different offers'' in Tab.~\ref{tab:df_false_edges_bpic17}. These are ``false positive'' directly-follows pairs as they do not describe a correct behavioral relation for an entity.

Choosing the offer id as case identifier and projecting the data to events related to an offer only, we calculated the total number of directly-follows pairs $(e_A,e_B)$ per offer; denoted as ``A df B local per offer'' in Tab.~\ref{tab:df_false_edges_bpic17}. This is the ground truth of the directly-following events wrt.\ the offer entities.

Table~\ref{tab:df_false_edges_bpic17}(1) shows how many directly-follows pairs in the data (``A df B for all events'') are \emph{false positives} (``A df B for events of different offers''), i.e., directly-follows pairs in the data which do not exist in reality (as they relate events of different offers).

Table~\ref{tab:df_false_edges_bpic17}(2) shows the share of ground truth directly-follows pairs in the data (``A df B local per offer'') over the directly-follows pairs in the sequential event log (``A df B for all events''). The discrepancy between both can be interpreted as \emph{false negatives}, i.e., directly-follows pairs which exist but have not been reported in the data.

\begin{table}
    \centering
    \includegraphics[angle=90,scale=0.6]{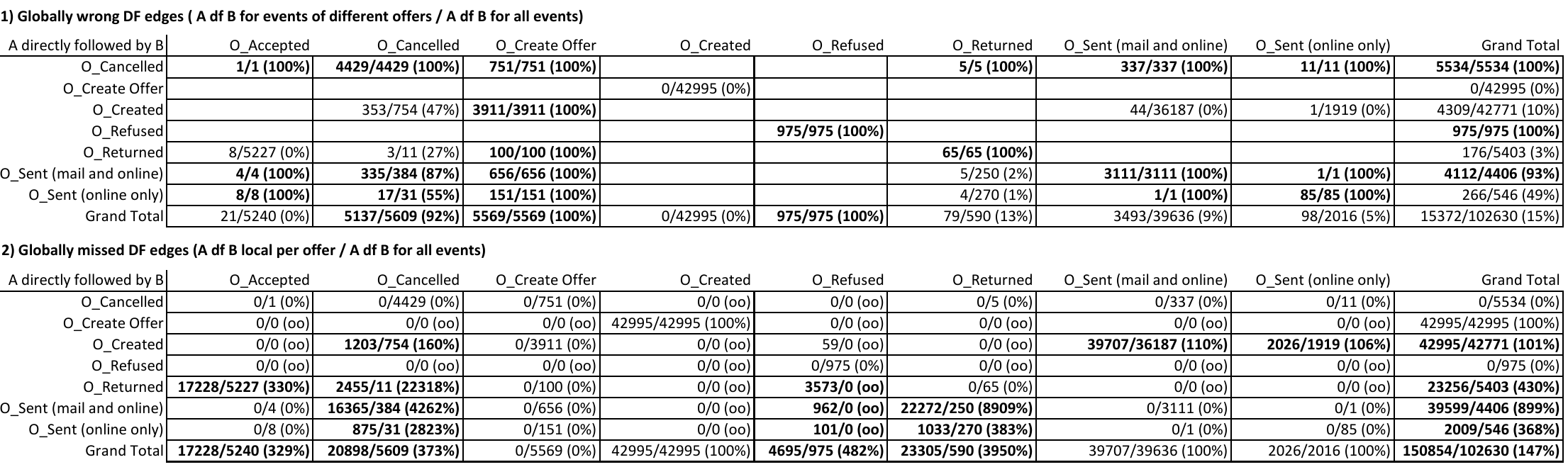}
    \caption{{Share of globally wrong directly-follows pairs in sequential event log of BPIC17 (1) and share of missing directly-follows pairs in sequential event log of BPIC17~\cite{BPIC2017} (2)}}
    \label{tab:df_false_edges_bpic17}
\end{table}

We observe several directly-follows pairs are reported always wrong, e.g., \emph{O\_Cancelled}-\emph{O\_Cancelled}, \emph{O\_Cancelled}-\emph{O\_Create Offer}, \emph{O\_Cancelled}-\emph{O\_Sent (mail and online)}, \emph{O\_Refused}-\emph{O\_Refused}, whereas other directly-follows pairs are always missing, e.g., \emph{O\_Returned}-\emph{O\_Refused}.

Overall, 15\% of all directly-follows pairs are false positive, with some activities having a 100\% false positive ratio. Further, the actual directly-follows relation has 47\% more directly-follows pairs than represented in the sequential event log, where some activities show up to 39 times more directly-follows pairs than in the sequential event log (\emph{O\_Returned}).

\end{document}